\documentclass{article}

\usepackage[english]{babel}

\usepackage[a4paper,top=5cm,bottom=4.5cm,left=4cm,right=4cm,marginparwidth=1.75cm]{geometry}

\usepackage{amsmath}
\usepackage{amsthm}
\usepackage{amsfonts}
\usepackage{amssymb}
\usepackage{bm}
\usepackage{mathrsfs}
\usepackage{graphicx}
\usepackage{comment}
\usepackage{xcolor}
\usepackage{stackengine} 
\usepackage{authblk} 

\theoremstyle{plain}
\newtheorem{theorem}{Theorem}[section]
\newtheorem{proposition}[theorem]{Proposition}
\newtheorem{corollary}[theorem]{Corollary}
\newtheorem{lemma}[theorem]{Lemma}

\theoremstyle{definition}
\newtheorem{definition}[theorem]{Definition}

\theoremstyle{remark}
\newtheorem{remark}[theorem]{Remark}

\newcommand{\R}{\mathbb R}
\newcommand{\N}{\mathbb N}
\renewcommand{\P}{\mathbb P}
\newcommand{\E}{\mathbb E}
\newcommand{\Rd}{\mathbb R^d}
\newcommand{\Ra}{\mathbb R^{d_1}}
\newcommand{\Rb}{\mathbb R^{d_2}}
\renewcommand{\d}{\mathrm d}
\newcommand{\nablax}{\nabla_{\!1}}
\newcommand{\nablay}{\nabla_{\!2}}
\newcommand{\laplax}{\nabla_{\!1}^2}
\newcommand{\laplay}{\nabla_{\!2}^2}

\newcommand{\epss}{\eta}
\newcommand{\Veff}{F}
\newcommand{\Vb}{V_\textup{b}}
\newcommand{\Vc}{V_\textup{c}}
\newcommand{\Veffb}{\Veff_\textup{b}}
\newcommand{\Veffc}{\Veff_\textup{c}}

\newcommand{\oast}{\stackMath\mathbin{\stackinset{c}{0ex}{c}{0ex}{\ast}{\bigcirc}}}
\newcommand{\Bigcirc}{\raisebox{-0.5mm}{\scalebox{1.35}{$\bigcirc$}}}
\newcommand{\oastb}{\stackMath\mathbin{\stackinset{c}{0ex}{c}{0ex}{\ast_2}{\Bigcirc}}}

\DeclareMathOperator{\LFP}{L*\!}
\DeclareMathOperator{\Ldual}{L}

\DeclareMathOperator{\supp}{supp}
\DeclareMathOperator{\osc}{osc}
\DeclareMathOperator{\mineigen}{min\;spec}

\DeclareMathOperator{\Ldualx}{L_1}
\DeclareMathOperator{\Ldualy}{L_2}
\DeclareMathOperator{\Ldualyt}{\langle L_2\rangle_\textit{t}}
\DeclareMathOperator{\Ldualys}{\langle L_2\rangle_\textit{s}}
\DeclareMathOperator{\M}{M}
\DeclareMathOperator{\I}{\mathcal I}

\DeclareMathOperator{\Hess}{Hess}
\DeclareMathOperator{\Hessx}{Hess_{11}}
\DeclareMathOperator{\Hessy}{Hess_{22}}
\DeclareMathOperator{\Hessxy}{Hess_{12}}

\DeclareMathOperator{\D}{D}
\DeclareMathOperator{\Da}{D_1}
\DeclareMathOperator{\Db}{D_2}
\DeclareMathOperator{\DKL}{D_\textup{KL}}
\DeclareMathOperator{\Di}{D_I}
\DeclareMathOperator{\Dai}{D_{1I}}
\DeclareMathOperator{\Dbi}{D_{2I}}

\DeclareMathOperator{\rhoK}{\rho_{\!_\textit{\scriptsize*}}}
\DeclareMathOperator{\rhoKa}{\rho_{\!_\textit{\scriptsize*}}^\textup{\tiny(1)}}
\DeclareMathOperator{\rhoKb}{\rho_{\!_\textit{\scriptsize*}}^\textup{\tiny(2)}}
\DeclareMathOperator{\rhot}{\rho_{\textit{t},\lambda}}
\DeclareMathOperator{\rhota}{\rho_{\textit{t},\lambda}^\textup{\tiny(1)}}
\DeclareMathOperator{\rhotb}{\rho_{\textit{t},\lambda}^\textup{\tiny(2)}}
\DeclareMathOperator{\rhoG}{\rho_\textup{G}}
\DeclareMathOperator{\rhoa}{\rho^\textup{\tiny(1)}}
\DeclareMathOperator{\rhob}{\rho^\textup{\tiny(2)}}
\DeclareMathOperator{\rhoT}{\rho_{\textit{t}}}
\DeclareMathOperator{\rhoTa}{\rho_{\textit{t}}^\textup{\tiny(1)}}
\DeclareMathOperator{\rhoTb}{\rho_{\textit{t}}^\textup{\tiny(2)}}
\DeclareMathOperator{\rhos}{\rho_{\textit{s}}}

\DeclareMathOperator{\rhosb}{\rho_{\textit{s}}^\textup{\tiny(2)}}
\DeclareMathOperator{\rhoS}{\rho_{\textit{s},\lambda}}
\DeclareMathOperator{\rhoSa}{\rho_{\textit{s},\lambda}^\textup{\tiny(1)}}
\DeclareMathOperator{\rhoSb}{\rho_{\textit{s},\lambda}^\textup{\tiny(2)}}

\DeclareMathOperator{\rhoi}{\rho_\textup{I}}
\DeclareMathOperator{\rhoia}{\rho_\textup{I}^\textup{\tiny(1)}}
\DeclareMathOperator{\rhoib}{\rho_\textup{I}^\textup{\tiny(2)}}

\newcommand{\mui}{\mu_\textup{I}}
\newcommand{\mutb}{\mu_\textit{t}^\textup{\tiny(2)}}
\newcommand{\musb}{\mu_\textit{s}^\textup{\tiny(2)}}
\newcommand{\muib}{\mu_\textup{I}^\textup{\tiny(2)}}

\title
{On the convergence to the non-equilibrium steady state 
of a Langevin dynamics with widely separated time scales and different temperatures}
\author[1]{Diego Alberici}
\author[2]{Nicolas Macris} 
\author[3]{Emanuele Mingione}
\affil[1]{Universit\`a dell'Aquila}
\affil[2]{\'Ecole Polytechnique F\'ed\'erale de Lausanne}
\affil[3]{Universit\`a di Bologna}

\begin{document}
\maketitle

\begin{abstract}

We study the solution of the two-temperatures Fokker-Planck equation and rigorously analyse its convergence towards an explicit non-equilibrium stationary measure for long time and two widely separated time scales. The exponential rates of convergence are estimated assuming the validity of  logarithmic Sobolev inequalities for the conditional and marginal distributions of the limit measure.
We show that these estimates are sharp in the exactly solvable case of a quadratic potential.
We discuss a few examples where the logarithmic Sobolev inequalities are satisfied through simple, though not optimal, criteria.
In particular we consider a spin-glass model with slowly varying external magnetic fields whose non-equilibrium measure 
corresponds to Guerra's hierarchical construction appearing in Talagrand's proof of the Parisi formula.

\end{abstract}

\section*{Introduction}

Langevin dynamics can be generalized to many time scales, satisfying a fluctuation-dissipation relation with a different temperature for each time scale. In this paper we consider the simplest incarnation of this \textit{multi-scale} (or \textit{multi-bath}) model, that is  a pair of Langevin equations in the overdamped limit
\begin{equation}\label{eq:SDEintro}
    \begin{cases}
    \,\lambda_1\,\d x_1 \,=\, -\nablax V(x_1,x_2)\,\d t \,+\, \sqrt{2 D_1}\,\d W_1 \\[6pt]
    \,\lambda_2\,\d x_2 \,=\, -\nablay V(x_1,x_2)\,\d t \,+\, \sqrt{2 D_2}\,\d W_2
    \end{cases}
\end{equation}
where $\lambda_1, \lambda_2 >0$ are friction coefficients that set two time scales, 
$-\nablax V(x_1,x_2)$, $-\nablay V(x_1,x_2)$ are driving forces arising from a \textit{confining} potential $V(x_1,x_2)$ coupling two families of degrees of freedom $x\equiv(x_1,x_2)\in\Ra\times\Rb$, and $W_1,\,W_2$ are independent Wiener processes on $\Ra,\,\Rb$ with diffusion coefficients $D_1,\,D_2>0\,$ respectively.
We adjust the friction and diffusion coefficients so that the fluctuation-dissipation relation with temperatures $\beta_1^{-1},\,\beta_2^{-1}>0$ holds for each degree of freedom separately. In other words we set
\begin{equation}\label{eq:SDEintro_diffusion}
   D_1 \equiv\, \beta_1^{-1}\lambda_1 \,,\quad
   D_2 \equiv\, \beta_2^{-1}\lambda_2 \,.
\end{equation}
By a standard application of Ito calculus, the probability density function $\rho=\rho_t(x_1,x_2)$ describing the system configuration at time $t$ verifies the following Fokker-Planck (FP) equation:
\begin{equation} \label{eq:FP_intro}
    \partial_t\rho \,=\, \frac{1}{\lambda_1}\; \nablax\cdot\left( \frac{1}{\beta_1}\nablax\rho \,+\, \rho\,\nablax  V \right) \,+\, \frac{1}{\lambda_2}\; \nablay\cdot\left( \frac{1}{\beta_2}\nablay\rho \,+\, \rho\,\nablay  V \right) \;.
\end{equation}
The goal of the present work is to study the probability distribution $\rho_t$ of the random vector $(x_1(t),x_2(t))$, with particular interest in its long time behavior  ($t\to\infty$) when the two time scales are widely separated ($\lambda_1/\lambda_2\to 0$), namely  when the $x_1$ variables thermalise much more quickly than the $x_2$ variables. We  mention that there exists a large literature on the analysis of the stochastic dynamics \eqref{eq:SDEintro} in the case  of two widely separated time scales for \textit{finite times} (see books \cite{FW,Pavliotis} and references therein for a rigorous approach using techniques of \textit{averaging} and \textit{homogeneization}), nevertheless we are not aware that these results cover the long-time behaviour and the stationary measure studied here.


For $\beta_1=\beta_2\equiv\beta$ the stationary solution of \eqref{eq:FP_intro} does not depend on $\lambda_1,\,\lambda_2\,$, since it is just the equilibrium measure $e^{-\beta\,V(x)}/Z$ where $Z\equiv\int\d x\, e^{-\beta\,V(x)}$.
For $\beta_1\neq\beta_2$ instead, the stationary solution of \eqref{eq:FP_intro} exists and does depend on $\lambda_1,\,\lambda_2\,$. In general its expression is unknown (except for exactly solvable cases such as quadratic potentials) but it admits an explicit limit $\rhoK$ when $\lambda_2\gg\lambda_1\,$. 
Higher order corrections to the measure $\rhoK$ would also be interesting to investigate. 
Let us summarize the heuristic argument \cite{Allahverdyan_2000pre, Contucci_2019} for the derivation of the limit non-equilibrium measure $\rhoK$. 
The $x_1$ variables quickly equilibrate on a time scale 
$t=O(\lambda_1)\ll\lambda_2$
so that the $x_2$ variables appear as quenched, thus on this time scale the conditional distribution $\rhoTa(x_1\vert x_2)$ tends to $\rhoKa(x_1\vert x_2) \equiv e^{-\beta_1 V(x_1,x_2)}/ Z_1(x_2)$ where $Z_1(x_2) \equiv \int \d x_1\,e^{-\beta_1 V(x_1,x_2)}$. For $t\gg \lambda_1$, the $x_2$ variables are subject to the average force under the measure $\rhoKa(x_1\vert x_2)$, that is
\begin{equation}
-\langle \nablay V(x_1, x_2)\rangle_{\rhoKa}\,=\, \beta_1^{-1}\,\nablay \log Z_1(x_2) \;.
\end{equation}
The associated \textit{effective potential} is $\Veff(x_2) \equiv -\beta_1^{-1} \log Z_1(x_2)$, therefore on a time scale $t=O(\lambda_2)$ the marginal distribution $\rhoTb(x_2)$ tends 
to 
$\rhoKb(x_2) \equiv e^{- \beta_2 \Veff(x_2)}/Z_2$ where $Z_2 \equiv \int \d x_2\, e^{- \beta_2 \Veff(x_2)}$. 
Putting pieces together this argument suggests that for $t\gg \lambda_2\gg\lambda_1$
\begin{equation}\label{stat}
    \rhoT(x_1,x_2) \,\to\, \rhoK(x_1,x_2) \equiv \rhoKa(x_1\vert x_2) \rhoKb(x_2) = \frac{e^{-\beta_1 V(x_1,x_2)}\,\big(\!\int \d x_1'\, e^{-\beta_1 V(x_1',x_2)}\big)^{\frac{\beta_2}{\beta_1}-1}}{\int \d x_2'\,  \big(\!\int \d x_1'\, e^{-\beta_1 V(x_1',x_2')}\big)^{\frac{\beta_2}{\beta_1}}}
\end{equation}
The measure $\rhoK(x)$ obtained for $t\gg\lambda_2\gg\lambda_1$ differs from the equilibrium measure $e^{-\beta\,V(x)}/Z$ unless $\beta_1=\beta_2=\beta\,$.
%
%
Notice that the change of variables $(t,\lambda_1,\lambda_2)\mapsto \big(\tfrac{t}{\lambda_1},\,1,\,\tfrac{\lambda_2}{\lambda_1}\big)$ does not affect the solution of equation \eqref{eq:FP_intro}, therefore we will assume without loss of generality
\begin{equation} \label{eq:lambda}
    \lambda_1 \equiv 1 \ ,\quad \lambda_2 \equiv \lambda
\end{equation}
and denote by $\rhot$ the time-dependent solution of \eqref{eq:FP_intro}.

The aim of the present paper is to provide a rigorous framework for the convergence \eqref{stat} under general hypothesis on the potential $V$. We base our analysis on \textit{logarithmic Sobolev inequalities} (LSI's), giving a method to estimate the exponential rates of convergence. The results are presented as follows.

In Section 1 a convergence result is stated under simplified hypothesis on the potential. For example, let us take $V$ as the sum of a strongly convex polynomial (of arbitrary degree) and any compactly supported smooth function.
The Kullback-Leibler divergence $\DKL(\rhot \Vert \rhoK)$ vanishes as $t/\lambda\to +\infty$, $\lambda \to +\infty$ and thanks to Pinsker's inequality the same holds true for the total variation distance $\Vert\rhot - \rhoK\Vert_\textup{TV}$.
More precisely, the KL divergence decomposes as $\DKL(\rhot \Vert \rhoK)=\Da(t, \lambda) + \Db(t, \lambda)\,$, where $\Da(t,\lambda)$ is the KL divergence between the conditional distributions $\rhota(\cdot\vert x_2)\,$, $\rhoKa(\cdot\vert x_2)$ averaged w.r.t. $\rhotb\,$, and $\Db(t,\lambda)$ is the KL divergence between the marginal distributions $\rhotb\,$, $\rhoKb$.
Separate estimates for $\Da(t, \lambda)$ and $\Db(t, \lambda)$ are  provided, as - a part from residuals of order $\lambda^{-1}$ - $\Da(t, \lambda)$ decays exponentially fast on a time scale of order $1$, while $\Db(t, \lambda)$ decays exponentially fast on a much larger time scale of order $\lambda\,$. 

In Section 2 the main Theorem \ref{th:main} generalises the hypothesis on $V$ and provides a method to estimate the exponential rates of convergence of $\Da(t,\lambda),\,\Db(t,\lambda)\,$. The potential $V$ is supposed to be smooth, with derivatives of any order having polynomial growth, and to satisfy suitable \textit{confining assumptions}. Assuming both the ``conditional potential'' $x_1\mapsto \beta_1 V(x_1,x_2)$ and the effective potential $\beta_2 F(x_2)$ (or more precisely the related measures $\rhoKa(x_1|x_2),\,\rhoKb(x_2)$) verify \textit{logarithmic Sobolev inequalities}, we can estimate the rates of convergence for $\Da(t,\lambda),\,\Db(t,\lambda)\,$ by the optimal logarithmic Sobolev constants.

In general, a probability measure $p_\infty(x) \equiv e^{-\beta\,U(x)}/Z$ verifies a LSI if there exists a constant $C>0$ such that for every suitably regular probability measure $\pi(x)$
\begin{equation} \label{eq:LSI_intro}
    \int\d x\;\pi(x)\,\log\frac{\pi(x)}{p_\infty(x)} \;\leq\; \frac{1}{2C}\, \int\d x\;\pi(x)\,\Big|\nabla\log\frac{\pi(x)}{p_\infty(x)}\Big|^2 \;.
\end{equation}
Logarithmic Sobolev inequalities are commonly used to analyse the approach to equilibrium of the one-temperature Fokker-Planck equation
\begin{equation} \label{eq:FP1_intro}
    \partial_t\,p_t(x) \,=\, \nabla\cdot \Big(\,\frac{1}{\beta}\,\nabla p_t(x) \,+\, p_t(x)\,\nabla U(x) \Big) \;.
\end{equation}
Taking as $\pi$ the time-dependent solution $p_t$ of \eqref{eq:FP1_intro}, the l.h.s. of \eqref{eq:LSI_intro} is the Kullback-Leibler divergence between $p_t$ and $p_\infty$, while the r.h.s. turns out to be $-\frac{\beta}{2C}\,\frac{\d}{\d t}\DKL(p_t\|p_\infty)\,$. Therefore \eqref{eq:LSI_intro} provides a first order differential inequality for $\DKL(p_t\|p_\infty)$, called \textit{free energy dissipation inequality}, which can be easily integrated giving exponential rate of convergence to equilibrium equal to $\frac{2C}{\beta}$. There exists a vast literature on LSI's, e.g., see the review \cite{MarkVillani} adopting this dynamical point of view.

The rest of Section 2 is devoted to the derivation of a simple sufficient criterion that guarantees the validity of LSI's for both measures $\rhoKa$ and $\rhoKb$. If $V$ is the sum of a strongly convex function and a bounded one, the classical Bakry-\'Emery-Holley-Stroock criterion provides a LSI for the measure $e^{-\beta\,V}/Z\,$. We show, by means of the Brascamp-Lieb concentration inequality \cite{BL,Brascamp2002A}, that the same condition on $V$ guarantees the validity of LSI's both for $\rhoKa$ and $\rhoKb$.
Unless the potential $V$ is strictly convex, the convergence rate provided by the Bakry-\'Emery-Holley-Stroock criterion may be exponentially slow in the size of the system (i.e., dimensions $d_1,\,d_2$). Investigating for further criteria for the two LSI's of interest is an important problem that we postpone to future work since it requires to focus on a specific potential.
%

In Section \ref{sec:examples} we analyse some examples where Theorem \ref{th:main} applies. First (Subsection \ref{subsec:gaussian}) we study the exactly solvable case of a quadratic potential $V$: the multi-bath Langevin dynamics boils down to a multi-dimensional Ornstein-Uhlenbeck process. The non-equilibrium stationary measure as well as interesting properties such as the associated rotational current densities have been discussed in detail in \cite{Dotsenko2013} (for any $\lambda$ and dimensions $d_1=d_2=1$). Here we provide explicit formulas for $\rho_{t, \lambda}$ and its approach to $\rho_*$ in the case $d_1 = d_2 =1$, showing that the convergence rates estimated by Theorem \ref{th:main} are sharp.
The second example (Subsection \ref{subsec:spinglass}) is a spin glass model where both the spin degrees of freedom and the external fields are given a dynamics in order to obtain the Guerra's measure \cite{Guerra_2003} in the spirit of \cite{Contucci_2019, Contucci_2021}.
The third example (Subsection \ref{subsec:inference}) comes from a paradigm in modern high-dimensional inference, namely the problem of estimating a rank-one matrix which is observed through a noisy channel (see \cite{Davenport-Romberg-2016} for a general review on the subject and a discussion of applications).

Section 4 is devoted to the proof of Theorem \ref{th:main}.
Integrating the Fokker-Planck equation \eqref{eq:FP_intro} w.r.t. $x_1$, one can see that marginal measure $\rhotb(x_2)$ satisfies a suitable Fokker-Planck equation itself (Subsection \ref{subsec:FPmarg}).
We use the two LSI's for the conditional and marginal measures $\rhoKa$, $\rhoKb$ to deduce \textit{dissipation inequalities} for the conditional and marginal KL divergences $\Da(t,\lambda)$, $\Db(t,\lambda)$ (Theorems \ref{th:conditional}, \ref{th:marginal}). Unlike the standard FP equation \eqref{eq:FP1_intro}, in the two-scales case LSI's are not enough to control the long-time behaviour, since additional terms appear in the time derivatives $\frac{\d}{\d t}\Da(t,\lambda)$, $\frac{\d}{\d t}\Db(t,\lambda)$ revealing the dependence of the stationary measure on $\lambda$ (unless $\beta_1=\beta_2$).
In order to control these additional terms we show that the measures $\rhot(x_1,x_2)$ and $\rhoKa(x_1|x_2)\rhotb(x_2)$ have uniformly bounded moments in $t,\,\lambda$ (Subsection \ref{subsec:expectation_poly}). These uniform bounds are obtained thanks to the confining assumptions on $V$ (see Appendix, Subsections \ref{subsec:proof_expectations}, \ref{subsec:proof_condexpectations}).
Finally, the core computations based on a careful application of integration by parts and Pinsker inequality are shown in Subsections \ref{subsec:KLderivatives}, \ref{subsec:convrho1}, \ref{subsec:convrho2}. In order to get a first idea of the proof, the reader can start from the latter subsections.

The Appendix (Subsections \ref{subsec:proof_expectations}, \ref{subsec:proof_condexpectations}) contains the proof of the uniform bounds on moments 
which is based on repeated use of Gronwall inequalities. 
The most technical aspects of the proof of Theorem \ref{th:main} are also postponed to the Appendix (Subsection \ref{subsec:proof_KLdiss}). In addition, the regularity of the solution $\rhot$ and its marginal $\rhotb$ are proven in Subsections \ref{subsec:proof_regularity}, \ref{subsec:proof_regularity2}.

\subsection*{Physical motivations} Dynamical systems containing fast and slow degrees of freedom are ubiquitous in physics. Actually in many physical problems the first issue is how to split the dynamics in fast and slow components and then construct an effective model by a reduction procedure also called \textit{adiabatic elimination}. In  a multi-bath dynamics like \eqref{eq:SDEintro}, such a division comes from the very definition of the model, in other words one has to assume that the above reduction has already taken place.

A different, to some extent, interpretation of   \eqref{eq:SDEintro} comes from Statistical Mechanics of disordered systems. In the context of spin glasses multi-scale models have been introduced  by Horner \cite{Horner1984A, Horner1984B}, Coolen, Penney, Sherrington \cite{Penney_1993, Coolen_1993}, Dotsenko, Franz, M\'ezard \cite{Dotsenko_1994}, and Allahverdyan, Nieuwenhuizen, Saakian \cite{Allahverdyan_1998, Allahverdyan_2000} as a mean to regularize the long-time behaviour of mean-field spin glasses dynamics, by giving the couplings a very slow dynamics instead of taking them \textit{quenched}. The dynamical system \eqref{eq:SDEintro} is studied per se in \cite{Allahverdyan_2000pre} and the stationary measure is heuristically derived in the limit of widely separated time scales (as well as corrections to this limit for particular potentials). In \cite{1999cond.mat.11086C} it is shown that the resulting (non-Gibbsian) stationary measure can be used to characterize out-of-equilibrium systems in the limit of small entropy production, displaying a natural interpretation in terms of effective temperatures.  

The concept of effective temperatures  in non-equilibrium systems was first discussed by Hohenberg and Shraiman \cite{Hohenberg1989ChaoticBO} for turbulent flows, by Cugliandolo, Kurchan and Peliti \cite{1999cond.mat.11086C, PhysRevE.55.3898} for spin glasses (see also \cite{Exartier}), and more recently for jamming \cite{PhysRevE.61.5464}. In these contexts the notion of effective temperature for each time-scale  arises through ratios of correlation and response functions generalizing the usual fluctuation-dissipation relation. We refer the interested reader to \cite{Crisanti_2003, Cugliandolo_2011,Bouchaud_1997, Maes20,Vulpiani17} for comprehensive reviews of theories of fluctuation-dissipation relations in non-equilibrium systems.

There are not many instances where one can compute stationary measures in out-of-equilibrium Statistical Mechanics. We mention that much work has been devoted to transport problems in extended systems coupled to many reservoirs at different temperatures (we refer to the review \cite{Derrida_2007} and references therein for this realm). The multi-bath model treated here is an easier problem that does not involve spatial transport (or heat flux), nevertheless it is of interest that it is amenable to an explicit solution.


Finally, we highlight the fact that the  heuristic argument used to obtain \eqref{stat} generalises to more than two families of degrees of freedom with widely separated time scales and different temperatures \cite{Contucci_2021} and the analogous of expression \eqref{stat} for $\rhoK$ is obtained through a hierarchy of conditional measures. As it turns out, the resulting measure appears in 
Guerra's interpolation scheme  \cite{Guerra_2003} which leads to Talagrand's proof \cite{Talapaper} of the celebrated Parisi formula \cite{1980,mezard1987spin} for the Sherringhton-Kirkpatrick model. 
Dynamical aspects of the connection between multi-bath models and Guerra's scheme have been recently investigated in   \cite{Contucci_2021}. In Example \ref{subsec:spinglass}  we give (non-optimal) criteria to rigorously establish this connection.

\section{Definitions and main result under very simple hypothesis} \label{sec:main_simple}
Let $V\!:\Ra\!\times\Rb\to\R$ be a potential coupling two groups of degrees of freedom denoted by $x\equiv(x_1,x_2)\in\Ra\!\times \Rb\equiv \R^d\,$.
In order to facilitate reading, in this section we assume that the potential has the simple form
\begin{equation}
    V = \Vc + \Vb
\end{equation}
where $\Vc$ is a strongly convex polynomial and $\Vb$ is a bounded smooth function with bounded derivatives of any order.
The reader may keep in mind the following example
\begin{equation}
    V(x) = x\cdot A\,x \,+\, \Vb(x)
\end{equation}
where $A$ is a positive definite matrix and $\Vb$ is any smooth modification on a compact support.
The hypothesis on the potential $V$ will be generalized in the next section.

Let $\beta_1,\beta_2,\lambda>0$. We want to study the time evolution of the system according to the SDE \eqref{eq:SDEintro} with diffusion coefficients \eqref{eq:SDEintro_diffusion} through a rigorous analysis of its probability flux $\rho:\Rd\times(0,\infty)\to(0,\infty)$ which is described by the following Fokker-Planck PDE
\begin{equation} \label{eq:FP}
\begin{cases}
    \;\partial_t\rhoT(x) \,=\, \nablax\cdot\left( \frac{1}{\beta_1}\nablax\rhoT(x) + \rhoT(x)\,\nablax  V(x) \right) \,+\, \frac{1}{\lambda}\, \nablay\cdot\left( \frac{1}{\beta_2}\nablay\rhoT(x) + \rhoT(x)\,\nablay  V(x) \right) \\[4pt]
    \;\rho_t \,\xrightarrow[t\to0_+]{L^1(\Rd)}\, \rhoi
\end{cases}
\end{equation}
We assume that the initial probability measure at time $t=0$ has a density $\rhoi$ with respect to the Lebesgue measure. Moreover we assume that $\rhoi$ is bounded, has bounded marginal density w.r.t. $x_2$, and finite moments of any order. 
All along the paper we will denote by $\rhoT(x)$ or $\rhot(x)$ the solution of \eqref{eq:FP}.

As explained in the Introduction, we are interested in the long time behaviour of the system when the families of degrees of freedom $x_1,\,x_2$ evolve on widely separated time scales ($x_2$ much slower than $x_1$). Let us introduce the following probability density $\rhoK:\Rd=\Ra\!\times\Rb\to(0,\infty)$
\begin{equation} \label{eq:rhoK}
\begin{split}
    \rhoK(x_1,x_2) \;&\equiv\; \frac{e^{-\beta_1 V(x_1,x_2)}}{Z_1(x_2)}\;\frac{e^{-\beta_2 F(x_2)}}{Z_2} \\[4pt]
    &=\, e^{-\beta_1 V(x_1,x_2)}\; Z_1(x_2)^{\frac{\beta_2}{\beta_1}-1}\, Z_2^{-1} \\[4pt]
    &=\, e^{-\beta_1 V(x_1,x_2)}\; e^{-(\beta_2-\beta_1) F(x_2)}\; Z_2^{-1}
\end{split}
\end{equation}
where
\begin{align} \label{eq:Z1}
    Z_1(x_2) \,&\equiv\, \int_{\Ra} e^{-\beta_1 V(x_1,x_2)}\,\d x_1 \\[4pt] \label{eq:Veff}
    \Veff(x_2) \,&\equiv\, -\frac{1}{\beta_1}\log Z_1(x_2) \\[4pt] \label{eq:Z2}
    Z_2 \,&\equiv\, \int_{\Rb} e^{-\beta_2 F(x_2)}\,\d x_2 \,=\, \int_{\Rb} Z_1(x_2)^{\frac{\beta_2}{\beta_1}}\,\d x_2 \;.
\end{align}
The first way of writing $\rhoK$ in  \eqref{eq:rhoK} corresponds to the product of its conditional and marginal probability densities:
\begin{equation}
    \rhoK(x_1,x_2) \,=\, \rhoKa(x_1|x_2)\,\rhoKb(x_2)
\end{equation}
where the conditional measure is the \textit{Gibbs measure} with potential $x_1\mapsto V(x_1,x_2)$ and inverse temperature $\beta_1$,
and the marginal measure is the \textit{Gibbs measure} with \textit{effective potential} $\Veff$ and inverse temperature $\beta_2$.
Notice that $\Veff(x_2)$ coincides with the \textit{free energy at equilibrium} of the \textit{fast} degrees of freedom for fixed $x_2$.

We will estimate the rate of convergence of $\rhot$ to $\rhoK$ studying the Kullback-Leibler (KL) divergence between the two measures. More precisely, we will analyse conditional and marginal KL-divergences separately. We set
\begin{equation} \label{eq:KL}
\begin{split}
    \D(t,\lambda) \,&\equiv\, \int_{\R^d} \rhot(x)\,\log\frac{\rhot(x)}{\rhoK(x)}\;\d x \\
    &=\, \int_{\R^d} \rhot(x)\;\log\frac{\rhota(x_1|x_2)}{\rhoKa(x_1|x_2)}\;\d x \,+\,
    \int_{\Rb} \rhotb(x_2)\;\log\frac{\rhotb(x_2)}{\rhoKb(x_2)}\;\d x_2 \\
    &\equiv\, \Da(t,\lambda) \,+\, \Db(t,\lambda)
\end{split}
\end{equation}
where 
\begin{equation}
    \rhot(x_1,x_2) \,=\, \rhota(x_1|x_2)\,\rhotb(x_2) 
\end{equation}
is the decomposition of $\rhot$ into conditional and marginal probability densities.
If $\rhot$ is replaced by $\rhoi$ into \eqref{eq:KL}, we denote the corresponding KL-divergences by $\Di,\Dai,\Dbi$ respectively.
Now we can state
\begin{theorem}[Main result under simple hypothesis] \label{th:main_simple}
Let $\beta_1,\beta_2>0$. Let $V=\Vc+\Vb$, where $\Vc$ is a strongly convex polynomial and $\Vb$ is smooth, bounded, with bounded derivatives of any order.
Let $\rhoi$ be a bounded probability density having finite moments of any order and bounded marginal density.\\
Then there exist positive constants $c_1=c_1(\beta_1,V)$, $c_2=c_2(\beta_2,V)$ such that:
\begin{itemize}
    \item[i)] for all $t>0\,$, $\lambda\geq1$
        \begin{equation} \label{eq:main_conditional}
            \Da(t,\lambda) \;\leq\; \Dai\;e^{-2c_1t} \,+\, R_1(t)\;\frac{1}{\lambda}
        \end{equation}
        where the remainder $R_1(t)$ is uniformly bounded in $t>0\,$;
    \item[ii)] for all $t>0\,$,
        $\lambda\geq\max(1, c_2/c_1)$, $\eta\in(0,1)\,$, $\epsilon>0$
        \begin{equation} \label{eq:main_marginal}
            \Db(t,\lambda) \;\leq\; \Dbi\;e^{-(1-\eta) 2c_2 \,t/\lambda} \,+\, R_2(t,\lambda,\eta)\; \frac{1}{\eta\,\epsilon\,\lambda} \,+\, R_3(t,\lambda,\eta)\,\frac{\epsilon}{\eta}
        \end{equation}
        where the remainders $R_2(t,\lambda,\eta),\,R_3(t,\lambda,\eta)$ are uniformly bounded in $t>0\,$, $\lambda>\max(1, c_2/c_1)\,$, $\eta\in(0,\frac{1}{2}]\,$.
    \end{itemize}
\end{theorem}

\begin{remark}
By inequalities \eqref{eq:main_conditional}, \eqref{eq:main_marginal} two different time scales emerge in the convergence of conditional and marginal KL-divergences:
\begin{align} \begin{split}
    &\Da(t,\lambda) \rightarrow 0 \quad\textrm{as }\ t\to\infty\,,\ \lambda\to\infty \\[4pt]
    &\Db(t,\lambda) \rightarrow 0 \quad\text{as }\  t\,\lambda^{-1}\to\infty\,,\ \lambda\to\infty \;.
\end{split} \end{align}
As a consequence, $\D(t,\lambda) \to 0$ as $t\,\lambda^{-1}\to\infty$ and $\lambda\to\infty\,$.
\end{remark}

\begin{remark}
The convergence in total variation distance can be deduced by Csisz\'ar-Kullback-Pinsker inequality: 
\begin{equation}
    \|\rhot-\rhoK\|_\textup{TV} \,\leq\, \sqrt{2 \D(t,\lambda)} \,=\, \sqrt{2}\,\sqrt{\Da(t,\lambda)+\Db(t,\lambda)} \;.
\end{equation}
\end{remark}

\begin{remark}
Let us consider the \textit{two-temperatures free energy functional} introduced in \cite{Allahverdyan_2000pre}:
\begin{equation}
    \mathcal F_{\beta_1,\beta_2}(\pi) \,\equiv\, \int_{\Rd} \pi(x)\,\Big(V(x) + \beta_1^{-1}\log\pi^\textup{\tiny(1)}(x_1|x_2) + \beta_2^{-1}\log\pi^\textup{\tiny(2)}(x_2)\Big)\,\d x
\end{equation}
for every probability density $\pi$ on $\Ra\times\Rb$ with conditional and marginal densities $\pi^\textup{\tiny(1)}$, $\pi^\textup{\tiny(2)}$ respectively.
It is easy to check that
\begin{equation}
     \mathcal F_{\beta_1,\beta_2}(\rhot) - \mathcal F_{\beta_1,\beta_2}(\rhoK) \,=\, \beta_1^{-1}\,\Da(t,\lambda) \,+\, \beta_2^{-1}\,\Db(t,\lambda)
\end{equation}
hence Theorem \ref{th:main_simple} immediately gives the convergence of $\mathcal F_{\beta_1,\beta_2}(\rhot)$.
\end{remark}

In fact, Theorem \ref{th:main_simple} is a particular case of the general result presented in the next section.

\section{Generalisation of the main result and logarithmic Sobolev inequalities} \label{sec:mainresults}

The upper bounds stated in the previous section hold true under more general hypothesis.
In fact, Theorem \ref{th:main_simple} is a particular case of a more general result based on \textit{logarithmic Sobolev inequalities}. In this section we will state the general result in Theorem \ref{th:main}, then we will show how to derive Theorem \ref{th:main_simple}.

We assume that the potential $V\!:\Ra\!\times\Rb\to\R$ is smooth, has polynomial growth together with all its derivatives, and is \textit{confining} in the following sense:
\begin{itemize}
    \item[(A1)] $V\in C^\infty(\R^d)\,$;
    \item[(A2)] for every multi-index $\nu\in\N^d$ there exist constants (depending on $\nu$) $r_1,r_2,\gamma_0,\gamma_1,\gamma_2\in[0,\infty)$ such that for all $x=(x_1,x_2)\in\R^d$
    \begin{equation}
    |D^\nu V(x)|\,\leq\,\gamma_0+\gamma_1\,|x_1|^{r_1}+\gamma_2\,|x_2|^{r_2}
    \end{equation}
    where $D^\nu$ is the usual multi-index notation for partial derivatives with respect to $x$;
    \item[(A3)] there exist $a_1,a_2\in(0,\infty)$, $a_0\in[0,\infty)$ such that for all $x\in\R^d$
    \begin{equation}
        V(x) \,\geq\, a_1\,|x_1|^2 \,+\, a_2\,|x_2|^2 \,- a_0 \;;
    \end{equation}
    \item[(A4)] there exist $a_1,a_2\in(0,\infty)$, $a_0\in[0,\infty)$ such that for all $x\in\R^d$
    \begin{equation}
        x\cdot \nabla V(x) \,\geq\, a_1\,|x_1|^2 \,+\, a_2\,|x_2|^2 \,- a_0 \;;
    \end{equation}
    \item[(A5)] there exist $a_1\in(0,\infty)$, $a_0,a_2,p\in[0,\infty)$ such that for all $x\in\R^d$
    \begin{equation}
        x_1\cdot \nablax V(x) \,\geq\, a_1\,|x_1|^2 \,-\, a_2\,|x_2|^p \,- a_0 \;.
    \end{equation}
\end{itemize}

\begin{remark} \label{rem:eVintegrability}
Assumption \textup{(A3)} guarantees that for every $r\in[0,\infty)$
\begin{align} \begin{split}
    &\int_{\Ra} |x_1|^r\, e^{-\beta_1 V(x_1,x_2)}\,\d x_1 \,<\infty \\
    &\int_{\Rb} |x_2|^r \left(\int_{\Ra} e^{-\beta_1 V(x_1,x_2)}\,\d x_1\right)^{\!\frac{\beta_2}{\beta_1}} \d x_2 \,< \infty
\end{split} \end{align}
hence the probability density $\rhoK$ introduced in \eqref{eq:rhoK} is suitably normalized and has finite moments of any order.
\end{remark}

We further assume that \textit{logarithmic Sobolev inequalities} for the conditional and marginal measures $\rhoKa$, $\rhoKb$ hold true. Precisely for all $\beta_1,\beta_2>0$ there exist finite positive constants $c_1=c_1(\beta_1,V)\,$, $c_2=c_2(\beta_2,V)$ such that 
\begin{itemize}
    \item[(LS1)]
    \begin{equation}
        \int_{\Ra}\!\pi_1(x_1)\,\log\frac{\pi_1(x_1)}{\rhoKa(x_1|x_2)}\,\d x_1 \,\leq\, \frac{1}{2\beta_1 c_1}\,
        \int_{\Ra}\!\pi_1(x_1)\,\Big|\nablax\log\frac{\pi_1(x_1)}{\rhoKa(x_1|x_2)}\Big|^2\,\d x_1
    \end{equation}
    for every $x_2\in\Rb$, for every positive probability density $\pi_1\in C^1(\Ra)$ with 
    $\pi_1\log\frac{\pi_1}{\rhoKa(\cdot|x_2)}\in L^1(\Ra)\,$;
    \item[(LS2)]
    \begin{equation}
        \int_{\Rb}\!\pi_2(x_2)\,\log\frac{\pi_2(x_2)}{\rhoKb(x_2)}\,\d x_2 \,\leq\, \frac{1}{2\beta_2 c_2 }\,
        \int_{\Rb}\!\pi_2(x_2)\,\Big|\nablay\log\frac{\pi_2(x_2)}{\rhoKb(x_2)}\Big|^2\,\d x_2
    \end{equation}
    for every positive probability density $\pi_2\in C^1(\Rb)$ with 
    $\pi_2\log\frac{\pi_2}{\rhoKb}\in L^1(\Rb)\,$.
\end{itemize}

Finally we assume that the initial probability measure $\mui$ describing the configuration of the system at time $t=0$ has density $\rhoi:\R^d\to[0,\infty)$ with respect to the Lebesgue measure. Let us write its decomposition into conditional and marginal probability densities
\begin{equation}
    \rhoi(x_1,x_2) \,=\, \rhoia(x_1|x_2)\,\rhoib(x_2)
\end{equation}
and assume the following conditions:
\begin{itemize}
    \item[(B1)] $\rhoi$ is bounded on $\Rd\,$;
    \item[(B2)] $\rhoib$ is bounded on $\Rb\,$;
    \item[(B3)] for every $r\in[0,\infty)$
    \begin{equation}
       \int_{\R^d}\rhoi(x)\,|x|^r\,\d x\,<\infty \;.
    \end{equation}
\end{itemize}

\begin{remark} \label{rem:integrability_logI}
The initial measure $\mui$ can vanish in some regions, but has to be absolutely continuous with respect to the Lebesgue measure in order to have finite Kullback-Leibler divergence from $\rhoK$. 
In fact, assumptions \textup{(B1)-(B3)} imply that $\rhoi,\,\rhoib$ have finite differential entropies:
\begin{equation}
    \int_{\R^d} \rhoi(x)\;|\log\rhoi(x)|\,\d x \,<\infty\;,\quad
    \int_{\Rb} \rhoib(x_2)\;\big|\log\rhoib(x_2)\big|\,\d x_2 \,<\infty
\end{equation}
(one can mimic the last part of the Proof of Proposition \ref{prop:integrability_log} in Subsection \ref{subsec:proof_expectations}). Therefore the integrals defining KL-divergences $\Di,\Dai,\Dbi$ are absolutely convergent.
From the results stated in Subsections \ref{subsec:expectation_poly}, \ref{subsec:FPmarg} it will be clear that the integrals defining $\D(t,\lambda),\Da(t,\lambda),\Db(t,\lambda)$ are also absolutely convergent.
\end{remark}

We can finally state 

\begin{theorem}[Main result under general hypothesis] \label{th:main}
    Let $\beta_1,\beta_2>0\,$. Suppose that $V$ verifies assumptions \textup{(A1)-(A5)}, logarithmic Sobolev inequalities  \textup{(LS1), (LS2)} hold true, and $\rhoi$ verifies assumptions \textup{(B1)-(B3)}.\\[2pt]
    Then the bounds \eqref{eq:main_conditional}, \eqref{eq:main_marginal} for $\Da,\,\Db$ hold true and the estimated rates of convergence $c_1,\,c_2$ are the optimal constants verifying the logarithmic Sobolev inequalities \textup{(LS1)}, \textup{(LS2)} respectively. 
    Moreover, the remainders are 
    \begin{align*} \label{eq:remainders} \begin{split}
    & R_1(t) \,\equiv\, \left(1-e^{-2c_1t}\right)\,\frac{c_0}{2c_1} \\[4pt]
    &R_2\!\left(t,\lambda,\epss\right) \,\equiv\, \big(1-e^{-(1-\eta)\,2c_2\,t/\lambda}\big)\,\frac{c_0}{(1-\eta)\,4 c_1 c_2} \,+\, \big(e^{-(1-\eta)\,2c_2\,t/\lambda}-e^{-2c_1 t}\big)\, \frac{\Dai -\, \frac{c_0}{2c_1\lambda}}{2 c_1 -\frac{(1-\eta)\,2c_2}{\lambda}} \\[4pt]
    & R_3\!\left(t,\lambda,\epss\right) \,\equiv\,     \big(1-e^{-(1-\eta)\,2c_2\,t/\lambda}\big)\, \frac{\tilde c_0}{(1-\eta)\,2c_2}
    \end{split} \end{align*}
    where $c_0,\,\tilde c_0$ are non-negative constants depending on $\beta_1,\,\beta_2,\,\rhoi,\,V\,$. Explicit expressions of $c_0,\,\tilde c_0$ are found in Theorems \ref{th:conditional}, \ref{th:marginal} and estimated in Remarks \ref{rem:compute_c0}, \ref{rem:compute_tildec0} respectively.
\end{theorem}

Section \ref{sec:main_proof} will be dedicated to the proof of Theorem \ref{th:main}. 
In the remaining part of the present section logarithmic Sobolev inequalities and the derivation of Theorem \ref{th:main_simple} from Theorem \ref{th:main} are discussed.

\begin{definition} 
A probability density $p:\R^n\to(0,\infty)$
is said to verify the \textit{logarithmic Sobolev inequality} with constant $C>0$ (briefly LSI$_n$($C$) or LSI($C$) or LSI) if
\begin{equation} \label{eq:logSobolev}
\int_{\R^n}\d x\; \pi(x)\, \log\frac{\pi(x)}{p(x)} \;\leq\; \frac{1}{2C}\,\int_{\R^n} \d x\; \pi(x)\;\Big|\nabla\log\frac{\pi(x)}{p(x)}\Big|^2
\end{equation}
for every probability density $\pi$ sufficiently regular, e.g., $\pi\in C^1(\R^n)$ such that $\pi>0\,$, $\int_{\R^n}\pi(x)\,\d x=1\,$, $\pi\log\frac{\pi}{p}\in L^1(\R^n)$.
\end{definition}

\begin{remark}
Setting  $f(r)\,\equiv\, r\,\log r - r +1\geq0$ for $r\geq0\,$, inequality \eqref{eq:logSobolev} also rewrites term by term as
\begin{equation} \label{eq:logSobolev_weak}
\int_{\R^n}\d x\; p(x)\;f\bigg(\frac{\pi(x)}{p(x)}\bigg) \;\leq\; \frac{2}{C}\,\int_{\R^n}\d x\; p(x)\;\bigg|\nabla\sqrt{\frac{\pi(x)}{p(x)}}\,\bigg|^2 
\end{equation}
The integrals in \eqref{eq:logSobolev_weak} are well defined for all $\pi\geq0$ such that $\int_{\R^n}\pi(x)\,\d x=1$ and $\sqrt{\pi/p}\in W^{1,2}_\textit{loc}(\R^n)\,$, allowing to extend the LSI to a larger class of densities $\pi$ (see \cite{AMTU}).
\end{remark}



LSI's were initially proved for Gaussian measures \cite{Stam, Gross} and many sufficient criteria have been subsequently introduced in order to guarantee their validity.
A fundamental property of LSI's is \textit{tensorisation}. It guarantees the independence of the optimal constant on the dimension $n$ of the system when the latter is composed of independent parts. Precisely:

\begin{proposition}[tensorisation \cite{Gross}] \label{prop:tensorisation}
    Let $U:\R^n\to\R\,$, 
    \begin{equation}
        U(x)=\sum_{i=1}^n U_i(x_i) \;.
    \end{equation}
    Suppose that the probability measure $e^{-\beta U_i}/Z_i$ verifies \textup{LSI$_1$($C$)} for every $i=1,\dots,n$.
    Then the probability measure $e^{-\beta U}/Z$ verifies \textup{LSI$_n$($C$)}.
\end{proposition}

The proof of Proposition \ref{prop:tensorisation} is straightforward. There exist several results extending this property to systems with weak interactions.
Beyond tensorisation, two fundamental criteria to establish LSI's are provided by Bakry-\'Emery theorem and Holley-Stroock perturbation lemma.

\begin{theorem}[Bakry-\'Emery \cite{BakryEmery}] \label{th:Bakry-Emery}
    Let $U\in C^2(\R^n)$ strongly convex, namely there exists $\alpha>0$ such that $x\in \R^n$
    \begin{equation}
        \Hess U(x) \,\succeq\, \alpha\,I_n \;.
    \end{equation}
    Then the probability measure $e^{-\beta U}/Z$ verifies \textup{LSI$_n$($C$)} with constant $C=\beta\alpha\,$.
\end{theorem}

\begin{theorem}[Holley-Stroock \cite{HolleyStroock}] \label{th:Holley-Stroock}
    Let $U:\R^n\to\R\,$,
    \begin{equation}
        U \,=\, U_0 \,+\, U_\textup{b} \,.
    \end{equation}
    Suppose that the probability measure $e^{-\beta U_0}/Z_0$ verifies \textup{LSI$_n$($C_0$)} and $U_\textup{b}$ is bounded.
    Then the probability measure $e^{-\beta U}/Z$ verifies \textup{LSI$_n$($C$)} with constant $C= C_0\,e^{-\beta\osc(U_\textup{b})}$, where $\osc(U_\textup{b})=\sup(U_\textup{b})-\inf(U_\textup{b})$.
\end{theorem}

We refer to \cite{AMTU,MarkVillani} for a proof of Theorems \ref{th:Bakry-Emery}, \ref{th:Holley-Stroock} adopting a dynamical point of view. 
Combining the previous three results it is possible to prove LSI's for a large class of potentials $U$.

We are interested in deriving inequalities \textup{(LS1)}, \textup{(LS2)} under simple hypothesis on the potential $V$. This can be done  combining the Bakry-\'Emery-Holley-Stroock criterion with Brascamp-Lieb concentration inequality \cite{BL, Brascamp2002A}. It will become clear that Theorem \ref{th:main_simple} is just a consequence of the more general Theorem \ref{th:main}.


\begin{proposition}\label{prop:logSobolev_condmarg}
Let $V\in C^2(\Rd)$, $V=\Vc+\Vb$ such that
\begin{itemize}
    \item[\textup{(C1)}] $\Vc\in C^2(\R^d)$ is strongly convex;
    \item[\textup{(C2)}] $\Vb$ is bounded;
    \item[\textup{(C3)}] the first and second order derivatives of $\Vc$ have polynomial growth.
\end{itemize}
Then the effective potential $\Veff\in C^2(\Rb)$ defined by \eqref{eq:Veff} writes as $\Veff=\Veffc+\Veffb$ with
\begin{align}
    \label{eq:Fc}
    \Veff_\textup{c}(x_2) \,&\equiv\, -\frac{1}{\beta_1}\,\log\int_{\Ra} e^{-\beta_1 \Vc(x_1,x_2)}\,\d x_1 \quad\textrm{strongly convex,}\\[4pt]
    \label{eq:Fb}
    \Veff_\textup{b}(x_2) \,&\equiv\, -\frac{1}{\beta_1}\,\log \left\langle e^{-\beta_1 \Vb} \right\rangle_{\textup c}(x_2)\quad\textrm{bounded} \;.
\end{align}
$\langle\,\cdot\,\rangle_\textup{c}(x_2)$ denotes the following expectation for any suitable observable $O(x_1,x_2)$
\begin{equation}
    \langle O\rangle_\textup{c}(x_2) \,\equiv\, \frac{\int_{\Ra}O(x_1,x_2)\,e^{-\beta \Vc(x_1,x_2)}\,\d x_1}{\int_{\Ra}e^{-\beta \Vc(x_1,x_2)}\,\d x_1} \;.
\end{equation}
In particular, for every $\beta_1,\beta_2>0$, logarithmic Sobolev inequalities \textup{(LS1), (LS2)} are verified with constants $c_1,c_2$ given by
\begin{align}
    \label{eq:c+b_c1}
    c_1 \,&=\, \alpha_1\; e^{-\beta_1 \osc_1} \\[4pt]
    \label{eq:c+b_c2}
    c_2 \,&=\, \alpha_2\; e^{-\beta_2 \osc_2}
\end{align}
where we set
\begin{align*}
        \alpha \,&\equiv\, \inf_{x\in\Rd}\,\mineigen \big(\Hess \Vc(x)\big)  \,>0  \\[4pt]
        \alpha_1 \,&\equiv\, \inf_{x\in\Rd}\, \mineigen \big(\Hessx \Vc(x_1,x_2) \big) \,\geq \,\alpha \\[4pt]
        \alpha_2 \,&\equiv\, \inf_{x_2\in\Rb} \mineigen \big(\Hessy \Veffc(x_2) \big) \,\geq\\
        &\geq\, \inf_{x\in\Rd}\, \mineigen\big(\Hessy \Vc - (\Hessxy \Vc)^T(\Hessx \Vc)^{-1}\Hessxy \Vc\big)(x) \,\geq\,\alpha
\end{align*}
and
\begin{align*}
        \osc \,&\equiv\, \sup_{\Rd}\Vb - \inf_{\Rd}\Vb \,<\infty\\[2pt]
        \osc_1 \,&\equiv\, \sup_{x_2\in\Rb} \Big( \sup_{x_1\in\Ra} - \inf_{x_1\in\Rb}\Big)\,\Vb(x_1,x_2) \,\leq\, \osc\\[4pt]
        \osc_2 \,&\equiv\, \sup_{\Rb}\Veffb - \inf_{\Rb}\Veffb\,\leq\, \osc \;.
\end{align*}
\end{proposition}



We split the proof of Proposition \ref{prop:logSobolev_condmarg} in two parts, corresponding to inequalities \textup{(LS1), (LS2)} respectively. The first one is straightforward. The second one requires an algebraic lemma and Brascamp-Lieb concentration inequality for log-concave measures.

\begin{proof}[Proof of Proposition \ref{prop:logSobolev_condmarg}. First part]
Bakry-\'Emery-Holley-Stroock criterion can be easily applied to prove inequality \textup{(LS1)}.
Recall that $\rhoKa(\cdot|x_2)$ is the Gibbs measure on $\Ra$ associated with potential $x_1\mapsto V(x_1,x_2)$ and inverse temperature $\beta_1$. 
For all $x_1,\,\phi\in\Ra$ we have
\begin{equation}
    \phi^T \,\Hessx\Vc(x_1,x_2)\; \phi \,=\,
    \tilde \phi^T \,\Hess \Vc(x_1,x_2)\; \tilde \phi \,\geq\, \alpha\, |\tilde \phi|^2 \,=\, \alpha\, |\phi|^2 
\end{equation}
setting $\tilde \phi\equiv\begin{pmatrix}\phi\\0\end{pmatrix}
\in\Ra\!\times\Rb$ and using hypothesis \textup{(C1)}. Thus $x_1\mapsto \Vc(x_1,x_2)$ is strongly convex and $\alpha_1\geq\alpha>0$.
On the other hand $x_1\mapsto \Vb(x_1,x_2)$ is clearly bounded and $\osc_1\leq\osc<\infty$.
Therefore inequality \textup{(LS1)} with constant $c_1$ given by \eqref{eq:c+b_c1} follows by Theorems \ref{th:Bakry-Emery}, \ref{th:Holley-Stroock}.
\end{proof}

\begin{lemma} \label{lem:Schur}
Let $A$ be a $d\times d$ invertible symmetric block matrix:
\begin{equation}
    A \,\equiv\, \begin{pmatrix} A_{11} & A_{12} \\ A_{12}^T & A_{22} \end{pmatrix} \;.
\end{equation}
The Schur complement of $A_{11}$ in $A$ is the $d_2\times d_2$ invertible symmetric matrix defined by 
\begin{equation}
    S \,\equiv\, A_{22} \,-\, A_{12}^T\,A_{11}^{-1}\,A_{12} \;.
\end{equation}
If there exists $\alpha>0$ such that $A \succeq \alpha I_{d}\,$,
then $S \succeq \alpha I_{d_2}$.
\end{lemma}

\begin{proof}
Matrix $A$ is similar to a block-diagonal matrix, indeed it is easy to check that
\begin{equation}
    A \,=\, P\,D\,P^T
\end{equation}
with
\begin{equation}
    P \,\equiv\, \begin{pmatrix} I_{d_1} & 0 \\ A_{12}^T\,A_{11}^{-1} & I_{d_2}  \end{pmatrix} \quad,\quad
    D \,\equiv\, \begin{pmatrix}  A_{11} & 0 \\ 0 & S\end{pmatrix} \;.
\end{equation}
Let $\psi\in\Rb$. Set $\tilde \psi \equiv \begin{pmatrix} -A_{11}^{-1}A_{12}\,\psi \\ \psi \end{pmatrix}\in\Rd$ observing that $P^T \tilde \psi = \begin{pmatrix} 0 \\ \psi \end{pmatrix}\in\Rd$. Then
\begin{equation}
    \tilde \psi^T A\,\tilde \psi \,=\, \tilde \psi^T P\,D\,P^T\tilde \psi \,=\, \psi^T S\,\psi 
\end{equation}
proving the thesis.
\end{proof}

\begin{proof}[Proof of Proposition \ref{prop:logSobolev_condmarg}. Second part]
Recall that $\rhoKb$ is the Gibbs measure on $\Rb$ with potential $\Veff$ defined by \eqref{eq:Veff} and inverse temperature $\beta_2$. 
$\Veff$ rewrites as $\Veff = \Veffc + \Veffb$ with $\Veffc,\Veffb$ defined by equations \eqref{eq:Fc}, \eqref{eq:Fb}.
Clearly from equation \eqref{eq:Fb}
\begin{equation}
    \sup_{\Rb} \Veff_{\textup b} - \inf_{\Rb} \Veff_{\textup b} \,=\, \osc_2 \,\leq\, \osc \;.
\end{equation}
On the other hand, from equation \eqref{eq:Fc} we can compute
\begin{equation} \label{eq:HessVeff}
    \Hessy\Veff_c \,=\, \left\langle\, \Hessy \!\Vc\,\right\rangle_{\textup c} -\, \beta_1\, \left\langle \nablay \Vc\;\nablay^T \Vc \right\rangle_{\textup c} \,+\, \beta_1\, \left\langle \nablay \Vc \right\rangle_{\textup c}\, \left\langle\nablay^T \Vc\right\rangle_{\textup c} 
\end{equation}
indeed hypothesis \textup{(C3)} guarantees that the derivatives w.r.t. $x_2$ can be taken inside the integral w.r.t. $x_1\,$.
Then for every $\psi\in\Rb$
\begin{equation} \label{eq:HessVeffu}
    \psi^T\Hessy\Veff_{\textup c}\; \psi \,=\,
    \left\langle \psi^T \Hessy \Vc\;\psi \right\rangle_{\textup c}
    -\, \beta_1 \left\langle \big(\psi^T\nablay \Vc\big)^2 \right\rangle_{\!\textup c}
    +\, \beta_1 \left\langle \psi^T\nablay \Vc \right\rangle_{\textup c}^2 \,.
\end{equation}
Brascamp-Lieb concentration inequality applies to the measure $\langle\,\cdot\,\rangle_{\textup c}$ since the potential $x_1\mapsto \Vc(x_1,x_2)$ is convex by hypothesis \textup{(C1)}, hence:
\begin{equation} \label{eq:BL} \begin{split}
    \left\langle \big(\psi^T\,\nablay \Vc\big)^2 \right\rangle_{\!\textup c}
    - \left\langle \psi^T\,\nablay \Vc \right\rangle_{\textup c}^2 \;&\leq\,
    \left\langle \nablax^T\big(\psi^T\nablay \Vc\big)\;\big(\beta_1\Hessx \Vc\big)^{-1} \;\nablax\big(\psi^T\nablay \Vc\big) \right\rangle_{\!\textup c} \\[4pt]
    &=\, \frac{1}{\beta_1} \left\langle \psi^T\,(\Hessxy \Vc)^T(\Hessx \Vc)^{-1}(\Hessxy \Vc)\;\psi \right\rangle_{\textup c} \,.
\end{split} \end{equation}
Plugging inequality \eqref{eq:BL} into \eqref{eq:HessVeffu} and using Lemma \ref{lem:Schur} together with hypothesis \textup{(C1)}, one obtains
\begin{equation} \begin{split}
    \psi^T\,\Hessy\Veff_{\textup c}\; \psi \;&\geq\, \left\langle \psi^T \left(\Hessy \Vc - (\Hessxy \Vc)^T(\Hessx \Vc)^{-1}\Hessxy \Vc\right) \psi \right\rangle_{\!\textup{c}} \\[4pt]
    &\geq\, \alpha\,|\psi|^2
\end{split}\end{equation}
for every $\psi\in\Rb$. Therefore $\Veffc$ is strongly convex and $\alpha_2\geq\alpha>0$.
Inequality \textup{(LS2)} with constant $c_2$ given by \eqref{eq:c+b_c2} then follows by Theorems \ref{th:Bakry-Emery}, \ref{th:Holley-Stroock}.
\end{proof}


\begin{proof}[Proof that Theorem \ref{th:main_simple} follows from Theorem \ref{th:main}]
Let $V=\Vc+\Vb$ with $\Vc$ strongly convex polynomial and $\Vb$ bounded, smooth with bounded derivatives of any order.
Let us show that the potential $V$ satisfies the hypothesis of Theorem \ref{th:main}.\\
Proposition \ref{prop:logSobolev_condmarg} guarantees that conditions \textup{(LS1)}, \textup{(LS2)} are verified (and gives an estimation of the constants $c_1,c_2$).
Conditions \textup{(A1)}, \textup{(A2)} are clearly verified.
Conditions \textup{(A3)}, \textup{(A4)} follow from the strong convexity of $\Vc$; indeed a second order expansion of $\Vc$ at its minimum point $x_\textup{c}$ shows that for all $x\in\R^d$
\begin{equation}
    \Vc(x)-\Vc(x_\textup{c}) \,\geq\, \frac{\alpha}{2}\,|x-x_\textup{c}|^2
\end{equation} 
while a first order expansion of $\nabla\Vc$ at $0$ shows that
\begin{equation}
    x\cdot\big(\nabla\Vc(x)-\nabla\Vc(0)\big) \,\geq\, \alpha\,|x|^2\;.
\end{equation}
Finally, since $\Vc$ is a polynomial, 
condition \textup{(A5)} is also guaranteed. Indeed, for every $k=1,\dots,d_1$ the leading order monomial containing $x_{1k}$ in $\Vc(x)$ is of type $a_k\,x_{1k}^{n_k}$ with $a_k>0$, $n_k\geq2$ even, and in $x\cdot\nablax\Vc(x)$ it becomes $a_k\,n_k\,x_{1k}^{n_k}\,$.  
\end{proof}

\section{Examples}\label{sec:examples}

\subsection{Quadratic potential in dimensions $d_1=d_2=1$: explicit computations}\label{subsec:gaussian}

Positive definite quadratic potentials in any dimensions $d_1,d_2$ clearly fit the hypothesis of Theorem \ref{th:main_simple}. Here we consider positive-definite quadratic potentials in dimensions $d_1=d_2=1$, so that explicit computations can be easily performed. We provide the solution of the two-temperatures Fokker-Planck equation for all times $t$ and compute explicit expansions for large $\lambda$. A comparison between explicitly computed convergence rates and estimates given by Theorem \ref{th:main} shows that the latter are sharp for quadratic potentials. 

Let $x\in\R$ be the \textit{fast} variable and $y\in\R$ be the \textit{slow} variable. We consider the two dimensional quadratic potential
\begin{equation}  \label{eq:quadratic_potential}
V(x,y) \,\equiv\, \frac{1}{2}\, ( x,\, y)\, A\, \begin{pmatrix} x\\ y\end{pmatrix}\,,
\end{equation}
where $A$ is a positive-definite symmetric matrix
\begin{equation} 
A \,\equiv\, \begin{pmatrix}a & c \\ c & b \end{pmatrix}\ ,\quad a>0\,,\ ab-c^2>0\,.
\end{equation}
Thanks to the quadratic nature of the potential, one can check that the effective potential \eqref{eq:Veff} is also quadratic
\begin{equation}
    \Veff(y) \,=\, \frac{\det A}{2\,a}\,y^2
\end{equation}
and the measure $\rho_*$ defined by \eqref{eq:rhoK} is a centered Gaussian distribution with covariance matrix
\begin{equation}\label{eq:Sigma}
\Sigma \,\equiv\,  \begin{pmatrix} \dfrac{1}{\beta_1 a}+\dfrac{1}{\beta_2}\,\dfrac{c^2}{a\,\det A} & \,-\dfrac{1}{\beta_2}\,\dfrac{c}{\det A} \\[12pt] -\dfrac{1}{\beta_2}\,\dfrac{c}{\det A} &\dfrac{1}{\beta_2}\,\dfrac{a}{\det A}  \end{pmatrix} \;.
\end{equation}

Clearly we have the following
\begin{proposition} 
    The quadratic potential \eqref{eq:quadratic_potential} verifies the hypothesis of Theorem \ref{th:main} with rates given by
    \begin{equation} \label{eq:quadratic_c1c2}
        c_1 \,=\, a\,,\quad 
        c_2 \,=\, \frac{\det A}{a} \;.
    \end{equation}
    Thus, for every $\beta_1,\beta_2>0$ the convergence of the two-temperatures dynamical measure $\rhot(s,y)$ to $\rhoK(s,y)$ is described by inequalities \eqref{eq:main_conditional},\eqref{eq:main_marginal}.
\end{proposition}

\begin{proof}
Hypothesis \textup{(A1)-(A5)} are clearly verified.
The logarithmic Sobolev inequalities \textup{(LS1)}, \textup{(LS2)} follow by Proposition \ref{prop:logSobolev_condmarg} with $\Vb=0$, $\alpha_1=a$, $\alpha_2=\frac{\det A}{a}$.
\end{proof}

In the following we will compute exactly the convergence rates of the dynamical measure $\rhot$ to $\rhoK$, in order to show that the estimates \eqref{eq:quadratic_c1c2} are sharp for a quadratic potential.
For a quadratic potential, the SDE \eqref{eq:SDEintro} with diffusion coefficients defined by \eqref{eq:SDEintro_diffusion}, \eqref{eq:lambda} is an \textit{Ornstein-Uhlenbeck process}, and  rewrites as
\begin{equation}  \label{OrnsteinUhlenbeck}
\begin{pmatrix} \d x(t)\\ \d y(t)\end{pmatrix} \,=\, -\,\Gamma \begin{pmatrix} x(t)\\ y(t)\end{pmatrix}\d t \,+\, \Delta\begin{pmatrix} \d W_1(t)\\ \d W_2(t)\end{pmatrix}
\end{equation}
with
\begin{equation} 
\Gamma \,\equiv
\begin{pmatrix} a & c \\ \frac{c}{\lambda} & \frac{b}{\lambda} \end{pmatrix}  
\quad,\quad  \Delta \,\equiv\, 
\begin{pmatrix} \sqrt{\frac{2}{\beta_1}} & 0 \\ 0 & \sqrt{\frac{2}{\lambda\beta_2}}  \end{pmatrix} \;.
\end{equation}
The fundamental solution $\rhoT(x,y)$ of  the Fokker -Planck equation associated to  
 \eqref{OrnsteinUhlenbeck},  is a Gaussian distribution of mean $\mu(t)$ and covariance matrix $\Omega(t)$ defined by
\begin{equation}\label{eq:OUprocess} 
\mu(t) \,\equiv\, e^{-t\,\Gamma}\begin{pmatrix} x_0\\ y_0\end{pmatrix} \quad,\quad 
\Omega(t) \,\equiv\, \int_0^t e^{-(t-s)\,\Gamma}\,\Delta\,\Delta^T\,e^{-(t-s)\,\Gamma^T}\,\d s
\end{equation}
where $(x_0,y_0)$ represents the initial condition. We can compute  \eqref{eq:OUprocess} exactly and deduce the behavior for large
$\lambda$. This is the object of the next proposition.

Let us denote by $\gamma_{1,2}$ the eigenvalues of the matrix $\Gamma\,$:
\begin{equation}
    \gamma_{1,2}\,\equiv\,\frac{(a+b\lambda^{-1}) \pm \sqrt{(a+b\lambda^{-1})^2-4\,\lambda^{-1}\det A}}{2} \;.
\end{equation}
We note that $\gamma_1,\gamma_2 > 0$ for all $\lambda>0$ and compute their expansions as $\lambda\to\infty\,$:
\begin{equation} \label{eq:ex_gaussian_expansionlambda}
    \gamma_1 = a + O(\lambda^{-1})\,, \quad
    \gamma_2 = \frac{\det A}{a}\,\lambda^{-1} + O(\lambda^{-2})\;.
\end{equation}

\begin{proposition} \label{Omegaexpansion}
Let $\Omega(t)=(\Omega_{ij}(t))_{i,j=1,2}$ and  $\mu(t)=(\mu_i(t))_{i=1,2}$ be defined by \eqref{eq:OUprocess}. Let $\Sigma=(\Sigma_{ij})_{i,j=1,2}$ defined by \eqref{eq:Sigma}. As $\lambda\to\infty$ we have the following expansions uniformly in $t>0\,$  
\begin{equation} \label{eq:Omegat} \begin{split}
    \Omega_{11}(t) \,&=\, \Sigma_{11} \,-\frac{1}{\beta_1 a}\, e^{-2\gamma_{1}t} \,-\frac{1}{\beta_2}\,\frac{c^2}{a^2}\,\frac{a}{\det A}\,e^{-2\gamma_{2}t} \,+\, O(\lambda^{-1}) \\[4pt]
    \Omega_{22}(t)\,&=\, \Sigma_{22}\,(1-e^{-2\gamma_{2}t}) \,+\, O(\lambda^{-1}) \\[6pt]
    \Omega_{12}(t)\,&=\,\Omega_{21}(t)\,=\, \Sigma_{12}\,(1-e^{-2\gamma_{2}t}) \,+\, O(\lambda^{-1})
\end{split} \end{equation}
and
\begin{equation} \label{eq:mut} \begin{split}
    \mu_1(t)\,&=\, \big(x_0+\frac{c}{a}y_0\big)\, e^{-\gamma_{1}t} \,+\,-\frac{c}{a}y_0\,e^{-\gamma_{2}t} \,+\,  O(\lambda^{-1}) \\[4pt]
    \mu_2(t)\,&=\,\,y_0\,e^{-\gamma_{2}t} \,+\, O(\lambda^{-1}) \;.
\end{split} \end{equation}
\end{proposition}

\begin{proof}
A direct computation shows that for $\tau\in\R$
\begin{equation}
e^{-\tau\,\Gamma} \,=\, \dfrac{1}{v_2-v_1} \begin{pmatrix} e^{-\gamma_{2}\tau}\,v_2-e^{-\gamma_{1}\tau}\,v_1 & e^{-\gamma_{1}\tau}-e^{-\gamma_{2}\tau} \\[8pt] (e^{-\gamma_{2}\tau}-e^{-\gamma_{1}\tau})\,v_1v_2 & e^{-\gamma_{1}\tau}\,v_2-e^{-\gamma_{2}\tau}\,v_1 \end{pmatrix}
\end{equation}
where $v_{1,2}\equiv\dfrac{\gamma_{1,2}-a}{c}$. Then using definition \eqref{eq:OUprocess} one obtains:
\begin{equation} \label{eq:processgauss2d} \begin{split}
    \Omega_{11}(t) \,&=\,\frac{r_1}{2\gamma_1}\,\big(1-e^{-2\gamma_{1}t}\big) +\, \frac{r_2}{2\gamma_2}\,\big(1-e^{-2\gamma_{2}t}\big) - \frac{2r_{12}}{\gamma_1+\gamma_2}\,\big(1-e^{-(\gamma_{1}+\gamma_2)t}\big) \\[4pt]
    \Omega_{12}(t) \,&=\, \frac{v_1\,r_1}{2\gamma_1}\,\big(1-e^{-2\gamma_{1}t}\big) +\, \frac{v_2\,r_2}{2\gamma_2}\,\big(1-e^{-2\gamma_{2}t}\big) - \frac{(v_1+v_2)\,r_{12}}{\gamma_1+\gamma_2}\,\big(1-e^{-(\gamma_1+\gamma_{2})t}\big) \\[4pt]
    \Omega_{22}(t) \,&=\, \frac{v_1^2\,r_1}{2\gamma_1}\,\big(1-e^{-2\gamma_{1}t}\big) +\, \frac{v_2^2\,r_2}{2\gamma_2}\,\big(1-e^{-2\gamma_{2}t}\big) - \frac{2\,v_1v_2\,r_{12}}{\gamma_1+\gamma_2}\,\big(1-e^{-(\gamma_1+\gamma_2)t}\big)
\end{split} \end{equation}
where
\begin{equation}
\begin{split}
r_1\,&\equiv\, \frac{2}{(v_2-v_1)^2}\,\bigg(\frac{v^2_2}{\beta_1}+\frac{1}{\lambda\beta_2}\bigg)\\[4pt]
r_2 \,&\equiv\, \frac{2}{(v_2-v_1)^2}\,   \bigg(\frac{v^2_1}{\beta_1}+\frac{1}{\lambda\beta_2}\bigg)\\[4pt]
r_{12} \,&\equiv\, \frac{2}{(v_2-v_1)^2}\,  \bigg(\frac{v_1v_2}{\beta_1}+\frac{1}{\lambda\beta_2}\bigg) \,.
\end{split}
\end{equation}
All in all we have obtained for  $\Omega(t)$ an expression of the form
\begin{equation}
\Omega(t) \,=\, M_0+M_1\,e^{-2\gamma_{1}t}+M_2\,e^{-2\gamma_{2}t}+M_{3}\,e^{-(\gamma_{1}+\gamma_2)t}
\end{equation}
where  $(M_{k})_{0\leq k \leq 3}$ are suitable time-independent $2\times 2$ matrices. Expansions of the matrices $M_k$ for large $\lambda$ can be obtained using expansions \eqref{eq:ex_gaussian_expansionlambda} of $\gamma_1,\gamma_2\,$. Identities \eqref{eq:Omegat} then follows using also the obvious bounds $0<e^{-\gamma_{1,2}\,t}< 1$ for any $t>0$.
Identities \eqref{eq:mut} for $\mu(t)$ are obtained in a similar way.
\end{proof}

The previous proposition can be used to obtain the behaviour of the KL divergence between the solution $\rhot$ and the measure $\rho_*$ for large $\lambda$. More precisely we are interested in the quantities  
$\Da(t,\lambda)\,$, $\Db(t,\lambda)$ defined by \eqref{eq:KL}.

\begin{proposition} \label{prop:KLgauss}
\begin{equation}
\Da(t,\lambda)\,=\, \frac{1}{2}\left\{-\log\left(1-e^{-2\gamma_{1}t}\right)-e^{-2\gamma_{1}t}\,+\,e^{-2\gamma_{1}t}\,\beta_1\,a\left(y_0+\frac{a}{c}\,x_0\right)^2\right\} \,+\,O(\lambda^{-1})
\end{equation}
\begin{equation}
\Db(t,\lambda)\,=\,\frac{1}{2}\left\{-\log\left(1-e^{-2\gamma_{2}t}\right)-e^{-2\gamma_{2}t}\,+\,e^{-2\gamma_{2}t}\,\beta_2\,\frac{a^2\,y_0^2}{c^2}\,\frac{\det A}{a}\right\}\,+\,O(\lambda^{-1})
\end{equation}
\end{proposition}

\begin{proof} 
Let $p_1,p_2$ be two $1$-dimensional Gaussian distributions with means $u_1,u_2$ and variances $\sigma_1,\sigma_2$ respectively. Their KL divergence reads
\begin{equation}\label{KLgauss}
\DKL(p_1\|\,p_2) \,=\, \frac{1}{2}\left\{\log\dfrac{\sigma_2}{\sigma_1}-1+ \dfrac{\sigma_1}{\sigma_2}+(u_2-u_1)\sigma_2^{-1}(u_2-u_1)\right\} \;.
\end{equation}    
Since $\rhoK$, $\rhoT$ are both $2$-dimensional Gaussian distributions, their conditional and marginal measures are Gaussian too and are easy to compute from \eqref{eq:Sigma},  \eqref{eq:processgauss2d} respectively:
\begin{itemize}
   \item $\rhoKa(\cdot|x_2)$ has mean $\Sigma_{11}\,\Sigma_{22}^{-1}\,x_2\,=\,\big(\frac{\beta_2}{\beta_1}\frac{\det A}{a^2} \,+\, \frac{c^2}{a^2}\big)\,x_2\,$ and variance $\Sigma_{11}-\Sigma_{12}\,\Sigma_{22}^{-1}\,\Sigma_{21}\,=\, \frac{1}{\beta_1 a}$
   \item $\rhoTa(\cdot|x_2)$ has mean $\mu_1(t)\,+\,\Omega_{11}(t)\,\Omega_{22}(t)^{-1}(x_2-\mu_2(t))$ and variance $\Omega_{11}(t)-\Omega_{12}(t)\,\Omega_{22}(t)^{-1}\Omega_{21}(t)$
   \item $\rhoKb$ has mean $0$ and variance $\Sigma_{22}\,=\, \frac{1}{\beta_2}\,\frac{a}{\det A}$
   \item $\rhoTb$ has mean $\mu_2(t)$ and variance $\Omega_{22}(t)\,$.
\end{itemize}
Then combining Proposition \ref{Omegaexpansion} with expression \eqref{KLgauss} we obtain the claim.
\end{proof}

Note that by Proposition \ref{prop:KLgauss} it follows that for every $\tau_0>0$ there exists finite positive constants $C_1(\tau_0),C_2(\tau_0)$ such that
\begin{align} \label{eq:quadratic_D1}
    \Da(t,\lambda) \,&\leq\, C_1(\tau_0)\; e^{-2\gamma_1 t}  + O(\lambda^{-1})\,, \quad \gamma_1\,t\geq\tau_0 \\[4pt]
    \label{eq:quadratic_D2}
    \Db(t, \lambda) \,&\leq\, C_2(\tau_0)\; e^{-2\gamma_2t} + O(\lambda^{-1})\,, \quad \gamma_2\,t\geq\tau_0\,.
\end{align}
The constants $C_{1,2}(\tau_0)$ blow up as $\tau_0\to 0$ indeed we have chosen the Dirac delta as initial distribution and its KL divergence from $\rhoK$ is infinite.
From equations \eqref{eq:quadratic_D1},\eqref{eq:quadratic_D2} one can see that for a quadratic potential the estimates of Theorem \ref{th:main} (together with Proposition \ref{prop:logSobolev_condmarg}) give sharp decay rates. It suffices compare the constants $c_1,c_2$ in \eqref{eq:quadratic_c1c2} with the expansion \eqref{eq:ex_gaussian_expansionlambda} of $\gamma_1,\gamma_2$. 

\subsection{Spin glass: dynamical approach to Guerra's scheme}\label{subsec:spinglass}
In this subsection we apply Theorem \ref{th:main} to a class of mean-field spin glass models. The aim is to put on  a solid ground some recent conjectures appearing in spin glass theory. Recently in \cite{Contucci_2019, Contucci_2021}  it has been shown by theoretical physics arguments that Guerra's hierarchical measure for $K$ steps of replica symmetry breaking ($K$-RSB) \cite{Guerra_2003} can be identified  with the stationary measure of a multi-bath Langevin dynamics of type \eqref{eq:SDEintro}-\eqref{eq:SDEintro_diffusion} for a suitable random potential $V$ involving $K+1$ families of degrees of freedom  with different time scales.
Thanks to this identification the authors push forward the well-known connection between static and dynamical properties of spin glass systems \cite{PMPF}. Here we provide sufficient conditions for such an identification to hold and a method to estimate the rates of convergence.

The assumptions required by Theorem \ref{th:main} - in particular the Bakry-\'Emery-Holley-Stroock criterion analysed in Section \ref{sec:mainresults} - are checked to hold with probability (with respect to the quenched variables) exponentially close to $1$ in a suitable region of parameters. In general we do not expect better than exponentially small convergence rate in the system's size. 
Nevertheless, 
better criteria for the validity of logarithmic Sobolev inequalities tailored on specific phases would yield more physical results. For example, at high temperatures one could use decay-of-correlations criteria to find convergence rates independent of the system's size (see, e.g., \cite{BB}). This aspect goes beyond the purpose of the present work. There exists a vast physics literature on mean-field spin glass dynamics in the large size limit, we refer the interested reader to  \cite{Bouchaud_1997} for a survey and we mention the recent works \cite{BenArous,Jagannath,Bodineau} for rigorous results on the subject.

For the readers who are not familiar with spin glasses, let us briefly recall Guerra's scheme and its role in equilibrium theory of glasses. Consider $N$ spin variables $s=(s_i)_{i=1,\,\ldots, N}$ taking values $-1,1$ interacting trough the random potential
\begin{equation}\label{SGsystem}
V_\textup{SK}(s)=  -\,\frac{1}{\sqrt{2N}}\sum_{i,j=1}^N J_{ij}\,s_is_j  
\end{equation}
where the couplings $J=(J_{ij})_{i,j=1\dots N}$ are independent standard Gaussian random variables. 
We assume that, for a given a realization of  $J$,  thermodynamic equilibrium of the system is the described by the random Boltzmann-Gibbs measure $\rho_\textup{SK}(s) \equiv e^{-\beta_1 V_\textup{SK}(s)}/Z_\textup{SK} \,$.
From a dynamical point of view one may think that the family $J$ is frozen (or \textit{quenched}) and does not play any role in the equilibration process.
Roughly speaking the idea of \textit{Guerra's scheme} \cite{Guerra_2003} is to construct a suitable interpolation between the system \eqref{SGsystem} and a simpler one. In its easier case, the so-called $1$-RSB, one introduces the  parameters  $\theta\in[0,1]$, $\bar q\in[0,1]$ and defines an interpolating potential
\begin{equation}\label{INpotential}
V_\theta(s,y) =  -\,\frac{\sqrt{\theta}}{\sqrt{2N}}\sum_{i,j=1}^N J_{ij}\,s_is_j 
- \sqrt{1-\theta\,} \sum_{i=1}^N\left(\sqrt{1-\bar q\,}\, y_i\,s_i \,+\, \sqrt{\bar q\,}\,h_i\,s_i \right)
\end{equation}
where $y=(y_i)_{i=1,\ldots,N}$ and $h=(h_i)_{i=1,\ldots,N}$ are  i.i.d. standard Gaussian random variables. 
Given $0<\beta_2<\beta_1$ and a realization of the \textit{quenched} variables  $J,\,h$, the $1$-RSB random measure associated to $V_\theta(s,y)$ is defined as 
\begin{equation}\label{eq: measure1RSB}
\rho_{\theta}(s,y)\,=\,\dfrac{e^{\frac{\beta_2}{\beta_1}\log Z_1(y)}}{Z_2} \dfrac{e^{-\beta_1V_{\theta}(s,y)}}{Z_1(y)}   \end{equation}
where $Z_1(y)=\sum_{s\in\{\pm1\}^N}e^{-\beta_1V_{\theta}(s,y)}$ and $Z_2=\mathbb{E}_y e^{\frac{\beta_2}{\beta_1}\log Z_1(y)} $.
%
%
Therefore $\rho_{\theta}$ interpolates between  $\rho_{1}(s,y)=\rho_\textup{SK}(s)$ and $\rho_{0}(s,y)$ which is a product measure on $(s_i,y_i)$ for $i=1,\ldots,N$. The latter is strictly related to the 1-RSB Parisi functional \cite{Talagrand_book}. 

Notice that, replacing hard spins $s_i$ with soft continuous spins, the measure  $ \rho_{\theta}$ in \eqref{eq: measure1RSB} can be seen as the stationary measure 
$\rhoK$ in \eqref{eq:rhoK} 
associated to the Langevin dynamics \eqref{eq:SDEintro}-\eqref{eq:SDEintro_diffusion} where $y$ is slowly varying. This observation is the starting point of the works \cite{Contucci_2019, Contucci_2021}. In the next paragraphs we will use Theorem \eqref{th:main} to formalize it. 
Let $s=(s_i)_{i=1,\dots, N}\in\R^N$ (\textit{fast variables}), $y=(y_i)_{i=1,\dots,N}\in\R^N$ (\textit{slow variables}) and consider the (slightly more general) potential
\begin{equation} \label{eq:guerra_potential}
\begin{split}
    V(s,y) \,=\, &-\frac{\sqrt{\Delta}}{\sqrt{2N}}\sum_{i,j=1}^N J_{ij}\,s_is_j \,- \sqrt{\Delta_0}\,\sum_{i=1}^N\, y_i\,s_i \,-\, \sqrt{\Delta_1}\,\sum_{i=1}^N\,h_i\,s_i\,+\\
    &+ \frac{A}{2}\,\sum_{i=1}^N\,(s_i^2-1)^2
    \,+\, \frac{B}{2}\,\sum_{i=1}^N\,y_i^2
\end{split}
\end{equation}
where $\Delta_0,\,\Delta_1,\,\Delta\geq0$ tune respectively the one-body and two-bodies spin interactions, $A>0$ pushes the \textit{soft spins} toward \textit{hard spins} $-1,1$, and $B>0$ tunes the variance of the external fields. 

\subsubsection{One-body system ($\Delta=0$)}
We first consider the one-body system where $\Delta=0$.
\begin{proposition} \label{prop:Guerra0}
    Let $\Delta=0$. Suppose $A B\geq \Delta_0$ and let $\beta_1,\beta_2>0$. Then the potential \eqref{eq:guerra_potential} verifies the hypothesis of Theorem \ref{th:main} with rates estimated by
    \begin{equation} \label{eq:onebody_c1c2}
        c_1 \,=\, A\,e^{-\frac{\beta_1 A}{2}}\,,\quad 
        c_2 \,=\, \big(B-\frac{\Delta_0}{A}\big)\,e^{-\frac{\beta_2 A}{2}}
    \end{equation}
    independently of the dimension $d_1=d_2=N$ of the system.
    Thus, the convergence of the two-temperatures dynamical measure $\rhot(s,y)$ to $\rhoK(s,y)$ is described by inequalities \eqref{eq:main_conditional},\eqref{eq:main_marginal}.
\end{proposition}

\begin{proof}
In order to prove the logarithmic Sobolev inequalities \textup{(LS1)}, \textup{(LS2)} with dimension-free constants, it is convenient to use the tensorisation property of potential \eqref{eq:guerra_potential} when $\Delta=0$. We have
\begin{equation} \label{eq:onebody_tensorisation}
    V(s,y) \,=\, \sum_{i=1}^N v(s_i,y_i)
\end{equation}
with
\begin{equation}
    v(s_i,y_i) \,\equiv - \sqrt{\Delta_0}\, y_i\,s_i \,- \sqrt{\Delta_1}\,h_i\,s_i\,+\, \frac{A}{2}\,(s_i^2-1)^2
    \,+\, \frac{B}{2}\,y_i^2 \,.
\end{equation}
Also the effective potential associated to $V$ tensorises:
\begin{equation}
    \Veff(y) \,\equiv\, -\frac{1}{\beta_1}\log\int_{\R^N}\d s\, e^{-\beta_1 V(s,y)} \,=\, \sum_{i=1}^N f(y_i)
\end{equation}
with
\begin{equation}
    f(y_i) \,\equiv\, -\frac{1}{\beta_1}\log\int_{\R}\d s_i\, e^{-\beta_1 v(s_i,y_i)} \;.
\end{equation}
Now, in dimension $1$ it is easy to establish a log-Sobolev inequality for the measures $s_i\mapsto e^{-\beta_1 v(s_i,y_i)}/z_1(y_i)$ and $y_i\mapsto e^{-\beta_2 f(y_i)}/z_2\,$ using Proposition \ref{prop:logSobolev_condmarg}. We want to write $v(s_i,y_i)\,=\, v_\textup{c}(s_i,y_i) + v_\textup{b}(s_i,y_i)$ with $v_\textup{c}$ strongly convex and $v_\textup{b}$ bounded.
For this purpose let us \textit{convexify} the double-well potential by introducing a convex function $\eta\in C^2(\R)$ such that
\begin{equation} \label{eq:eta}
    \eta(x) \,=\, \begin{cases} x^2 & \textrm{for }|x|< x_0-\delta \\[4pt]
    (x^2-1)^2 & \textrm{for }|x|> x_0+\delta
    \end{cases}
\end{equation}
where $x_0\equiv\frac{1+\sqrt{5}}{2}$ is the largest point where the above quadratic and quartic curves intersect and $\delta\equiv\frac{1}{10}\,$. It is possible to determine $\eta$ on the interval $[x_0-\delta,\,x_0+\delta]$ by an interpolating polynomial of degree $5$ and it turns out that for all $x\in\R$
\begin{equation} \label{eq:eta_deriv}
    \eta'(x)\,\geq\,2x\,,\quad \eta''(x)\,\geq\, 2\,.
\end{equation}
Then we set
\begin{equation}
    v_\textup{b}(s_i) \,\equiv\, \frac{A}{2}\left((s_i^2-1)^2 \,-\, \eta(s_i)\right) \;.
\end{equation}
$v_\textup{b}\in C^2(\R)$ is bounded since it has compact support, and $\osc(v_\textup{b}) \,=\, \frac{A}{2}\,$.
$v_\textup{c}(s_i,y_i) \,\equiv\, v(s_i,y_i) - v_\textup{b}(s_i)$ is strongly convex, indeed
\begin{equation}
    \Hess v_\textup{c}(s_i,y_i) \,=\, \begin{pmatrix} \frac{A}{2}\,\eta''(s_i) & -\sqrt{\Delta_0}\\[2pt] -\sqrt{\Delta_0} & B \end{pmatrix}
\end{equation}
and its fist entry and its Schur complement are bounded below by $A$, $B-\frac{\Delta_0}{A}$ respectively.
By Proposition \ref{prop:logSobolev_condmarg}, it follows that the $1$-dimensional measures $s_i\mapsto e^{-\beta_1 v(s_i,y_i)}/z_1(y_i)\,$, $y_i\mapsto e^{-\beta_2 f(y_i)}/z_2$ verify inequalities LSI$_1$($\beta_1c_1$) and LSI$_1$($\beta_2c_2$) respectively for $c_1,c_2$ given by \eqref{eq:onebody_c1c2}.
The tensorisation property (see Proposition \ref{prop:tensorisation}) entails that the $N$-dimensional measures $s\mapsto e^{-\beta_1 V(s,y)}/Z_1(y)\,$, $y\mapsto e^{-\beta_2 F(y)}/Z_2$ also verify LSI$_N$($\beta_1c_1$), LSI$_N$($\beta_2c_2$) respectively.\\
Finally, assumptions \textup{(A1),(A2)} are obviously true and assumptions \textup{(A3)-(A5)} follow by strong convexity of $v_\textup{c}$ and tensorisation \eqref{eq:onebody_tensorisation}.
\end{proof}

\subsubsection{Interacting system ($\Delta>0$)}

\begin{proposition} \label{prop:Guerra1}
    Suppose $\big(A-\sqrt\Delta\big)\,B>\Delta_0$. Let $\beta_1,\beta_2>0$ and $\tau\in\big(0,A-\sqrt{\Delta}-\frac{\Delta_0}{B}\big)$. Then, with probability larger than
    \begin{equation}
        1- e^{-\frac{\tau^2}{4\Delta} N}
    \end{equation}
    the random potential \eqref{eq:guerra_potential} verifies the hypothesis of Theorem \ref{th:main} with rates estimated by
        \begin{equation} \label{eq:guerra_c1c2}
            c_1\,=\,\alpha_\tau\,e^{-\beta_1\frac{A}{2}N} \,,\quad c_2\,=\,\alpha_\tau\, e^{-\beta_2\frac{A}{2}\,N}
        \end{equation}
    and $\alpha_\tau>0$ smallest eigenvalue of the positive-definite matrix $\begin{pmatrix}A-\sqrt\Delta-\tau & -\sqrt{\Delta_0}\\ -\sqrt{\Delta_0} & B \end{pmatrix}$.
    Thus, with the same probability, the convergence of the two-temperatures dynamical random measure $\rhot$ to the random measure $\rhoK$ is controlled by inequalities \eqref{eq:main_conditional},\eqref{eq:main_marginal}.
\end{proposition}

For $\Delta>0$, the potential \eqref{eq:guerra_potential} does not tensorise. 
For small values of $\Delta$ we would expect rates $c_1,c_2$ independent of the dimension $N$, but this requires a more careful analysis of LSI's which is beyond the purposes of the present paper and is postponed to future work.

\begin{proof}
In order to verify logarithmic Sobolev inequalities \textup{(LS1), (LS2)} we can write $V=\Vc+\Vb$ with $\Vc$ strongly convex and $\Vb$ bounded. Using the convex function $\eta\in C^2(\R)$ defined in the previous Subsection, we set 
\begin{equation}
    \Vb(s) \,\equiv\, \frac{A}{2}\,\sum_{i=1}^N\left((s_i^2-1)^2 -\eta(s_i)\right)
\end{equation}
which is bounded with $\osc(\Vb)=\frac{A}{2}N$. Now we have to show that $\Vc(s,y) \,\equiv\, V(s,y) \,-\, \Vb(s)$ is strongly convex with large probability. Let $(\phi,\psi)=(\phi_1,\dots,\phi_N,\psi_1,\dots,\psi_N)\in\R^{2N}$ and compute the quadratic form
\begin{equation} \label{eq:ex_spinglass_hess} \begin{split}
    &(\phi,\psi)^T\, \Hess \Vc(s,y)\;(\phi,\psi) \;=\\[4pt]
    &=\, \frac{A}{2}\,\sum_i \eta''(s_i)\,\phi_i^2 \,-\, \frac{\sqrt{\Delta}}{\sqrt{2N}}\,\sum_{i, j}J_{ij}\,\phi_i\phi_j \,-\, 2\sqrt{\Delta_0}\,\sum_i\phi_i\psi_i \,+\,B\sum_i\,\psi_i^2 \;.
\end{split} \end{equation}
Observe that
\begin{equation}
    \sum_{i,j} J_{ij}\,\phi_i\,\phi_j \,\leq\, \sigma_{\max}(J)\,\sum_i\phi_i^2 
\end{equation}
where $\sigma_{\max}(J)$ denotes the largest singular value of the random matrix $J$, and by general concentration inequalities (see, e.g., Theorem 3.1.1 in \cite{Vershynin2012}) we have:
\begin{equation}
    \P\bigg(\frac{1}{\sqrt N}\,\sigma_{\max}(J)\,\leq \sqrt{2}\,+ \tau\,\sqrt{\frac{1}{2\Delta}}\,\bigg) \,\geq\, 1-e^{-\frac{\tau^2}{4\Delta} N} \,\equiv\,p_{\tau,N}
\end{equation}
for every $\tau>0\,$. 
Therefore, with probability greater than $p_{\tau,N}$ the following lower bound is valid for all $(\phi,\psi)\in\R^{2N}$
\begin{equation} \label{eq:ex_spinglass_hess2} \begin{split}
    &(\phi,\psi)^T\, \Hess \Vc(s,y)\;(\phi,\psi) \;\geq\\[4pt]
    &\geq\, \big(A-\sqrt\Delta-\tau\big)\,\sum_i \phi_i^2 \,-\, 2\sqrt{\Delta_0}\,\sum_i\phi_i\psi_i \,+\,B \sum_i\psi_i^2 \,\geq\\
    &\geq\, \alpha_\tau\;|(\phi,\psi)|^2 \;.
\end{split} \end{equation}
By Proposition \ref{prop:logSobolev_condmarg}, it follows that with probability larger that $p_{N,\tau}$ the random measures $s\mapsto e^{-\beta_1 V(s,y)}/Z_1(y)$, $y\mapsto e^{-\beta_2\Veff(y)}$ verify inequalities LSI($\beta_1 c_1$), LSI($\beta_2 c_2)$ respectively, with $c_1,c_2$ defined in \eqref{eq:guerra_c1c2}.\\
Finally, assumptions \textup{(A1),(A2)} are clearly verified, assumptions \textup{(A3)-(A4)} follow from the fact that $\Vc(s,y)$ is strongly convex, and assumption \textup{(A5)} also needs the control \eqref{eq:eta_deriv} on the first derivative.
\end{proof}

%

\subsection{High-dimensional inference (rank-one matrix estimation)}\label{subsec:inference}

An important paradigm in modern high-dimensional inference is the problem of low rank-matrix estimation \cite{Davenport-Romberg-2016}. In its simplest rank-one incarnation, the statistician is given a noisy version of a large rank-one $N_1\times N_2$ matrix with elements $u_i^*\,v_j^*$ determined by latent (hidden) vectors $u^*\in \mathbb{R}^{N_1}$, $v^*\in\mathbb{R}^{N_2}$ with independent sub-Gaussian random components. Given the data matrix 
\begin{equation} \label{eq:ex_inference_obs}
    J_{ij} \,=\, \sqrt{\frac{\Delta}{N}}\,u_i^*\,v_j^* \,+\, Z_{ij}\;,
\end{equation}
where $Z_{ij}$ are independent standard Gaussian variables, $N=N_1+N_2$, and $\Delta >0$ is the signal-to-noise ratio, the task of the statistician is to estimate the latent matrix $u^*v^*$. When prior distributions of latent vectors are known, estimates given by averages under the Bayes posterior distribution are optimal estimators (in the sense that they minimize the mean-square-error). As it turns out, the Bayes posterior in nothing else than the Gibbs distribution of a spin-glass with a replica symmetric solution (we refer to \cite{miolane2018fundamental, Barbier-Luneau-Macris-ISIT, barbier-miolane-macris-Layered} for the rigorous analysis of such  models). However, when priors are unknown to the statistician, other estimators may be better
and it is of interest to depart from the standard Bayes posterior or Gibbs distributions. The Langevin multibath dynamics may serve as a useful estimation algorithm based on a non-Bayesian (or non-Gibbsian) distribution. Whether the two-temperatures Langevin dynamics is a useful algorithmic approach to such problems is still an interesting open question in itself.
Without entering into more details here, this motivates the study of the multibath dynamics \eqref{eq:SDEintro} for the following spin-glass potential
\begin{equation} \label{eq:inference_potential} \begin{split}
    V(u,v) \,=\;\,& \frac{1}{2}\,\sum_{i=1}^{N_1}\,\sum_{j=1}^{N_2} \bigg(J_{ij}-\sqrt{\frac{\Delta}{N}}\,u_iv_j\bigg)^{\!2} \;+\\[4pt] &+\sum_{i=1}^{N_1}\bigg(\frac{a}{2}\,u_i^2 \,+\,\frac{A}{12}\, u_i^4\bigg) \,+\, \sum_{j=1}^{N_2}\bigg(\frac{b}{2}\,v_j^2 \,+\, \frac{B}{12}\, v_j^4\bigg)\;,
\end{split} \end{equation}
where the random data matrix $J_{ij}$ is generated according to \eqref{eq:ex_inference_obs} and $a,b,A,B>0$.
In this setting the data matrix is entirely quenched and degrees of freedom $u\in \mathbb{R}^{N_1}$, $v\in \mathbb{R}^{N_2}$ undergo Langevin dynamics with different temperatures and timescales. 
In a low signal-to-noise ratio regime, we show that the assumptions required by Theorem \ref{th:main} - in particular the convexity criterion analysed in Section \ref{sec:mainresults} - hold with probability exponentially close to one in the size of the system and the convergence rate is independent of the size of the system.
It is an open problem to investigate convergence in the high signal-to-noise ratio regime.

\begin{proposition}
    Suppose $A B\geq\Delta^2\,\gamma(1-\gamma)$ and $a b> K_0\,$, setting $\gamma\equiv\frac{N_1}{N}\,$,
    \begin{equation} \label{eq:inference_Ktau}
        K_{\tau} \,\equiv\, \Delta\,\gamma\,(1-\gamma)\,(\sigma_u+\tau_1)(\sigma_v+\tau_2)\,+\,\sqrt{\Delta}\,\big(\sqrt{\gamma}\,+\sqrt{1-\gamma}\,+\tau_0\big)
    \end{equation}
    for $\tau=(\tau_0,\tau_1,\tau_2)\in[0,\infty)^3$, and
    $\sigma_u \equiv \sqrt{\E (u_1^*)^2}\,$, $\sigma_v \equiv \sqrt{\E (v_1^*)^2}\,$.\\[2pt]
    Let $\tau_0,\tau_1,\tau_2>0$ such that $ab>K_\tau$.
    Then, with probability larger than
    \begin{equation} \label{eq:inference_pN}
        p_{\tau,N} \,\equiv\, 1-e^{-\frac{1}{2}\tau_0^2N}-e^{-k_u\tau_1^2N_1}-e^{-k_v\tau_2^2N_2}
    \end{equation}
    where the constants $k_u,k_v>0$ depend only on the sub-Gaussian distributions of $u_1^*,v_1^*\,$,
    the random potential \eqref{eq:inference_potential} verifies the hypothesis of Theorem \ref{th:main} with rates independent of $N$ and estimated by
        \begin{equation} \label{eq:inference_c1c2}
            c_1\,=\,c_2\,=\,\textrm{smallest eigenvalue of the positive-definite matrix } \begin{pmatrix}a & -K_\tau\\-K_\tau  & b \end{pmatrix} \;.
        \end{equation}
    Thus, with the same probability, the convergence of the two-temperatures dynamical random measure $\rhot$ to the random measure $\rhoK$ is controlled by inequalities \eqref{eq:main_conditional},\eqref{eq:main_marginal}.
\end{proposition}

\begin{proof}
We claim that the potential $V$ is a strongly convex polynomial in the variables $u,v$.
For $(\phi,\psi)=(\phi_1,\dots,\phi_{N_1},\psi_1,\dots,\psi_{N_2})\in\R^N$ let us compute:
\begin{equation} \label{eq:ex_inference_hess1} \begin{split}
    (\phi,\psi)^T\,\Hess V(u,v)\;(\phi,\psi) \,=\;\,& \frac{\Delta}{N}\,\big(\,|\phi|^2\,|v|^2 + |u|^2\,|\psi|^2 + 4\,(u\cdot\phi)\,(v\cdot\psi) \big) \;+\\[4pt]
    &- 2\,\sqrt{\frac{\Delta}{N}}\,\sum_{i,j}J_{ij}\,\phi_i\psi_j \;+\\[4pt]
    &+\,
    \sum_i \big(a + A\,u_i^2\big)\,\phi_i^2 \,+\,
    \sum_j \big(b + B\,v_j^2\big)\,\psi_j^2 \;.
\end{split} \end{equation}
It is convenient to rewrite the random interaction term using expression \eqref{eq:ex_inference_obs}:
\begin{equation} \begin{split}
    \sqrt{\frac{\Delta}{N}}\,\sum_{i,j}J_{ij}\,\phi_i\psi_j \,&=\,
    \frac{\Delta}{N}\,\sum_{i,j}u_i^*\,\phi_i\,v_j^*\,\psi_j \,+\,\sqrt{\frac{\Delta}{N}}\,\sum_{i,j}Z_{ij}\,\phi_i\psi_j \\[4pt]
    &\leq\, \frac{\Delta}{N}\, |u^*|\,|v^*|\,|\phi|\,|\psi| \,+\, \sqrt{\frac{\Delta}{N}}\;\sigma_{\max}(Z)\,|\phi|\,|\psi|\;,
\end{split} \end{equation}
where $\sigma_{\max}(Z)$ denotes the largest singular value of the $N_1\times N_2$ random matrix $Z$.
By general concentration inequalities (see, e.g., Corollary 3.35 in \cite{Vershynin2012} and Theorem  3.1.1 in \cite{Vershynin_book}), for any $\tau_0,\tau_1,\tau_2>0$ we have:
\begin{align}
    \P\bigg(\,\frac{1}{\sqrt{N_1}}\,|u^*| \,\leq\, \sigma_u+\tau_1\bigg) \,&\geq\, 1- e^{-k_u\,\tau_1^2\,N_1} \\[4pt]
    \P\bigg(\,\frac{1}{\sqrt{N_2}}\,|v^*| \,\leq\, \sigma_v+\tau_2\bigg) \,&\geq\, 1- e^{-k_v\,\tau_2^2\,N_2} \\[4pt]
    \P\bigg(\,\frac{1}{\sqrt{N}}\,\sigma_{\max}(Z) \,\leq\, \sqrt{\gamma}+\sqrt{1-\gamma}+\tau_0)\bigg) \,&\geq\, 1- e^{-\frac{1}{2}\,\tau_0^2\,N}
\end{align}
where the constants $k_u,k_v>0$ depend only on the sub-Gaussian distributions of $u_1^*,v_1^*$.
Therefore with probability larger than $p_{\tau,N}$ defined by \eqref{eq:inference_pN}, the following lower bound holds true for every $(\phi,\psi)\in\R^N$
\begin{equation} \label{eq:ex_inference_hess2} \begin{split}
    (\phi,\psi)^T\,\Hess V(u,v)\;(\phi,\psi) \,\geq\;\,& \frac{\Delta}{N}\,\big(\,|\phi|^2\,|v|^2 + |u|^2\,|\psi|^2 + 4\,(u\cdot\phi)\,(v\cdot\psi) \big) \;+\\[4pt]
    &- 2\,K_{\tau}\,|\phi|\,|\psi| \;+\\[4pt]
    &+\,\sum_i \big(a + A\,u_i^2\big)\,\phi_i^2 \,+\,
    \sum_j \big(b + B\,v_j^2\big)\,\psi_j^2 
\end{split} \end{equation}
where $K_\tau$ is defined by \eqref{eq:inference_Ktau}.
The r.h.s. of \eqref{eq:ex_inference_hess2} splits in three terms. The first one is
\begin{equation} \label{eq:ex_inference_lb1}
\frac{\Delta}{N}\,\big(\,|\phi|^2\,|v|^2 + |u|^2\,|\psi|^2 + 2\,(u\cdot\phi)\,(v\cdot\psi) \big) \,\geq\, 
\frac{\Delta}{N}\,\big(\,|\phi|\,|v| - |u|\,|\psi| \big)^2 \,\geq\,0\;.
\end{equation}
The second one is
\begin{equation} \label{eq:ex_inference_lb2} \begin{split}
    &\frac{\Delta}{N}\,2\,(u\cdot\phi)\,(v\cdot\psi)
    +\,\sum_i A\,u_i^2\,\phi_i^2 \,+\,
    \sum_j B\,v_j^2\,\psi_j^2 \;=\\
    &=\frac{1}{N}\,\sum_{i,j}\bigg(2\Delta\,u_i\,\phi_i\,v_j\,\psi_j \,+\, \frac{A}{1-\gamma}\,u_i^2\,\phi_i^2 \,+\, \frac{B}{\gamma}\,v_j^2\,\psi_j^2 \bigg) \,\geq\,0
\end{split} \end{equation}
provided
\begin{equation} \label{eq:ex_inference_cond1}
    AB\,\geq\,\gamma\,(1-\gamma)\,\Delta^2\;.
\end{equation}
The last term is
\begin{equation} \label{eq:ex_inference_lb3}
-2 K_\tau\,|\phi|\,|\psi| \,+\, a\,|\phi|^2 \,+\, b\, |\psi|^2 \,\geq\, \alpha_\tau\,(|\phi|^2+|\psi|^2)
\end{equation}
where $\alpha_\tau>0$ is the smallest eigenvalue 
of the positive-definite matrix $\begin{pmatrix}a & -K_\tau\\ -K_\tau & b\end{pmatrix}$, provided
\begin{equation} \label{eq:ex_inference_cond2}
    ab \,>\, K_\tau^2 \;.
\end{equation}
Plugging inequalities \eqref{eq:ex_inference_lb1}, \eqref{eq:ex_inference_lb2}, \eqref{eq:ex_inference_lb3} into \eqref{eq:ex_inference_hess2} we conclude that with probability larger that $p_{\tau,N}$ 
\begin{equation}
    \Hess V(u,v) \,\succeq\, \alpha_\tau\, I \;.
\end{equation}
Inequalities \textup{(LS1),(LS2)} follow by Proposition \ref{prop:logSobolev_condmarg}. Assumptions \textup{(A1),(A2)} are clearly true, assumptions \textup{(A3)-(A5)} also follow. 
\end{proof}

\section{Proof of Theorem \ref{th:main} and further results} \label{sec:main_proof}

The strategy to prove Theorem \ref{th:main} is based on the estimation of free energies dissipation along the two-temperatures dynamics. Logarithmic Sobolev inequalities \textup{(LS1)}, \textup{(LS2)} play a fundamental role in this approach. On the other hand, differently from the single-temperature case, several remainder terms appear and one needs to control them by showing they uniformly bounded in $t,\lambda$. With the help of Csizar-Kullback-Pinsker inequality, one finally obtains a system of two differential inequalities for $\Da(t,\lambda)$, $\Db(t,\lambda)$ that can be easily integrated leading to inequalities \eqref{eq:main_conditional},\eqref{eq:main_marginal}.\\
The core of the proof is developed in Subsections \ref{subsec:convrho1}, \ref{subsec:convrho2}. Previous subsections contain auxiliary results about the Fokker-Planck equation \eqref{eq:FP}.
A bracket notation for conditional expectations will be used from now on: for any suitable observable $O$
\begin{align} \label{eq:bracket} \begin{split} 
    \langle O \rangle_{t,\lambda}\, (x_2) \,&\equiv\, \int_{\Ra} O(x_1,x_2)\, \rhota(x_1|x_2)\, \d x_1 \\[4pt]
    \langle O \rangle_{*}\, (x_2) \,&\equiv\, \int_{\Ra} O(x_1,x_2)\, \rhoKa(x_1|x_2)\, \d x_1 \;.
\end{split} \end{align}

\subsection{Existence, uniqueness and regularity of the solution}
Given $\beta_1,\beta_2,\lambda>0\,$, the Fokker-Planck differential operator defining equation \eqref{eq:FP} and its formal adjoint are
\begin{align} \label{eq:L_FP}
    \LFP\rho \;&\equiv\;  \nablax\cdot\left( \frac{1}{\beta_1}\,\nablax\rho \,+\, \rho\,\nablax  V \right) \;+\; \frac{1}{\lambda}\; \nablay\cdot\left( \frac{1}{\beta_2}\,\nablay\rho \,+\, \rho\,\nablay  V \right) \\[4pt]
\label{eq:Ldual}
   \Ldual\varphi \;&\equiv\; \left( \frac{1}{\beta_1}\,\laplax\varphi \,-\, \nablax\varphi\cdot\nablax  V \right) \;+\; \frac{1}{\lambda}\; \left( \frac{1}{\beta_2}\,\laplay\varphi \,-\, \nablay\varphi\cdot\nablay  V \right) \;.
\end{align}
They rewrite in more compact form as
\begin{align} 
 \label{eq:L_FP_matrix} 
    \LFP\rho \;&=\; \nabla \cdot\, \left(\,\Lambda^{-1}\beta^{-1}\,\nabla\rho \,+\, \Lambda^{-1}\,\nabla V\,\rho\,\right)  \\[4pt]
\label{eq:Ldual_matrix} 
   \Ldual\varphi \;&=\; \nabla\cdot\left(\Lambda^{-1}\beta^{-1}\,\nabla\varphi\right) \,-\, \Lambda^{-1}\, \nabla V\, \cdot \nabla\varphi 
\end{align}
where $\Lambda$ and $\beta$ are the following $d\times d$ diagonal matrices
\begin{equation} \label{eq:matricesLB}
    \Lambda \,\equiv\, \begin{pmatrix}
    I_{d_1} & 0 \\[2pt]
    0 & \lambda\,I_{d_2}
    \end{pmatrix} \;, \quad
    \beta \,\equiv\, \begin{pmatrix}
    \beta_1\,I_{d_1} & 0 \\[2pt]
    0 & \beta_2\,I_{d_2}
    \end{pmatrix} \;.
\end{equation}
Sometimes we will write $\Ldual_\lambda,\,\LFP_\lambda$ in order to stress the dependence of these operators on the friction coefficient $\lambda\,$.

From the general theory of Fokker-Planck PDEs we have the following 

\begin{theorem}[Existence, uniqueness and regularity of the solution of Fokker-Planck equation \eqref{eq:FP}] \label{th:existence_uniq_reg}
    Suppose that the potential $V$ verifies assumptions \textup{(A1)-(A4)} and the initial probability measure $\mui$ has a density $\rhoi$ verifying assumption \textup{(B3)}.\\
    Then there exists a unique collection $\mu\equiv(\mu_t)_{t>0}$ of Borel probability measures on $\R^d$ such that
    \begin{equation}\label{eq:FPdef}
            \int_{\R^d}\varphi(x)\;\mu_t(\d x) - \int_{\R^d}\varphi(x)\;\mui(\d x) \;=\; \int_0^t\!\int_{\R^d} \Ldual\varphi(x)\;\mu_s(\d x)\;\d s
    \end{equation}
    for every $t>0$ and every $\varphi\in C^2_c(\Rd)$.
    Moreover there exists a 
    continuous positive density flux $\rho:\Rd\times(0,\infty)\to(0,\infty)$ such that for every $t>0$
    \begin{equation} \label{eq:density}
        \mu_t(\d x) \,=\, \rho_t(x) \,\d x \;.
    \end{equation}
    In fact, 
    \begin{itemize}
        \item[i.] $\rho\in C^\infty(\R^d\!\times\!(0,\infty))\,$, hence it is a strong solution of
        \begin{equation} \label{eq:FP_strong}
            \partial_t \rho \,=\, \LFP\rho 
        \end{equation}
        \item[ii.] $\|\rhoT-\rhoi\|_{L_1(\R^d)}\to0$ as $t\to0_+\,$.
\end{itemize}
\end{theorem}

\begin{remark}
Equation \eqref{eq:FPdef} is a weak formulation of \eqref{eq:FP}.
Let $\varphi\in C^2_c(\Rd)$. The function $s\mapsto\int_{\R^d}\Ldual\varphi(x)\rhos(x)\,\d x\,$ is continuous on $(0,\infty)$ by continuity of $\rho$ and dominated convergence. 
The Fundamental Theorem of Calculus applies and identity \eqref{eq:FPdef} entails that
$t\mapsto\int_{\R^d} \varphi(x)\,\rho_t(x)\,\d x$ belongs to $C^1(0,\infty)$,
\begin{equation} \label{eq:FPdef2}
    \frac{\d}{\d t} \int_{\R^d} \varphi(x)\,\rho_t(x)\,\d x \,=\, \int_{\R^d} \Ldual\varphi(x)\,\rho_t(x)\,\d x \;,
\end{equation}
and
\begin{equation} \label{eq:FPdef_continuity}
    \int_{\R^d} \varphi(x)\,\rho_t(x)\,\d x \,\xrightarrow[t\to0]{}\, \int_{\R^d} \varphi(x)\,\rho_I(x)\,\d x \;.
\end{equation}
Finally, as $\rhoT\in C^2(\Rd)$, integrating by parts one finds
\begin{equation} \label{eq:FPbyparts}
    \int_{\R^d} \Ldual\varphi(x)\,\rho_t(x)\,\d x \,=\,
    \int_{\R^d} \varphi(x)\,\LFP\rho_t(x)\,\d x
\end{equation}
Identities \eqref{eq:FPdef2}, \eqref{eq:FPbyparts} for every $\varphi\in C^2_c(\Rd)$ give equation \eqref{eq:FP_strong}.
\end{remark}

Theorem \ref{th:existence_uniq_reg} is standard. We refer, e.g., to \cite{BKRS} for existence and uniqueness of the solution and to \cite{JKO} for its regularity.

In \cite{BKRS} the problem is studied under great generality: see in particular Sections 6.1, 6.3, 6.4, 6.6, 9.1, 9.3, and 9.4 therein.
For the interested reader, it may be worth remarking that in our case, thanks to the existence of a continuous density flux $\rho$, equation \eqref{eq:FPdef} holds true for every $t>0$, not only for almost every $t$. 
We also notice that if $\mu_t$ are sub-probability measures satisfying equation \eqref{eq:FPdef} then $\mu_t$ are in fact probability measures. 
Several conditions guaranteeing the uniqueness of a (sub-)probability solution are provided in \cite{BKRS}; among them the condition $\Lambda^{-1}\nabla V\in L^1(\R^d\!\times\!(0,T),\mu_t(\d x)\d t)$ for all $T>0$, which we discuss here in Subsection \ref{subsec:expectation_poly}.



The regularity of solution can be obtained adapting the bootstrap argument sketched in \cite{JKO}. A complete proof is postponed to the Appendix, Subsection \ref{subsec:proof_regularity}.

\subsection{Integrability of $D^\nu V$, $D^\nu F$, $\log\rho$, $\nabla\log\rho$}
\label{subsec:expectation_poly}

The results stated in this Subsection will be extensively used.
A crucial fact in order to control remainders is that any product of derivatives of the potential $V$ has uniformly bounded expectations for all $t>0\,$, $\lambda\geq1\,$, where expectations may be either taken either with respect to the dynamical measure $\rhot$ or to the measure $\rhoKa\cdot\rhotb\,$.
As a consequence, expectations of derivatives of the effective potential $\Veff=-\beta_1^{-1}\log\int e^{-\beta_1 V}\d x_1$ with respect to the measure $\rhotb$ turn out to be also uniformly bounded for all $t>0\,$, $\lambda\geq1\,$. 

\begin{theorem} \label{th:expectation_norm_lapla}
Suppose that $V$ verifies assumptions \textup{(A1)-(A5)} and $\rhoi$ verifies assumption \textup{(B3)}.
Then for every finite collection of multi-indices $\nu_1,\dots,\nu_n\in\N^d$ and exponents $s_1,\dots,s_n\in[0,\infty)$
\begin{equation} \label{eq:expectation_DV}
    \sup_{t>0,\,\lambda\geq1}\;\int_{\R^d} \,\prod_{i=1}^n\big|D^{\nu_i} V(x)\big|^{s_i}\,\rhot(x)\,\d x \;<\infty  \;.
\end{equation}
\end{theorem}

\begin{theorem} \label{th:condexpectation_norm}
Suppose that $V$ verifies assumptions \textup{(A1)-(A4)} and $\rhoi$ verifies assumption \textup{(B3)}.
Then for every finite collection of multi-indices $\nu_1,\dots,\nu_n\in\N^{d}$ and exponents $s_1,\dots,s_n\in[0,\infty)$
\begin{equation} \label{eq:expectation_DV_rhoKrhot}
    \sup_{t>0,\,\lambda\geq1}\;\int_{\R^d} \,\prod_{i=1}^n\big|D^{\nu_i} V(x_1,x_2)\big|^{s_i}\,\rhoKa(x_1|x_2)\,\rhotb(x_2)\;\d x_1\,\d x_2 \;<\infty  \;.
\end{equation}
As a consequence, for every $\nu\in\N^{d_2}$, 
$s\in[0,\infty)$
\begin{equation} \label{eq:expectation_DF}
    \sup_{t>0,\,\lambda\geq1}\; \int_{\Rb}\big|D_{x_2}^\nu\Veff(x_2)\big|^s\,\rhotb(x_2)\;\d x_2 \;<\, \infty \;.
\end{equation}
\end{theorem}

The proofs of Theorems \ref{th:expectation_norm_lapla}, \ref{th:condexpectation_norm} are postponed to the Appendix, Subsections \ref{subsec:proof_expectations}, \ref{subsec:proof_condexpectations}. In fact, by a non trivial use of Gronwall inequality, it will be shown that the expectation of any polynomial is uniformly bounded in $t>0$, $\lambda\geq1$. Explicit bounds will be provided.

As a consequence of Theorems \ref{th:expectation_norm_lapla}, \ref{th:condexpectation_norm} we have the following results.

\begin{corollary} \label{cor:expectations_log*}
Suppose that $V$ verifies assumptions \textup{(A1)-(A5)} and $\rhoi$ verifies assumption \textup{(B3)}. Then the supremum over $t>0,\,\lambda\geq1$ of the following integrals are finite:
\begin{equation*} \begin{split}
    \int_{\R^d}\big|\log\rhoKa(x_1|x_2)\big|^2\,\rhot(x)\,\d x\,,\ 
    \int_{\Rb}\big|\log\rhoKb(x_2)\big|^2\,\rhotb(x_2)\,\d x_2 \,,\\[4pt]
    \int_{\R^d}\big|\nabla\log\rhoKa(x_1|x_2)\big|^2\,\rhot(x)\,\d x \,,\ 
    \int_{\R^d}\big|\nablay\log\rhoKb(x_2)\big|^2\,\rhotb(x_2)\,\d x_2 \;.
\end{split} \end{equation*}
\end{corollary}

\begin{proof}
It follows by Theorems \ref{th:expectation_norm_lapla}, \ref{th:condexpectation_norm} since  $\log\rhoKa(x_1|x_2) = -\beta_1 (V(x_1,x_2)-\Veff(x_2))$
and $\log\rhoKb(x_2) = -\beta_2 \Veff(x_2) -\log Z_2\,$.
\end{proof}

\begin{proposition}[Log integrability] \label{prop:integrability_log}
Suppose that $V$ verifies assumptions \textup{(A1)-(A4)} and $\rhoi$ verifies assumptions \textup{(B1), (B3)}. Then for every $T>0$ we have $\|\rho\|_{L^\infty(\R^d\times(0,T))}<\infty$ and
\begin{equation}
    \sup_{t\in(0,T)}\;\int_{\R^d} |\log\rhoT(x)|^2\,\rhoT(x)\,\d x \,<\infty \;.
\end{equation}
\end{proposition}

\begin{theorem}[Logarithmic gradient integrability] \label{th:integrability_gradlog}
Suppose that $V$ verifies assumptions \textup{(A1)-(A4)} and $\rhoi$ verifies assumption \textup{(B1), (B3)}.
Then $\rhoT\in W^{1,1}(\R^d)$ for a.e. $t>0$ and
\begin{equation} \label{eq:integrability_gradlog}
    \int_0^T \!\!\int_{\R^d} \big|\nabla\log\rhoT(x)\big|^2\,\rhoT(x)\,\d x\,\d t \,<\infty
\end{equation}
for all $T\in(0,\infty)\,$. As a consequence there exists a subset $N=N(\lambda,\beta_1,\beta_2,\rhoi,V)$ of $(0,\infty)$ of zero Lebesgue measure such that for all $t\in(0,\infty)\setminus N$
\begin{equation} \label{eq:integrability2_gradlog}
    \int_{\R^d} \big|\nabla\log\rhoT(x)\big|^2\,\rhoT(x)\,\d x \,<\infty\;.
\end{equation}
\end{theorem}

The proofs of Proposition \ref{prop:integrability_log} and Theorem \ref{th:integrability_gradlog} are postponed to the Appendix, Subsection \ref{subsec:proof_expectations} and will mainly refer to general results contained \cite{BKRS}, Chapter 7.




\subsection{Fokker-Planck equation for the marginal $\rhotb$} \label{subsec:FPmarg}

In this Subsection the marginal probability density $\rhoTb(x_2)\equiv\int_{\R^d}\rhoT(x_1,x_2)\,\d x_1$ is proved to satisfy in turn a Fokker-Planck equation with time-dependent drift:
\begin{equation} \label{eq:FPstrong_marg}
    \partial_t\rhob\,=\, \frac{1}{\lambda}\,\nablay\cdot\left(\frac{1}{\beta_2}\,\nablay\rhob \,+\, \rhob\;\langle\nablay V\rangle_t \right)\;,
\end{equation}
where the bracket $\langle\cdot\rangle_t$ has been defined by \eqref{eq:bracket} as the conditional expectation with respect to $\rhoTa(x_1|x_2)$.
Regularity and integrability results for $\rhob$ and $\rhoa$ will be stated.
Let us rewrite the Fokker-Planck operator \eqref{eq:Ldual} as
\begin{equation}
    \Ldual \,=\, \Ldualx \,+\; \frac{1}{\lambda}\,\Ldualy
\end{equation}
where we set
\begin{align}
    \Ldualx \varphi \,&\equiv\, \frac{1}{\beta_1}\,\nablax^2\varphi \,-\, \nablax V\cdot\nablax\varphi \\[4pt]
    \Ldualy \varphi \,&\equiv\, \frac{1}{\beta_2}\,\nablay^2\varphi \,-\, \nablay V\cdot\nablay\varphi \;.
\end{align}
We introduce a time-dependent Fokker-Planck operator on $\Rb\,$:
\begin{equation}
    \Ldualyt \,\psi \,\equiv\, \frac{1}{\beta_2}\,\nablay^2\psi \,-\, \langle\nablay V\rangle_t\cdot\nablay\psi\;.
\end{equation}

\begin{theorem} \label{th:FP_marg}
In the hypothesis of Theorem \ref{th:existence_uniq_reg}, the marginal measure $\mutb\equiv\rhoTb \d x_2$ verifies
\begin{equation} \label{eq:FP_marg}
    \int_{\Rb}\psi(x_2)\,\mutb(\d x_2) \,-\, \int_{\Rb}\psi(x_2)\,\muib(\d x_2) 
    \,=\, \frac{1}{\lambda}\,\int_0^t\!\int_{\Rb}\!\Ldualys\psi(x_2)\;\musb(\d x_2)\,\d s
\end{equation}
for every $\psi\in C^2_c(\Rb)$, every $t>0\,$, setting $\muib\equiv\rhoib\d x_2\,$.
\end{theorem}

\begin{remark}
We can view equation \eqref{eq:FPstrong_marg} as a linear Fokker-Planck equation for $\rhob$, with $\rhoa$ given. 
Equation \eqref{eq:FP_marg} is a weak formulation of \eqref{eq:FPstrong_marg}.
\end{remark}

\begin{proof}[Proof of Theorem \ref{th:FP_marg}]
Let $\psi\in C^2_c(\Rb)\,.$ Consider a sequence $\zeta_N\in C^2_c(\Ra)$ such that $\zeta_N(x_1)=1$ for $|x_1|\leq N$ and $\zeta_N\,$, $\nablax\zeta_N\,$, $\Hessx\zeta_N$ are uniformly bounded for all $N\in\N\,$. Set
\begin{equation}
    \varphi_N(x_1,x_2)\,\equiv\, \zeta_N(x_1)\,\psi(x_2) \;.
\end{equation}
Now $\varphi_N\in C^2_c(\R^d)$, hence by equation \eqref{eq:FPdef}
\begin{equation} \label{eq:appo_FPmarg}
    \int_{\R^d}\varphi_N(x)\rhoT(x)\,\d x
    - \int_{\R^d}\varphi_N(x)\rhoi(x)\,\d x 
    \,=\, \int_0^t\!\int_{\R^d} \Ldual\varphi_N(x)\,\rhos(x)\,\d x\,\d s \;.
\end{equation}
For every $x=(x_1,x_2)\in\R^d$ we have $\varphi_N(x)\to\psi(x_2)$ as $N\to\infty$ and
\begin{equation}
\Ldual\varphi_N(x) \,=\, (\Ldualx\zeta_N)(x)\;\psi(x_2) \,+\, \frac{1}{\lambda}\,\zeta_N(x_1)\,(\Ldualy\psi)(x) \;\rightarrow\; \frac{1}{\lambda}\,(\Ldualy\psi)(x) \;.
\end{equation}
Since $\nablax V\in L^1(\R^d\!\times(0,t),\rhos(x)\d x\d s)$ by Theorem \ref{th:condexpectation_norm} (actually Corollary \ref{cor:expectation_xt} suffices), there is dominated convergence. Therefore letting $N\to\infty$ identity \eqref{eq:appo_FPmarg} gives
\begin{equation} \label{eq:appo2_FPmarg}
    \int_{\R^d}\psi(x_2)\rhoT(x)\,\d x
    - \int_{\R^d}\psi(x_2)\rhoi(x)\,\d x 
    \,=\, \frac{1}{\lambda}\,\int_0^t\!\int_{\R^d} (\Ldualy\psi)(x)\rhos(x)\,\d x\,\d s \;.
\end{equation}
Now integrate with respect to $x_1$ first, as $\psi$ depends on $x_2$ only. Since $\nablay V\in L^1(\R^d\times(0,t),\,\rhos(x)\,\d x\,\d s)$ by Theorem \ref{th:condexpectation_norm} (Proposition \ref{prop:expectation_vlambda}), Fubini theorem applies. Therefore equation \eqref{eq:appo2_FPmarg} rewrites as \eqref{eq:FP_marg}.
\end{proof}

\begin{theorem}[Marginal regularity] \label{th:regularity2}
In the hypothesis of Theorem \ref{th:existence_uniq_reg} we have
\begin{itemize}
    \item[i.] $\rhob\in C^\infty(\Rb\!\times(0,\infty))\,$, hence it is a strong solution of Fokker-Planck equation \eqref{eq:FPstrong_marg} ;
    \item[ii.] $\|\rhoTb-\rhoib\|_{L^1(\Rb)}\to0$ as $t\to0\,$.
\end{itemize}
\end{theorem}

Theorem \ref{th:regularity2} has been obtained modifying the bootstrap argument in \cite{JKO} and relies on the integrability results of Subsection \ref{subsec:expectation_poly}. 
Its proof is postponed to the Appendix, Subsection \ref{subsec:proof_regularity2}.

\begin{proposition}[Marginal log integrability] \label{prop:integrability_log2}
Suppose that $V$ verifies assumptions \textup{(A1)-(A4)} and $\rhoi$ verifies assumptions \textup{(B2), (B3)}. Then for every $T>0$ we have $\|\rhob\|_{L^\infty(\Rb\times(0,T))}<\infty$ and
\begin{equation}
    \sup_{t\in(0,T)} \int_{\Rb} \big|\log\rhoTb(x_2)\big|^2\,\rhoTb(x_2)\;\d x_2 \,<\infty \;.
\end{equation}
\end{proposition}

\begin{theorem}[Marginal logarithmic gradient integrability] \label{th:integrability_gradlog2}
Suppose that $V$ verifies assumptions \textup{(A1)-(A4)} and $\rhoi$ verifies assumptions \textup{(B2), (B3)}.
Then $\rhoTb\in W^{1,1}(\Rb)$ for a.e. $t>0$ and
\begin{equation} \label{eq:integraibility_gradlog2}
    \int_0^T \!\!\int_{\Rb} \big|\nablay\log\rhoTb(x_2)\big|^2\,\rhoTb(x_2)\;\d x_2\,\d t \,<\infty
\end{equation}
for all $T\in(0,\infty)\,$. As a consequence there exists a subset $N=N(\lambda,\beta_1,\beta_2,\rhoi,V)$ of $(0,\infty)$ of zero Lebesgue measure such that for all $t\in(0,\infty)\setminus N$
\begin{equation} \label{eq:integraibility2_gradlog2}
    \int_{\Rb} \big|\nablay\log\rhoTb(x_2)\big|^2\,\rhoTb(x_2)\,\d x_2 \,<\infty\;.
\end{equation}
\end{theorem}

Proofs of Proposition \ref{prop:integrability_log2} and Theorem \ref{th:integrability_gradlog2} are postponed to the Appendix, Subsection \ref{subsec:proof_expectations} and will refer to general results of \cite{BKRS}, since $\rhob$ is itself solution of a suitable Fokker-Planck equation.

\begin{corollary}[Conditional regularity] \label{cor:regularity1}
In the hypothesis of Theorem \ref{th:existence_uniq_reg}, the conditional density $\rhoa\in C^\infty(\Ra\!\times\!\Rb\!\times\!(0,\infty))\,$.
\end{corollary}

\begin{proof}
It follows immediately by Theorems \ref{th:existence_uniq_reg}, \ref{th:regularity2} since $\rho>0$ and $\rhoa(x_1|x_2)\equiv\frac{\rho(x_1,x_2)}{\rhob(x_2)}$.
\end{proof}

\begin{corollary}[Conditional log integrability] \label{cor:integrability_log1}
Suppose that $V$ verifies assumptions \textup{(A1)-(A4)} and $\rhoi$ verifies assumptions \textup{(B1)-(B3)}. Then for every $T>0$
\begin{equation}
    \sup_{t\in(0,T)} \int_{\R^d} \big|\log\rhoTa(x_1|x_2)\big|^2\,\rhoT(x_1,x_2)\;\d x_1\,\d x_2 \,<\infty \;.
\end{equation}
\end{corollary}

\begin{proof}
It follows immediately combining Propositions \ref{prop:integrability_log}, \ref{prop:integrability_log2}.
\end{proof}

\begin{corollary}[Conditional logarithmic gradient integrability] \label{cor:integrability_gradlog1}
Suppose that $V$ verifies assumptions \textup{(A1)-(A4)} and $\rhoi$ verifies assumption \textup{(B1)-(B3)}.
Then $\rhoTa\in W^{1,1}\big(\R^d,\rhoTb\d x\big)$ for a.e. $t>0$ and
\begin{equation} \label{eq:integrability_gradlog1}
    \int_0^T \!\!\int_{\R^d} \big|\nabla\log\rhoTa(x_1|x_2)\big|^2\,\rhoT(x_1,x_2)\,\d x_1\,\d x_2\,\d t \,<\infty\;.
\end{equation}
As a consequence $T\in(0,\infty)$ and there exists a subset $N=N(\lambda,\beta_1,\beta_2,\rhoi,V)$ of $(0,\infty)$ of zero Lebesgue measure such that for all $t\in(0,\infty)\setminus N$
\begin{equation} \label{eq:integrability2_gradlog1}
    \int_{\R^d} \big|\nabla\log\rhoTa(x_1|x_2)\big|^2\,\rhoT(x_1,x_2)\,\d x_1\,\d x_2 \,<\infty\;.
\end{equation}
\end{corollary}

\begin{proof}
It follows immediately combining Theorems \ref{th:integrability_gradlog}, \ref{th:integrability_gradlog2}.
\end{proof}

\subsection{Joint and marginal entropy dissipation} \label{subsec:KLderivatives}

The proof of Theorem \ref{th:main} is based on the estimation of time-derivatives of the KL-divergences $\D(t,\lambda),\,\Db(t,\lambda)$.  Formally one differentiate the integrals in \eqref{eq:KL} with respect to time $t$, takes the derivative inside the integral, applies the Fokker-Planck equation, and integrates by parts.
The following propositions provide a rigorous integral version of this initial steps.

\begin{proposition} \label{prop:KLdissipation}
Suppose that $V$ verifies assumptions \textup{(A1)-(A4)} and $\rhoi$ verifies assumptions \textup{(B1), (B3)}.
Then for every $t,\lambda>0$
\begin{equation} \label{eq:KLdissipation}
\begin{split}
    &\D(t,\lambda) \,-\, \Di \,=\\ &=-\int_0^t\!\int_{\R^d}  \Lambda^{-1}\big(\beta^{-1}\nabla\log\rhoS \,+\,\nabla V\big)(x)\,\cdot\,\nabla\log\frac{\rhoS}{\rhoK}(x)\,\rhoS(x)\,\d x\,\d s \;.
\end{split} \end{equation}
As a consequence, the function $t\mapsto\D(t,\lambda)$ extended with $\Di$ at $t=0$ is absolutely continuous on compact subsets of $[0,\infty)$ and there exists a subset of zero Lebesgue measure $N_\lambda=N_\lambda(\beta_1,\beta_2,\rhoi,V)\subseteq(0,\infty)$ such that for all $t\in(0,\infty)\setminus N_\lambda$
\begin{equation} \label{eq:KLdissipation_der}
    \frac{\d}{\d t}\D(t,\lambda) \,=\, \int_{\R^d}  \Lambda^{-1}\big(\beta^{-1}\nabla\log\rhot +\,\nabla V\big)(x)\,\cdot\,\nabla\log\frac{\rhot}{\rhoK}(x)\,\rhot(x)\,\d x \;.
\end{equation}
\end{proposition}

\begin{proposition} \label{prop:KL2dissipation}
Suppose that $V$ verifies assumptions \textup{(A1)-(A4)} and $\rhoi$ verifies assumptions \textup{(B2), (B3)}.
Then for every $t,\lambda>0$
\begin{equation} \label{eq:KL2dissipation} \begin{split}
    &\Db(t,\lambda) \,-\, \Dbi \,=\\ &=-\frac{1}{\lambda}\,\int_0^t\!\int_{\Rb} \! \bigg(\frac{1}{\beta_2}\,\nablay\log\rhoSb \,+\,\langle\nablay V\rangle_{s,\lambda}\bigg)(x_2)\cdot\nablay\log\frac{\rhoSb}{\rhoKb}(x_2)\,\rhoSb\,(x_2)\;\d x_2\,\d s\;.
\end{split} \end{equation}
As a consequence, the function $t\mapsto\Db(t,\lambda)$ extended with $\Dbi$ at $t=0$ is absolutely continuous on compact subsets of $[0,\infty)$ and there exists a subset of zero Lebesgue measure $N_\lambda=N_\lambda(\beta_1,\beta_2,\rhoi,V)\subseteq(0,\infty)$ such that for all $t\in(0,\infty)\setminus N_\lambda$
\begin{equation} \label{eq:KL2dissipation_der}
    \frac{\d}{\d t}\Db(t,\lambda) \,= -\,\frac{1}{\lambda}\,\int_{\Rb}\! \bigg(\frac{1}{\beta_2}\,\nablay\log\rhotb \,+\,\langle\nablay V\rangle_{t,\lambda}\bigg)(x_2)\,\cdot\,\nablay\log\frac{\rhotb}{\rhoKb}(x_2)\,\rhotb\,(x_2)\;\d x_2 \;.
\end{equation}
\end{proposition}

The proofs of Propositions \ref{prop:KLdissipation}, \ref{prop:KL2dissipation} are postponed to the Appendix, Subsection \ref{subsec:proof_KLdiss}. They rely on the integrability results of Subsection \ref{subsec:expectation_poly} and on the regularity guaranteed by Theorems \ref{th:existence_uniq_reg}, \ref{th:regularity2}.

\subsection{Convergence of the conditional measure $\rhota$ to $\rhoKa\,$: proof of Theorem \ref{th:main}, first part} \label{subsec:convrho1}

We are ready to prove our main Theorem \ref{th:main}. This Subsection is devoted to study of the conditional measure, while the next one will devoted to the marginal measure.
In both Subsections $N_\lambda$ will denote a zero Lebesgue measure subset of $(0,\infty)$ which may depend on $\lambda$, $\beta_1$, $\beta_2$, $\rhoi$, $V$: one may take the intersection of the two sets introduced in Propositions \ref{prop:KLdissipation}, \ref{prop:KL2dissipation}.

\begin{theorem} \label{th:conditional}
    Suppose that $V$ verifies assumptions \textup{(A1)-(A5)}, the logarithmic Sobolev inequality \textup{(LS1)} holds true with constant $c_1=c_1(\beta_1,V)$, and $\rhoi$ verifies assumptions \textup{(B1)-(B3)}.
    Then there exists a finite non-negative constant $c_0=c_0(\beta_1,\beta_2,\rhoi,V)$ such that for all $\lambda\geq1\,$, $t\in(0,\infty)\setminus N_\lambda$
    \begin{equation} \label{eq:dt_conditional}
        \frac{\d}{\d t}\Da(t,\lambda) \,\leq\, -2c_1 \Da(t,\lambda) \,+\, \frac{c_0}{\lambda} \;.
    \end{equation}
    A suitable choice for $c_0$ is to take the maximum among $0$ and the supremum over $t>0\,$, $\lambda\geq1$ of the following integral
    \begin{equation} \label{eq:c0_upperbound} 
        \int_{\R^d}\! \bigg(\Big(\frac{\beta_2}{4}-\beta_1\Big)\,|\nablay V|^2 \,+\,\frac{\beta_1^2}{4\beta_2}\,|\nablay \Veff|^2 \,+\, \frac{\beta_1}{2}\,\nablay V\cdot\nablay\Veff \,
+\,\frac{\beta_1}{\beta_2}\,\big(\nablay^2V -\nablay^2\Veff\big)\bigg) \rhot(x)\,\d x \,.
    \end{equation}
\end{theorem}

\begin{remark} \label{rem:compute_c0}
According to Theorem \ref{th:conditional}, $c_0$ can be evaluated providing an upper bound for expression \eqref{eq:c0_upperbound} independent of $t>0\,$, $\lambda\geq1\,$.
In practice this can be done as follows.
Bound $|\nablay V|^2$, $\nablay^2V$ by expressions of type $\gamma_0+\gamma_1\,|x_1|^{r_1}+\gamma_2\,|x_2|^{r_2}$, which is possible by assumption \textup{(A2)}.
Observe that
\begin{align}
    \nablay \Veff \,&=\, \langle\nablay V\rangle_* \\[4pt]
    \nablay^2\Veff \,&=\, \langle\nablay^2 V\rangle_* \,-\, \beta_1\,\langle|\nablay V|^2\rangle_* \,+\, \beta_1\,|\langle\nablay V\rangle_*|^2 \;,
\end{align}
and use Proposition \ref{prop:condexpectation_poly} to bound
$\langle |x_1|^{r_1} \rangle_*\leq C_0+C_1\,|x_2|^{s}$.
Then use Corollary \ref{cor:expectation_y} and Proposition \ref{prop:expectation_x} to bound all the terms of type $\int |x_2|^{r_2}\rhot(x)\,\d x\,$, $\int |x_2|^{s}\rhot(x)\,\d x\,$, $\int |x_1|^{r_1}\rhot(x)\,\d x$ uniformly with respect to $t,\,\lambda\,$.
Finally plug the obtained bounds in (a suitable rearrangement of) expression \eqref{eq:c0_upperbound}.
\end{remark}

\begin{proof}[Proof of Theorem \ref{th:conditional}]
Let $\lambda\geq1\,$, $t\in(0,\infty)\setminus N_\lambda\,$. It is a standard fact that the joint KL divergence \eqref{eq:KL} splits into the sum of a conditional and a marginal contribution, hence
\begin{equation} \label{eq:appo1_condconvergence}
    \frac{\d}{\d t}\Da(t,\lambda) \,=\, \frac{\d}{\d t}\D(t,\lambda) \,-\, \frac{\d}{\d t}\Db(t,\lambda) \;.
\end{equation}
On the other hand we can express $\frac{\d}{\d t}\D$ and $\frac{\d}{\d t}\Db$ according to Propositions \ref{prop:KLdissipation}, \ref{prop:KL2dissipation}.
It takes a moment to convince oneself that the integral on the r.h.s. of identity \eqref{eq:KLdissipation_der} splits into five contributions:
\begin{equation} \label{eq:appo2_condconvergence}
    \frac{\d}{\d t}\D(t,\lambda) \,=\, -I_1(t,\lambda) \,-\,\frac{1}{\lambda}\,\Big(I_2^{(11)} +\, I_2^{(12)} + I_2^{(21)} + I_2^{(22)}\Big)(t,\lambda)
\end{equation}
where
\begin{equation}
    I_1(t,\lambda) \equiv \int_{\R^d} \bigg(\frac{1}{\beta_1}\,\nablax\log\rhota(x_1|x_2) \,+\,\nablax V(x)\bigg)\,\cdot\,\nablax\log\frac{\rhota}{\rhoKa}(x_1|x_2)\,\rhot(x)\,\d x\,,
\end{equation}
\begin{equation}
    I_2^{(11)}(t,\lambda) \equiv \int_{\R^d} \bigg(\frac{1}{\beta_2}\,\nablay\log\rhota(x_1|x_2) \,+\,\nablay V(x)\bigg)\,\cdot\,\nablay\log\frac{\rhota}{\rhoKa}(x_1|x_2)\,\rhot(x)\,\d x\,,
\end{equation}
\begin{equation}
    I_2^{(22)}(t,\lambda) \equiv \int_{\R^d} \bigg(\frac{1}{\beta_2}\,\nablay\log\rhotb(x_2) \,+\,\nablay V(x)\bigg)\,\cdot\,\nablay\log\frac{\rhotb}{\rhoKb}(x_2)\,\rhot(x)\,\d x\,,
\end{equation}
\begin{equation}
    I_2^{(12)}(t,\lambda) \equiv \int_{\R^d} \frac{1}{\beta_2}\,\nablay\log\rhota(x_1|x_2) \,\cdot\, \nablay\log\frac{\rhotb}{\rhoKb}(x_2)\,\rhot(x)\,\d x\,,
\end{equation}
\begin{equation}
    I_2^{(21)}(t,\lambda) \equiv \int_{\R^d} \frac{1}{\beta_2}\,\nablay\log\rhotb(x_2) \,\cdot\, \nablay\log\frac{\rhota}{\rhoKa}(x_1|x_2)\,\rhot(x)\,\d x \,.
\end{equation}
The subscripts $_1,\,_2$ are used to indicate contributions given respectively by the first $d_1$ and last $d_2$ terms of the dot product in \eqref{eq:KLdissipation_der}. Since $\log\rho = \log\rhoa + \log\rhob\,$ the superscripts $^{(11)}$, $^{(12)}$, $^{(21)}$, $^{(22)}$ denote the four possible combinations of conditional and marginal densities in the two logarithmic gradient terms appearing in \eqref{eq:KLdissipation_der}. Only $\nablax\log\rhoa$ appears in $I_1$ since $\nablax\log\rhob=0\,$.
Notice also that by Theorems \ref{th:expectation_norm_lapla}, \ref{th:condexpectation_norm}, \ref{th:integrability_gradlog}, \ref{th:integrability_gradlog2}, and Corollaries \ref{cor:expectations_log*}, \ref{cor:integrability_gradlog1} all the previous integrals are absolutely convergent. 
Now, by Proposition \ref{prop:KL2dissipation} we have
\begin{equation} \label{eq:appo3_condconvergence}
    \frac{\d}{\d t}\Db(t,\lambda) \,= -\,\frac{1}{\lambda}\,I_2^{(22)}\!(t,\lambda)\;.
\end{equation}
Therefore identities \eqref{eq:appo1_condconvergence}, \eqref{eq:appo2_condconvergence}, \eqref{eq:appo3_condconvergence} entail that
\begin{equation} \label{eq:appo4_condconvergence}
    \frac{\d}{\d t}\Da(t,\lambda)  \,=\, -I_1(t,\lambda) - \frac{1}{\lambda}\,\Big(I_2^{(11)} + I_2^{(12)} + I_2^{(21)}\Big)(t,\lambda) \;.
\end{equation}
The first integral can be easily estimated by the logarithmic Sobolev inequality \textup{(LS1)} for $\rhoKa$. Indeed $I_1$ rewrites as
\begin{equation}
    I_1(t,\lambda) \,=\, \frac{1}{\beta_1}\, \int_{\R^d} \Big|\nablax\log\frac{\rhota}{\rhoKa}(x_1|x_2)\Big|^2\,\rhot(x)\;\d x\,.
\end{equation}
Integrating with respect to $x_1$ first and using inequality \textrm{(LS1)} with $\pi\equiv\rhota\,(\cdot|x_2)$ one obtains:
\begin{equation} \label{eq:I1_bound}
    I_1(t,\lambda) \,\geq\, 2c_1 \Da(t,\lambda) \;.
\end{equation}
Now we have to control the remainder $I_2^{(11)} \!+ I_2^{(12)}\! + I_2^{(21)}$ uniformly for $t\in(0,\infty)\setminus N_\lambda$ and $\lambda\geq1\,$.
We rewrite
\begin{equation} \label{eq:I2(11)_appo} \begin{split}
    I_2^{(11)}(t,\lambda)\,=\,& \int_{\R^d}\!\bigg(\frac{1}{\beta_2}\,\nablay\log\rhota(x_1|x_2)\,+\,\nablay V(x)\bigg)\,\cdot\\[4pt]
    &\ \cdot\Big(\nablay\log\rhota(x_1|x_2)\,+\,\beta_1\,\big(\nablay V(x)-\nablay\Veff(x_2)\big)\Big) \rhot(x)\,\d x  \;,
\end{split} \end{equation}
and we claim
\begin{equation} \label{eq:I2(12)_zero}
    I_2^{(12)}(t,\lambda)\,=\,0\,,\\[4pt]
\end{equation}
\begin{equation} \label{eq:I2(21)_byparts}
    I_2^{(21)}(t,\lambda) \,=\, -\frac{\beta_1}{\beta_2}\,\int_{\R^d}\!\Big(\nablay^2V(x)\,+\,\nablay V(x)\,\cdot\,\nablay\log\rhota(x_1|x_2)\,-\,\nablay^2 F(x_2)\Big) \rhot(x)\,\d x \,.
\end{equation}
The sum of equations \eqref{eq:I2(11)_appo}, \eqref{eq:I2(12)_zero}, \eqref{eq:I2(21)_byparts} then gives
\begin{equation} \label{eq:I2(11)(12)(21)_sum}
    \Big(I_2^{(11)} + I_2^{(12)} + I_2^{(21)}\Big)(t,\lambda) \,=\, \int_{\R^d} R_{t,\lambda}(x)\,\rhot(x)\,\d x
\end{equation}
with
\begin{equation} \begin{split}
R_{t,\lambda} \,\equiv\ \, &\frac{1}{\beta_2}\,\big|\nablay\log\rhota\big|^2 \,+\, \nablay\log\rhota\,\cdot\,\nablay V \,-\,\frac{\beta_1}{\beta_2}\,\nablay\log\rhota\,\cdot\,\nablay\Veff \;+\\[2pt]
&+\, \beta_1\,|\nablay V|^2 \,-\, \beta_1\,\nablay V \cdot \nablay\Veff \,-\,\frac{\beta_1}{\beta_2}\,\big(\nablay^2V -\nablay^2\Veff\big) \;.
\end{split} \end{equation}
The terms containing $\nablay\log\rhoSa$ can be bounded from below by using the expansion of the square $\big|a\,\nablay\log\rhoSa \,+\, b\,\nablay V \,+\, c\,\nablay \Veff\big|^2 \geq0$ with $a\equiv\beta_2^{-\frac{1}{2}}\,$, $b\equiv\frac{1}{2}\,\beta_2^{\frac{1}{2}}\,$, $c\equiv-\frac{1}{2}\,\beta_1\,\beta_2^{-\frac{1}{2}}\,$. In this way we obtain
\begin{equation} \label{eq:R_lb} \begin{split}
R_{t,\lambda} \,\geq\; &\Big(\beta_1-\frac{\beta_2}{4}\Big)\,|\nablay V|^2 \,-\,\frac{\beta_1^2}{4\beta_2}\,|\nablay \Veff|^2 \,-\, \frac{\beta_1}{2}\,\nablay V\cdot\nablay\Veff \,
-\,\frac{\beta_1}{\beta_2}\,\big(\nablay^2V -\nablay^2\Veff\big)
\end{split} \end{equation}
and the latter lower bound does not depend on $t,\lambda$ anymore.
Plugging \eqref{eq:R_lb} into \eqref{eq:I2(11)(12)(21)_sum} and using Theorems \ref{th:expectation_norm_lapla}, \ref{th:condexpectation_norm} proves that there exists a finite non-negative constant $c_0=c_0(\beta_1,\beta_2,\rhoi,V)$ such that
\begin{equation} \label{eq:I2(11)(12)(21)_bound}
    \inf_{\lambda\geq1,\,t>0,\,t\notin N_\lambda} \Big(I_2^{(11)} + I_2^{(12)} + I_2^{(21)}\Big)(t,\lambda) \,\geq\, -c_0 \;.
\end{equation}
Finally, plugging inequalities \eqref{eq:I1_bound}, \eqref{eq:I2(11)(12)(21)_bound} into expression \eqref{eq:appo4_condconvergence} gives
\begin{equation}
    \frac{\d}{\d t}\Da(t,\lambda) \,\leq\, -2c_1 \Da(t,\lambda) \,+\, \frac{c_0}{\lambda} \;.
\end{equation}
To conclude the proof it remains to prove claims \eqref{eq:I2(12)_zero}, \eqref{eq:I2(21)_byparts}.
First we have:
\begin{equation} \begin{split}
    I_2^{(12)}(t,\lambda) \,&=\, 
    \frac{1}{\beta_2}\,\int_{\Rb} \bigg[\int_{\Ra} \!\nablay\rhota(x_1|x_2)\,\d x_1\bigg] \cdot\, \nablay\log\frac{\rhotb}{\rhoKb}(x_2)\,\rhotb(x_2)\,\d x_2 \\[4pt]
    &=\, 0\;,
\end{split} \end{equation}
where the first identity is due to Fubini theorem and the last one holds true because the term inside square brackets vanishes for almost every $x_2$ (see Lemma \ref{lem:int_der} in the Appendix, Subsection \ref{subsec:proof_KLdiss}). 
Secondly, we have:
\begin{equation} \label{eq:I2(21)_bound} \begin{split}
    &I_2^{(21)}(t,\lambda) \,=\\[4pt] &= \frac{1}{\beta_2}\,\int_{\Rb}\! \nablay\rhotb(x_2) \cdot \bigg[\int_{\Ra}\!\!\nablay\rhota(x_1|x_2)\,\d x_1\,-\int_{\Ra}\!\!\nablay\log\rhoKa(x_1|x_2)\,\rhota(x_1|x_2)\,\d x_1\bigg]\,\d x_2 \\[4pt]
    &= \frac{\beta_1}{\beta_2}\,\int_{\Rb}\! \nablay\rhotb(x_2) \cdot \bigg[\int_{\Ra}\!\!\Big(\nablay V(x)-\nablay \Veff(x_2)\Big)\,\rhota(x_1|x_2)\,\d x_1\bigg]\,\d x_2 \\[4pt]
    &=
    -\,\frac{\beta_1}{\beta_2}\,\int_{\Rb} \!\rhotb(x_2)\; \bigg[\int_{\Ra}\!\!\big(\nablay^2 V(x)\,\rhota(x_1|x_2)\, +\, \nablay V(x)\cdot\nablay\rhota(x_1|x_2) \big)\,\d x_1 - \nablay^2\Veff(x_2)\bigg]\,\d x_2 \;,
\end{split} \end{equation}
where the first identity is due to Fubini theorem, the second one is due to Lemma \ref{lem:int_der}, and the last one is essentially integration by parts (justified by Lemma \ref{lem:div_int} in the Appendix, Subsection \ref{subsec:proof_KLdiss}).
Using again Fubini theorem identity, \eqref{eq:I2(21)_byparts} is finally obtained and both claims are proven.
\end{proof}

\begin{proof}[Proof of Theorem \ref{th:main}. First part]
Inequality \eqref{eq:main_conditional} follows by Theorem \ref{th:conditional} and Gronwall Lemma \ref{lem:Gronwall}, since the map $t\mapsto\Da(t,\lambda)$ is absolutely continuous on compact subsets of $[0,\infty)$ (Proposition \ref{prop:KLdissipation}).
\end{proof}

\subsection{Convergence of the marginal measure $\rhotb$ to $\rhoKb\,$: proof of Theorem \ref{th:main}, second part} \label{subsec:convrho2}

\begin{theorem}\label{th:marginal}
    Suppose that $V$ verifies assumptions \textup{(A1)-(A5)}, logarithmic Sobolev inequality \textup{(LS2)} holds true with constant $c_2=c_2(\beta_2,V)$ and $\rhoi$ verifies assumptions \textup{(B1)-(B3)}.
    Then there exists a finite non-negative constant $\tilde c_0=\tilde c_0(\beta_1,\beta_2,\rhoi,V)$ such that for all $\lambda\geq1\,$, $t\in(0,\infty)\setminus N_\lambda\,$, $\eta\in(0,1)\,$, $\epsilon>0$ 
    \begin{equation} \label{eq:dt_marginal}
        \lambda\;\frac{\d}{\d t}\Db(t,\lambda) \,\leq\, -\,(1-\eta)\,2c_2\,\Db(t,\lambda) \,+\, \frac{1}{\eta\,\epsilon}\, \Da(t,\lambda) \,+\, \frac{\epsilon}{\eta}\,\tilde c_0 \;.
    \end{equation}
    A suitable choice for $\tilde c_0$ is to take the supremum over $t>0\,$, $\lambda\geq1$ of the following quantity:
    \begin{equation} \label{eq:c0tilde_upperbound}
        \beta_2^2\,\int_{\R^d} |\nablay V|^4\,\rhot(x)\,\d x \,+\, \beta_2^2\,\int_{\R^d} |\nablay V|^4\,\rhoKa(x_1|x_2)\,\rhotb(x_2)\,\d x \;.
    \end{equation}   
\end{theorem}

\begin{remark} \label{rem:compute_tildec0}
According to Theorem \ref{th:marginal}, $\tilde c_0$ can be evaluated providing an upper bound for expression \eqref{eq:c0tilde_upperbound} independent of $t>0\,$, $\lambda\geq1\,$.
In practice we can compute a bound of type 
\begin{equation}
    |\nablay V|^4 \,\leq\, \gamma_0+\gamma_1\,|x_1|^{r_1}+\gamma_2\,|x_2|^{r_2} \,,
\end{equation}
which is possible by assumption \textup{(A2)}. 
We use Proposition \ref{prop:condexpectation_poly} to bound
$\langle |x_1|^{r_1} \rangle_*\leq C_0+C_1\,|x_2|^{s}$.
Then we use Corollary \ref{cor:expectation_y} in order to bound
$\int |x_2|^{\sigma}\rhot(x)\,\d x \,\leq\, M_\sigma\,$
for $\sigma=r_2,s\,$, and we use Proposition \ref{prop:expectation_x} to bound $\int |x_1|^{r_1}\rhot(x)\,\d x \,\leq\,  M'_{r_1}\,$.
Finally we can plug the obtained bounds into expression \eqref{eq:c0tilde_upperbound}.
\end{remark}

\begin{proof}
Let $\lambda\geq1\,$, $t\in(0,\infty)\setminus N_\lambda\,$. 
Proposition \ref{prop:KL2dissipation} provides expression \eqref{eq:KL2dissipation_der} for $\frac{\d}{\d t}\Db\,$. Adding and subtracting a term $-\frac{1}{\beta_2}\,\nablay\log\rhoKb(x_2) = \nablay \Veff(x_2)$ shows that
\begin{equation} \label{eq:J1+J2}
    \lambda\;\frac{\d}{\d t}\Db(t,\lambda) \,= -\, (J_1+J_2)(t,\lambda)
\end{equation}
where
\begin{equation}
    J_1(t,\lambda) \,\equiv\, \frac{1}{\beta_2}\,\int_{\Rb}\Big|\nablay\log\frac{\rhotb}{\rhoKb}(x_2)\Big|^2\,\rhotb(x_2)\,\d x_2 
\end{equation}
\begin{equation}
    J_2(t,\lambda) \,\equiv\, \int_{\Rb}\!\big(\langle\nablay V\rangle_{t,\lambda}-\nablay F\big)(x_2)\,\cdot\,\nablay\log\frac{\rhotb}{\rhoKb}(x_2)\;\rhotb(x_2)\,\d x_2 \;.
\end{equation}
A further term of type $J_1$ comes from $J_2$ using Cauchy-Schwarz inequality. Precisely
\begin{equation} \label{eq:J2_bound}
    |J_2(t,\lambda)| \,\leq\, \sqrt{\beta_2\,J_1(t,\lambda)\, J_3(t,\lambda)\,}\;\leq\; \eta\,J_1(t,\lambda)\,+\, \frac{\beta_2}{4\eta}\,J_3(t,\lambda)
\end{equation}
for every $\eta\in(0,1)$, where
\begin{equation} \label{eq:J3def}
    J_3(t,\lambda)\,\equiv\, \int_{\Rb}\!\big|\langle\nablay V\rangle_{t,\lambda}-\nablay F\big|^{2\,}\!(x_2)\;\rhotb(x_2)\,\d x_2 \;.
\end{equation}
Plugging \eqref{eq:J2_bound} into \eqref{eq:J1+J2} gives
\begin{equation} \label{eq:J1+J3}
    \lambda\;\frac{\d}{\d t}\Db(t,\lambda)\,\leq -(1-\eta)\,J_1(t,\lambda) \;+\; \frac{\beta_2}{4\eta}\;J_3(t,\lambda) \;.
\end{equation}
Now, $J_1$ can be estimated using the logarithmic Sobolev inequality \textup{(LS2)} for $\rhoKb\,$. Indeed taking $\pi\equiv\rhotb$ in  \textup{(LS2)} we find
\begin{equation} \label{eq:J1bound}
    J_1(t,\lambda) \,\geq\, 2c_2\,\Db(t,\lambda) \;.
\end{equation}
In order to estimate $J_3$, let us start by observing that $\nablay\Veff=\langle\nablay V\rangle_*\,$, hence
\begin{equation} \label{eq:nablaV-nablaF}
    \big|\langle\nablay V\rangle_{t,\lambda}-\nablay F\big|(x_2) \,\leq\, \int_{\Ra}\! |\nablay V(x)|\;\big|\rhota-\rhoKa\big|(x_1|x_2)\;\d x_1\;.
\end{equation}
The Csiz\'ar-Kullback-Pinsker inequality provides an upper bound for the total variation distance in terms of the KL divergence:
\begin{equation} \label{eq:CKPineq}
    \int_{\Ra}\! \big|\rhota-\rhoKa\big|(x_1|x_2)\;\d x_1  \;\leq\; \bigg(2\, \int_{\Ra}\! \log\frac{\rhota}{\rhoKa}(x_1|x_2)\;\rhota(x_1|x_2)\;\d x_1\bigg)^{\frac{1}{2}} \,,
\end{equation}
then squaring and integrating w.r.t. $\rhotb(x_2)\;\d x_2$ on both sides gives
\begin{equation} \label{eq:CKPineq2}
    \int_{\Rb}\! \left(\int_{\Ra}\! \big|\rhota-\rhoKa\big|(x_1|x_2)\;\d x_1\right)^{\!2}\,\rhotb(x_2)\,\d x_2 \;\leq\; 2\,\Da(t,\lambda) \;. 
\end{equation}
In order to be able to apply the latter inequality, we split $J_3$ in three terms and use the uniform integrability of $|\nablay V|\,$. Let $\epsilon>0$ and let us split the r.h.s. of \eqref{eq:nablaV-nablaF} in two parts:
\begin{equation} \label{eq:T1+T2+T3}
\begin{split}
    &\big|\langle\nablay V\rangle_{t,\lambda}-\nablay F\big|(x_2) \,\leq\, 
    \int_{\big\{|\nablay V(x)|^2\leq\epsilon^{-1}\big\}}\! |\nablay V(x)|\;\big|\rhota-\rhoKa\big|(x_1|x_2)\;\d x_1 \;+\\[2pt]
    &\phantom{\big|\langle\nablay V\rangle_{t,\lambda}-\nablay F\big|(x_2) \,\leq\,}+ \int_{\big\{|\nablay V(x)|^2>\epsilon^{-1}\big\}}\! |\nablay V(x)|\;\big(\rhota(x_1|x_2)+\rhoKa(x_1|x_2)\big)\;\d x_1 \leq \\[2pt]
    &\leq\, \epsilon^{-\frac{1}{2}} \int_{\Ra}\! \big|\rhota-\rhoKa\big|(x_1|x_2)\;\d x_1\,+\, \epsilon^{\frac{1}{2}}\int_{\Ra}\! |\nablay V(x)|^2\,\big(\rhota(x_1|x_2)+\rhoKa(x_1|x_2)\big)\;\d x_1 \,.
\end{split}
\end{equation}
Taking the square on both sides of  \eqref{eq:T1+T2+T3} and integrating w.r.t. $\rhotb(x_2)\,\d x_2$,  the Csiz\'ar-Kullback-Pinsker inequality \eqref{eq:CKPineq2} together with Jensen inequality give
\begin{equation} 
    J_3(t,\lambda) \,\leq\, 4\epsilon^{-1}\Da(t,\lambda) \,+\;
    4\epsilon\int_{\R^d}|\nablay V(x)|^4\;\big(\rhot(x)\,+\,\rhoKa(x_1|x_2)\rhotb(x_2)\big)\;\d x \,.
\end{equation}
By Theorems \ref{th:expectation_norm_lapla}, \ref{th:condexpectation_norm} we know that
\begin{equation}
    \sup_{t>0,\,\lambda\geq1}\int_{\R^d}|\nablay V(x)|^4\,\big(\rhot(x)\,+\,\rhoKa(x_1|x_2)\rhotb(x_2)\big)\;\d x \,\equiv\, \tilde c_0 \,<\infty 
\end{equation}
hence
\begin{equation} \label{eq:J3bound}
    J_3(t,\lambda) \,\leq\, 4\,\epsilon^{-1}\,\Da(t,\lambda) \,+\, 4\,\tilde c_0\,\epsilon \;.
\end{equation}
Finally, plugging estimates \eqref{eq:J1bound}, \eqref{eq:J3bound} into \eqref{eq:J1+J3} we obtain
\begin{equation}
    \lambda\;\frac{\d}{\d t}\Db(t,\lambda) \,\leq -\,(1-\eta)\,2c_2\,\Db(t,\lambda)\,+\, \frac{\beta_2}{\eta}\,\epsilon^{-1}\Da(t,\lambda) \,+\, \frac{\beta_2\,\tilde c_0}{\eta}\,\epsilon \;.
\end{equation}
Renaming $\beta_2^{-1}\epsilon\equiv\epsilon'\,$, and $\beta_2^2\,\tilde c_0\equiv \tilde c_0'$ we find the desired inequality \eqref{eq:dt_marginal}.
\end{proof}

\begin{proof}[Proof of Theorem \ref{th:main}. Second part]
Inequality \eqref{eq:main_marginal} follows by Theorems \ref{th:conditional}, \ref{th:marginal} using an extended Gronwall lemma for systems of differential inequalities (see Lemma \ref{lem:GronwallSYS} in the Appendix).
Inequalities \eqref{eq:dt_conditional}, \eqref{eq:dt_marginal} rewrite as:
\begin{equation}
    \frac{\d}{\d t}\begin{pmatrix}
    \Da(t,\lambda) \\[2pt] \Db(t,\lambda)
    \end{pmatrix} \,\preceq\, -\, C_\lambda\, \begin{pmatrix}
    \Da(t,\lambda) \\[2pt] \Db(t,\lambda)
    \end{pmatrix} + B_\lambda
\end{equation}
for a.e. $t>0\,$, where $\preceq$ denotes componentwise inequality and
\begin{equation}
    C_\lambda\,\equiv\, \begin{pmatrix}
    2\,c_1 & 0 \\[4pt] 
    -\frac{1}{\eta\,\epsilon\,\lambda} & \frac{(1-\eta)\,2c_2}{\lambda}
    \end{pmatrix} \quad,\quad 
    B_\lambda\,\equiv\, \begin{pmatrix}
    \frac{c_0}{\lambda} \\[4pt] \frac{\epsilon\,\tilde c_0}{\eta\,\lambda}
    \end{pmatrix} \;.
\end{equation}
Therefore by Lemma \ref{lem:GronwallSYS} in the Appendix, since the maps $t\mapsto\Da(t,\lambda)\,$, $t\mapsto\Db(t,\lambda)$ are absolutely continuous on compact subsets of $[0,\infty)$ (Propositions \ref{prop:KLdissipation}, \ref{prop:KL2dissipation}), we have
\begin{equation} \label{eq:proofthmain_appo}
    \begin{pmatrix}
    \Da(t,\lambda) \\[2pt] \Db(t,\lambda)
    \end{pmatrix} \,\preceq\,
    e^{-t\,C_\lambda} \begin{pmatrix}
    \Dai \\[2pt] \Dbi
    \end{pmatrix} \,+\, \big(I_2-e^{-t\,C_\lambda}\big)\, C_\lambda^{-1}B_\lambda
\end{equation}
for all $t>0\,$.
For $\lambda\geq (1-\eta)\,\frac{c_2}{c_1}$ standard computations show that
\begin{equation}
    e^{-t\,C_\lambda} \,=\, \begin{pmatrix}
    e^{-2c_1 t} & 0 \\[4pt] \frac{e^{-(1-\eta)\,2c_2\,t/\lambda}\,-\,e^{-2c_1 t}}{\eta\,\epsilon\,2\,(c_1\lambda\,-\,(1-\eta)\,c_2)} & e^{-(1-\eta)\,2c_2\,t/\lambda}
    \end{pmatrix}
\end{equation}
\begin{equation}
    C_\lambda^{-1}B_\lambda \,=\, \begin{pmatrix}
    \frac{c_0}{2c_1\lambda} \\[4pt] \frac{c_0}{\eta\,\epsilon\,\lambda\,4(1-\eta)c_1c_2} + \frac{\epsilon\,\tilde c_0}{\eta\,2(1-\eta)c_2}
    \end{pmatrix}
\end{equation}
hence the second component of inequality \eqref{eq:proofthmain_appo} gives the desired inequality \eqref{eq:main_marginal}.
\end{proof}

\begin{remark}
Since any positive power of $|\nablay V|$ is integrable (Theorems \ref{th:expectation_norm_lapla}, \ref{th:condexpectation_norm}), inequality \eqref{eq:dt_marginal} holds true also if we replace the term $\tilde c_0\,\epsilon$ therein by $\tilde c_0(r)\,\epsilon^r$ for any $r>0\,$.
Indeed the proof of Theorem \ref{th:marginal} can be modified using the following upper bound in \eqref{eq:T1+T2+T3} :
\begin{equation} \begin{split}
\int_{\big\{|\nablay V(x)|^2>\epsilon^{-1}\big\}}\! |\nablay V(x)|\;\big(\rhota(x_1|x_2)+\rhoKa(x_1|x_2)\big)\;\d x_1 \\
\leq\,\epsilon^{\frac{r}{2}}\,\int_{\Ra}\! |\nablay V(x)|^{r+1}\,\big(\rhota(x_1|x_2)\,+\,\rhoKa(x_1|x_2)\big)\;\d x_1\;.
\end{split}\end{equation}
Then following the same proof one obtains the desired result with
\begin{equation}
    \tilde c_0(r) \,\equiv\, \beta_2^2
    \sup_{t>0,\,\lambda\geq1}\int_{\R^d}|\nablay V(x)|^{2r+2}\,\big(\rhot(x)+\rhoKa(x_1|x_2) \rhotb(x_2)\big)\;\d x \,<\infty \;.
\end{equation}
As a consequence also inequality \eqref{eq:main_marginal} in Theorem \ref{th:main} modifies replacing the term $R_3(t,\lambda,\eta)\,\epsilon\,$ therein by $R_3(t,\lambda,\eta,r)\,\epsilon^r\,$, 
\begin{equation}
    R_3\!\left(t,\lambda,\eta,r\right) \,\equiv\, \big(1-e^{-(1-\eta)\,2c_2\,t/\lambda}\big)\, \frac{\tilde c_0(r)}{(1-\eta)\,2c_2} \;.
\end{equation}
\end{remark}

\section{Appendix} \label{sec:appendix}

\subsection{Gronwall-type inequalities}

The first Subsection of the Appendix is devoted to Gronwall-type inequality for the Fokker-Planck equation.
In particular we show how the expectation of an observable $v$ with respect to $\rhot$ can be controlled by means of a simple \textit{Lyapunov condition} on $\Ldual v$, in the spirit of Section 7.1 in \cite{BKRS}.
A generalization of Gronwall inequality to systems of differential inequalities is also discussed.
These results are extensively used in the next Subsections. 

Let us start by the following
\begin{proposition} \label{prop:GronwallL0}
Let $v\in C^2(\Rd)\cap L^1(\rhoi)\,$, 
$C_0,\,C_1\in\R$ such that for all $x\in\Rd$
\begin{equation} \label{eq:GronwallL_bound}
    \Ldual v(x) \,\leq\, C_0 \,+\, C_1\, v(x) \;.
\end{equation}
Suppose that for all $T>0$
\begin{equation} \label{eq:finite_int} \begin{split}
    &\sup_{t\in(0,T)}\,\int_{\R^d}|v(x)| \,\rhoT(x)\,\d x \,<\infty \ , \quad \sup_{t\in(0,T)}\,\int_{\R^d}|\Ldual v(x)|\, \rhoT (x)\,\d x \,<\infty \ ,\\
    &\sup_{t\in(0,T)}\,\int_{\R^d}|\nabla v(x)|\, \rhoT(x)\,\d x \,<\infty \;.
\end{split} \end{equation}
Then for all $t>0$ we have:
\begin{equation} \label{eq:GronwallL_thesis}
    \int_{\R^d} v(x)\,\rhoT(x)\,\d x \,\leq\, e^{C_1 t}\int_{\R^d} v(x)\,\rhoi(x)\,\d x \,+\, \left(e^{C_1 t}-1\right)\,\frac{C_0}{C_1} \;.
\end{equation}
\end{proposition}

\begin{proof}
Consider a sequence $\psi_N\in C^2(\R^d)$ such that
\begin{equation}
\psi_N(x)\,=\,\begin{cases}
\,1 & \textrm{for }|x|\leq N\\
\,0 & \textrm{for }|x|\geq N+1
\end{cases}
\end{equation}
and $\psi_N\,$, $\nabla\psi_N\,$, $\Hess\psi_N$ are uniformly bounded. We have $\varphi_N\equiv v\,\psi_N\in C^2_c(\R^d)$, hence by identity \eqref{eq:FPdef}
\begin{equation} \label{eq:FPN}
    \int_{\R^d}\varphi_N(x)\rhoT(x)\,\d x \,- \int_{\R^d}\varphi_N(x)\rhoi(x)\,\d x \;=\; \int_0^t\!\int_{\R^d} \Ldual\varphi_N(x)\rhoS(x)\,\d x\,\d s \;.
\end{equation}
Now,
\begin{equation}
    \Ldual\varphi_N \,=\, \Ldual v\;\psi_N \,+\, v\,\Ldual\psi_N \,+\, 2\,\nabla v\cdot(\Lambda\beta)^{-1}\nabla\psi_N 
\end{equation}
hence hypothesis \eqref{eq:finite_int} guarantees that there is dominated convergence in \eqref{eq:FPN}. Precisely letting $N\to\infty$ we find
\begin{equation}
    \int_{\R^d}v(x)\rhoT(x)\,\d x \,- \int_{\R^d}v(x)\rhoi(x)\,\d x \;=\; \int_0^t\!\int_{\R^d} \Ldual v(x)\rhoS(x)\,\d x\,\d s \;.
\end{equation}
As a consequence $t\mapsto\int_{\R^d}v(x)\rhoT(x)\,\d x$ is absolutely continuous on compact subsets of $[0,\infty)$ and there exists
\begin{equation}
    \frac{\d}{\d t}\int_{\R^d}v(x)\rhoT(x)\,\d x
    \;=\, \int_{\R^d}\Ldual v(x)\rhoT(x)\,\d x
\end{equation}
for a.e. $t>0\,$.
Then by hypothesis \eqref{eq:GronwallL_bound} we have
\begin{equation}
    \frac{\d}{\d t}\int_{\R^d}v(x)\rhoT(x)\,\d x
    \;\leq\; C_0 + C_1 \int_{\R^d} v(x)\rhoT(x)\,\d x \;.
\end{equation}
Thesis \eqref{eq:GronwallL_thesis} finally follows applying Gronwall inequality (see Lemma \ref{lem:Gronwall}).
\end{proof}

The previous result can be extended to observables $v$ that are not known to be integrable a priori:
\begin{proposition} \label{prop:GronwallL}
Let $v\in C^2(\Rd)\cap L^1(\rhoi)\,$, $C_0,\,C_1\in\R$ such that inequality \eqref{eq:GronwallL_bound} holds true for all $x\in\R^d\,$.
Suppose $v\geq0$ and
\begin{equation}
    v(x)\to\infty \ \ \textrm{as } |x|\to\infty \;.
\end{equation}
Then inequality \eqref{eq:GronwallL_thesis} holds true for all $t>0\,$.
\end{proposition}

Proposition \ref{prop:GronwallL} is essentially a reinterpretation of Theorem 7.1.1 in \cite{BKRS} for the Fokker-Planck operator $\Ldual$ defined in \eqref{eq:Ldual}.
Note in particular that thanks to the regularity of our setting, we do not need any assumption on the signs of $C_0,\,C_1$ unlike in \cite{BKRS}.

\begin{proof}
Let $\zeta_N\in C^2([0,\infty))$ such that
\begin{equation}
    \zeta_N(r) \,=\, \begin{cases}
    \,r & \textrm{for }r\leq N-1 \\[2pt]
    \,N & \textrm{for }r\geq N+1
    \end{cases}\;,
\end{equation}
$0\leq\zeta_N'\leq1$ and $\zeta_N''\leq0\,$. 
Observe that $\varphi_N\,\equiv\,\zeta_N\circ v - N\,\in C^2_c(\Rd)$ since $v(x)\to\infty$ as $|x|\to\infty$, hence applying identities \eqref{eq:FPdef_continuity}-\eqref{eq:FPdef2} to $\varphi_N$ we find:
\begin{equation} \label{eq:GronwallL_fp2}
    \frac{\d}{\d t}\,\int_{\R^d} \zeta_N(v(x)) \rhoT(x)\,\d x
    \,=\, \int_{\R^d} \Ldual(\zeta_N\circ v)(x) \rhoT(x)\,\d x 
\end{equation}
for all $t>0$ and
\begin{equation} \label{eq:GronwallL_fpcont}
    \int_{\R^d} \zeta_N(v(x)) \rhoT(x)\,\d x \,\xrightarrow[t\to0]{}\, \int_{\R^d} \zeta_N(v(x)) \rhoi(x)\,\d x \;.
\end{equation}
%
Now observe that, since $\nabla (\zeta_N\circ v) = (\zeta_N'\circ v)\,\nabla v$ and $\Hess(\zeta_n\circ v) = (\zeta_N''\circ v)\,\nabla v\,\nabla^T\!v \,+\, (\zeta_N'\circ v)\Hess v\,$, by concavity of $\zeta_N$ we have
\begin{equation} \label{eq:GronwallL_appo}
    \Ldual(\zeta_N\circ v) \,\leq\, (\zeta_N'\circ v)\,\Ldual v \;.
\end{equation}
Using the properties of $\zeta_N$ one can check that $\zeta_N'(r)\,r\leq\zeta_N(r)$ for all $r\geq0$. Therefore using hypothesis \eqref{eq:GronwallL_bound} it follows that
\begin{equation} \label{eq:GronwallL_appo2}
    \Ldual(\zeta_N\circ v) \,\leq\, C_0 \,+\, C_1\, (\zeta_N\circ v) \;.
\end{equation}
Substituting estimate \eqref{eq:GronwallL_appo2} into \eqref{eq:GronwallL_fp2}  we find
\begin{equation} \label{eq:GronwallL_appo3}
    \frac{\d}{\d t}\,\int_{\R^d} \zeta_N(v(x)) \rhoT(x)\,\d x \;\leq\; C_0\,+\, C_1\,\int_{\R^d} \zeta_N(v(x)) \rhoT(x)\,\d x \;.
\end{equation}
Therefore applying Gronwall inequality (see Lemma \ref{lem:Gronwall}), from \eqref{eq:GronwallL_appo3}, \eqref{eq:GronwallL_fpcont} we obtain
\begin{equation}
    \int_{\R^d} \zeta_N(v(x)) \rhoT(x)\,\d x \;\leq\; e^{C_1t}\,\int_{\R^d} \zeta_N(v(x)) \rhoi(x)\,\d x \,+\, \left(e^{C_1t}-1\right)\,\frac{C_0}{C_1} \;.
\end{equation}
Finally, the thesis \eqref{eq:GronwallL_thesis} follows by monotone convergence letting $N\to\infty\,$.
\end{proof}

\begin{remark} \label{rem:GronwallL}
The hypothesis of Propositions \ref{prop:GronwallL0}, \ref{prop:GronwallL} can be weakened. Instead of bound \eqref{eq:GronwallL_bound}, it suffices to assume:
\begin{equation}
    \Ldual v(x) \,\leq\, u_0(x) \,+\, C_1\, v(x)
\end{equation}
where the function $u_0\in \cap_{t>0}L^1(\rhoT)\,$ is such that
\begin{equation}
    \sup_{t>0}\int_{\R^d} u(x) \rhoT(x)\,\d x \,\leq\, C_0 \;.
\end{equation}
\end{remark}

For completeness, let us briefly recall the classical Gronwall inequality used in the previous proofs. 

\begin{lemma}[Gronwall inequality] \label{lem:Gronwall}
Let $f\!:[0,\infty)\to\R$ be absolutely continuous on compact sets. Let $a,b\!:[0,\infty)\to\R$ continuous such that
\begin{equation} \label{eq:Gronwall_ip}
    f'(t) \,\leq\, a(t)\,f(t) \,+\, b(t)
\end{equation}
for almost every $t>0$.
Then setting
\begin{equation}
    e(t) \,\equiv\, \exp\int_0^t a(s)\,\d s \;,
\end{equation}
we have for all $t\geq0$
\begin{equation} \label{eq:Gronwall}
    f(t) \,\leq\, e(t)\,f(0) \,+\, e(t) \int_0^t \frac{b(s)}{e(s)}\,\d s \;.
\end{equation}
In particular if $a,b$ are constant, inequality \eqref{eq:Gronwall} becomes:
\begin{equation} \label{eq:Gronwall_const}
    f(t) \,\leq\, e^{a t}\,f(0) \,+\, \big(e^{a t}-1\big)\,\frac{b}{a} \;.
\end{equation}
\end{lemma}

The next lemma deals with a system of differential inequalities, extending the classical Gronwall inequality. This result was essentially due to \cite{Wazewski}. A proof can be obtained also by adapting Lemma E.4 in \cite{Terrell} to the two-dimensional case with two absolutely continuous functions. 
The symbol $\preceq$ will denote a componentwise inequality, namely we write $(u_1,u_2)\preceq(v_1,v_2)$ for ``$\,u_1\leq v_1$ and $u_2\leq v_2\,$''.

\begin{lemma}[Comparison lemma for systems of differential inequalities] \label{lem:GronwallSYS}
Let $f_1,\,f_2\!:[0,\infty)\to\R$ be absolutely continuous functions on compact sets; let $f\equiv(f_1,f_2)\,$. Let $A\equiv(A_1,A_2):\R\times\R^2\to\R^2$ be a Lipschitz continuous function such that
\begin{equation} \label{eq:GronwallSYS_hyp}
    f'(t) \,\preceq\, A\big(t,f(t)\big)
\end{equation}
for almost every $t>0$.
Suppose that:
\begin{itemize}
    \item[i.] the map $\xi_2\mapsto A_1(t,\xi_1,\xi_2)$ is non-decreasing on $\R$, for every $(t,\xi_1)\in[0,\infty)\times\R\,$;
    \item[ii.] the map $\xi_1\mapsto A_2(t,\xi_1,\xi_2)$ is non-decreasing on $\R$, for every $(t,\xi_2)\in[0,\infty)\times\R\,$.
\end{itemize}
Then we have
\begin{equation}\label{eq:GronwallSYS}
    f(t) \,\preceq\, g(t)
\end{equation}
for all $t\geq0\,$, where $g$ is the unique solution of the Cauchy problem
\begin{equation}
    \begin{cases}\label{eq:GronwallSYS_Cauchy}
    \,g'(t) = A(t,g(t)) \\[2pt]
    \,g(0) = f(0)
    \end{cases} \;.
\end{equation}
In particular if $A$ is linear and time-independent, namely $A(t,\xi)=A\,\xi+B$ for a suitable $2\times2$ real matrix $A\equiv\begin{pmatrix}A_{11} & A_{12} \\ A_{21} & A_{22} \end{pmatrix}$ with $A_{12},A_{21}\geq0$ and a suitable vector $B\equiv\begin{pmatrix}B_1 \\ B_2 \end{pmatrix}\in\R^2$, then inequality \eqref{eq:GronwallSYS} becomes:
\begin{equation} \label{eq:GronwallSYS_linconst}
    f(t) \,\preceq\, e^{tA}\,f(0) \,+\, \big(e^{tA}-I_2\big)\, A^{-1} B \;.
\end{equation}
\end{lemma}

\subsection{Expectations of polynomials with respect to $\rhot\,$: proof of Theorem \ref{th:expectation_norm_lapla}}
\label{subsec:proof_expectations}
Assumption \textup{(A2)} provides polynomial bounds on the derivatives of $V$. Theorem \ref{th:expectation_norm_lapla} will follow if we show that any positive power of the variables $|x_1|\,$, $|x_2|$ has uniformly bounded expectations in the measure $\rhot$ for all $t>0\,$, $\lambda\geq1\,$.
Because of the different time scales in the evolution of the variables $x_1,\,x_2$, we have to start by the following auxiliary 

\begin{proposition} \label{prop:expectation_vlambda}
For $x=(x_1,x_2)\in\R^d$ let
\begin{equation}
    v_{\lambda}(x) \;\equiv\; \frac{1}{\lambda}\,|x_1|^{2} +\, |x_2|^{2} \;.
\end{equation}
In the hypothesis of Theorem \ref{th:existence_uniq_reg}, for every $r\in[1,\infty)$ there exists a finite non-negative constant $M_{2r}=M_{2r}(\beta_1,\beta_2,\rhoi,V)$ such that for all $t>0$, $\lambda\geq1$
\begin{equation} \label{eq:Phip_statement}
   \int_{\R^d} v_{\lambda}(x)^r\rhot(x)\;\d x \;\leq\, M_{2r} \;.
\end{equation}
In particular by assumption \textup{(A4)} there exist $a\in(0,\infty)$, $a_0\in[0,\infty)$ such that for every $x\in\Rd$
\begin{equation}
    x\cdot\nabla V(x) \,\geq\, a\,|x|^2 - a_0 \;,
\end{equation}
then a suitable choice for $M_{2r}$ is given by
\begin{equation} \label{eq:M} \begin{split}
    M_{2r} \,\equiv\,
    \int_{\R^d}|x|^{2r}\rhoi(x)\,\d x
    \,+\,\frac{2^r (r-1)^{r-1}}{a^r r^r}\,\bigg(a_0\,+\,\frac{d_1}{\beta_1}+\frac{d_2}{\beta_2}+\frac{2(r-1)}{\beta_1\land\beta_2}\bigg)^{\!r}\;.
\end{split} \end{equation}
\end{proposition}


\begin{proof}
We are going to apply Proposition \ref{prop:GronwallL} to $v_\lambda^r$.
Computing first and second order derivatives one finds
\begin{equation} \label{eq:vlambda_L} \begin{split}
    \Ldual_\lambda\left(v_\lambda(x)^r\right) \,&=\, -\,\frac{2r}{\lambda}\;v_\lambda(x)^{r-1}\,
    x \cdot\nabla V(x) 
    \,+\, \frac{2r}{\lambda}\;v_\lambda(x)^{r-1}\, \bigg(\frac{d_1}{\beta_1}+\frac{d_2}{\beta_2}\bigg) \; +\\[4pt]
    &\quad\;+\, \frac{2r}{\lambda}\; v_\lambda(x)^{r-2}\;2(r-1)\, 
    \bigg(\frac{|x_1|^2}{\lambda\beta_1}+\frac{|x_2|^2}{\beta_2}\bigg)\\[8pt]
    &\leq\, -\,\frac{2r}{\lambda}\; v_\lambda(x)^{r-1}\left(
    x\cdot\nabla V(x) \,-\, k_r\right)\;,
\end{split} \end{equation}
where we set $k_r\equiv \frac{d_1}{\beta_1}+\frac{d_2}{\beta_2}+\frac{2(r-1)}{\beta_1\land\beta_2}\,$.
Using assumption \textup{(A4)} and taking $\lambda\geq1$ we have for all $x\in\R^d$
\begin{equation}
    x\cdot\nabla V(x)  \;\geq\; a\,|x|^2 
    - a_0 \;\geq\; a\,v_\lambda(x) - a_0 \;.
\end{equation}
Hence:
\begin{equation} \label{eq:vlambda_Lbound} \begin{split}
    \Ldual_\lambda \left(v_\lambda(x)^r\right) \;&\leq\, -\,\frac{2r}{\lambda}\;v_\lambda(x)^{r-1}\,\big(a\, v_\lambda(x) -a_0 -k_r\big) \\[4pt]
    &\leq\, -\,\frac{2r}{\lambda}\, \bigg(\frac{a}{2}\; v_\lambda(x)^r - m_r\bigg)
\end{split} \end{equation}
where we set $-m_r\equiv\min_{\xi\geq0}\left(\frac{a}{2}\,\xi^r-(a_0+k_r)\,\xi^{r-1}\right) = -\big(\frac{2\,(r-1)}{a}\big)^{r-1}\big(\frac{a_0+k_r}{r}\big)^r$.
Finally, from inequality \eqref{eq:vlambda_Lbound} and Proposition \ref{prop:GronwallL} it follows that for all $t>0\,$, $\lambda\geq1$
\begin{equation} \begin{split}
    \int_{\R^d} v_\lambda(x)^r\rhot(x)\,\d x \;&\leq\; 
    e^{-\frac{ra\,t}{\lambda}}\int_{\R^d} v_\lambda(x)^r\rhoi(x)\,\d x \;+\, \big(1-e^{-\frac{r\alpha \,t}{2\lambda}}\big)\,\frac{2\,m_r}{a} \\
    &\leq\; \int_{\R^d} |x|^{2r} 
    \rhoi(x)\,\d x \;+\, \frac{2\, m_r}{a} \;. 
\end{split} \end{equation}
\end{proof}

\begin{corollary} \label{cor:expectation_y}
Let $r\in[0,\infty)\,$. In the hypothesis of Theorem \ref{th:existence_uniq_reg} we have:
\begin{equation}
    \sup_{t>0,\,\lambda\geq1}\,\int_{\R^d} 
    |x_2|^r \rhot(x)\;\d x \;\leq\; M_r \;, 
\end{equation}
where for $r\geq2\,$
$M_{r}$ is defined by Proposition \ref{prop:expectation_vlambda}, while for $r<2$ we set $M_r\equiv1-\frac{r}{2}+\frac{r}{2}\,M_2\,$.
\end{corollary}

\begin{proof}
It follows by Proposition \ref{prop:expectation_vlambda} since
\begin{equation}
    |x_2|^r \leq\, v_\lambda(x_1,x_2)^\frac{r}{2}
\end{equation}
for all $(x_1,x_2)\in\Rd$. In addition if $r\in[0,2)$, one has $\xi^r\leq 1-\frac{r}{2}+\frac{r}{2}\,\xi^2$ for all $\xi\geq0\,$.
\end{proof}

\begin{corollary} \label{cor:expectation_xt}
Let $r\in[0,\infty)\,$.
In the hypothesis of Theorem \ref{th:existence_uniq_reg} we have for all $\lambda\geq1$
\begin{equation}
    \sup_{t>0}\;\int_{\R^d} 
    |x_1|^r\rhot(x)\;\d x \;\leq\; \lambda\;M_r \;,
\end{equation}
where $M_{r}$ is defined as in Corollary \ref{cor:expectation_y}.
\end{corollary}

\begin{proof}
It follows by Proposition \ref{prop:expectation_vlambda} since
$
|x_1|^r \leq \lambda\,v_\lambda(x_1,x_2)^\frac{r}{2} \,$.
\end{proof}

\begin{proposition} \label{prop:expectation_x}
In the hypothesis of Theorem \ref{th:existence_uniq_reg} plus \textup{(A5)}, for every $r\in[0,\infty)$ there exists a finite non-negative constant $M_r'=M_r'(\beta_1,\beta_2,\rhoi,V)$ such that
\begin{equation} \label{eq:x_bound}
    \sup_{t>0,\,\lambda\geq1}\,\int_{\R^d} 
    |x_1|^r\rhot(x)\;\d x \;\leq\; M_r' \;.
\end{equation}
In particular by assumption \textup{(A5)} there exist $a_1\in(0,\infty)$, $a_0,a_2,p\in[0,\infty)$ such that for every $(x_1,x_2)\in\Rd$ 
\begin{equation}
    x_1\cdot\nablax V(x_1,x_2) \,\geq\, a_1\,|x_1|^2 - a_2\,|x_2|^p -a_0 \;,
\end{equation}
then a suitable choice for $M_r'$ is given by:
\begin{equation}
    M_r' \,\equiv\, \begin{cases} \,\int_{\R^d}|x_1|^r\rhoi(x)\,\d x \,+\, \frac{1}{(r-1)\,a_1}\,\big(2\,a_2\,M_{p(r-1)}+r\,m_r\big) & \textrm{if }r>2 \\[8pt]
    \,\int_{\R^d}|x_1|^2\rhoi(x)\,\d x \,+\, \frac{1}{a_1}\,\big(a_2\,M_{p}+a_0+\frac{d_1}{\beta_1}\big) & \textrm{if }r=2 \\[8pt]
    1-\frac{r}{2}+\frac{r}{2}\,M'_2 & \textrm{if }r<2
    \end{cases}\;,
\end{equation}
where $M_{p(r-1)}$ is defined by Corollary \ref{cor:expectation_y} and $m_r\equiv-\min_{\xi\geq0}\big(\frac{a_1}{2}\,\xi^r - a_2\,\frac{r-2}{r-1}\,\xi^{r-1} - (a_0+\frac{d_1+r-2}{\beta_1})\,\xi^{r-2}\big)\,$.
\end{proposition}

\begin{proof}
Let $r>2$.
We want to apply Proposition \ref{prop:GronwallL0} to \begin{equation}
   v(x)\,\equiv\, |x_1|^r \;.  
\end{equation} 
First of all observe that $v$ satisfies the integrability hypothesis \eqref{eq:finite_int} thanks to Corollary \ref{cor:expectation_xt} and assumption \textup{(A2)} for $\nabla V$.
Now, computing first and second derivatives we have
\begin{equation}
    \Ldual v(x) \,=\, -r\,|x_1|^{r-2}
    \left(x_1\cdot\nablax V(x) - k_r\right)\;,
\end{equation}
setting $k_r\equiv\frac{d_1+r-2}{\beta_1}\,$.
By assumption \textup{(A5)}
\begin{equation}
    x_1\cdot\nablax V(x) \,\geq\, a_1\,|x_1|^2 - a_2\,|x_2|^p - a_0 \;,
\end{equation}
hence
\begin{equation} \label{eq:Lbound_expx0}
    \Ldual v(x) \,\leq -r\,
    \big(a_1\,|x_1|^r - a_2\,|x_1|^{r-2}\,|x_2|^p - (a_0+k_r)\,|x_1|^{r-2} \big) \;.
\end{equation}
Young's inequality guarantees that
\begin{equation}
    |x_1|^{r-2}\,|x_2|^p \,\leq\, \frac{1}{\sigma}\,|x_1|^{(r-2)\,\sigma} + \frac{1}{\tau}\,|x_2|^{p\,\tau}
\end{equation}
for every $\sigma,\tau>1\,$, $\sigma^{-1}+\tau^{-1}=1\,$. For example, $\sigma\equiv\frac{r-1}{r-2}$ ensures $(r-2)\,\sigma<r$ and enforces $\tau\equiv r-1\,$. For this choice of $\sigma,\tau$ we obtain:
\begin{equation} \label{eq:Lbound_expx} \begin{split}
    \Ldual v(x) \,&\leq -r\,
    \bigg(a_1\,|x_1|^r - a_2\,\frac{r-2}{r-1}\,|x_1|^{r-1} - \frac{a_2}{r-1}\,|x_2|^{p(r-1)} - (a_0+k_r)\,|x_1|^{r-2} \bigg) \\[4pt]
    &\leq\, -r \,    \bigg(\frac{a_1}{2}\,|x_1|^r - \frac{a_2}{r-1}\,|x_2|^{p(r-1)} - m_r \bigg)
\end{split} \end{equation}
where we set $-m_r\equiv\min_{\xi\geq0}\big(\frac{a_1}{2}\,\xi^r - a_2\,\frac{r-2}{r-1}\,\xi^{r-1} - (a_0+k_r)\,\xi^{r-2}\big)\,$. 
By Corollary \ref{cor:expectation_y} we know that for all $t>0\,$, $\lambda\geq1$
\begin{equation}
    \int_{\R^d}|x_2|^{p(r-1)}\rhot(x_1,x_2)\,\d x_1\,\d x_2 \;\leq\, M_{p(r-1)} \;.
\end{equation}
Therefore by Proposition \ref{prop:GronwallL0} and Remark \ref{rem:GronwallL} we have for all $t>0\,$, $\lambda\geq1$ 
\begin{equation} \begin{split}
    \int_{\R^d}|x_1|^r\rhot(x)\,\d x \;&\leq\; e^{-\frac{r\,a_1}{2}t} \int_{\R^d}|x_1|^r\rhoi(x)\,\d x \,+\, \big(1-e^{-\frac{r\,a_1}{2}t}\big)\,\Big(\frac{2a_2\,M_{p(r-1)}}{(r-1)\,a_1}+\frac{2\,m_r}{a_1}\Big) \\[4pt]
    &\leq\; \int_{\R^d}|x_1|^r\rhoi(x)\,\d x \,+\, \frac{2a_2\,M_{p(r-1)}}{(r-1)\,a_1}+\frac{2\,m_r}{a_1}
\end{split} \end{equation}
which concludes the proof in the case $r>2$.
If $r=2$, Young inequality is not needed and inequality \eqref{eq:Lbound_expx0} suffices to apply Proposition \ref{prop:GronwallL0} and Remark \ref{rem:GronwallL}, obtaining for all $t>0\,$, $\lambda\geq1$
\begin{equation} \begin{split}
    \int_{\R^d}|x_1|^2\rhot(x)\,\d x \;&\leq\; e^{-2a_1\,t} \int_{\R^d}|x_1|^2\rhoi(x)\,\d x \,+\, \big(1-e^{-2a_1\,t}\big)\,\frac{a_2\,M_{p}+k_2'}{a_1} \\[4pt]
    &\leq\; \int_{\R^d}|x_1|^2\rhoi(x)\,\d x \,+\, \frac{a_2\,M_{p}+k_2'}{a_1}\;.
\end{split} \end{equation}
Finally, if $r<2$ we can just use the bound $|x_1|^r\leq 1-\frac{r}{2}+\frac{r}{2}|x_1|^2$ and come back to the previous case.
\end{proof}

\begin{proof}[Proof of Theorem \ref{th:expectation_norm_lapla}]
It follows by combining assumption \textup{(A2)} with Corollary \ref{cor:expectation_y} and Proposition \ref{prop:expectation_x}.
\end{proof}

\begin{proof}[Proof of Proposition \ref{prop:integrability_log}]
We rely on Corollary 7.3.8 in \cite{BKRS} that ensures 
\begin{equation} \label{eq:rho_bounded}
    \|\rho\|_{L^\infty(\R^d\!\times(0,T))} \,<\infty  \;,
\end{equation}
provided $\rhoi$ is bounded on $\R^d$ (assumption \textup{(B1)}) and
\begin{equation}
    \int_0^T \!\!\int_{\R^d} |\nabla V(x)|^{d+3}\rhoT(x)\,\d x\,\d t \,<\infty\:,
\end{equation}
which in turn follows from Corollaries \ref{cor:expectation_y}, \ref{cor:expectation_xt} together with assumption \textup{(A2)}.
We also make use of the inequality $|\!\log\xi|^2\,\xi\leq \sqrt[4]{\xi}\,$ for $\xi\in[0,1]\,$. 
Therefore:
\begin{equation} \label{eq:logint_appo1}
\int_{\{\rho_t\geq1\}} |\log\rhoT(x)|^2\,\rhoT(x)\,\d x \;\leq\,  \left(\,\log\|\rho\|_{L^\infty(\R^d\!\times(0,T))}\right)^2 \;;
\end{equation}
\begin{equation} \label{eq:logint_appo2}
\int_{\{\rho_t\leq\, e^{-|x|}\}} |\log\rhoT(x)|^2\,\rhoT(x)\,\d x \;\leq\, \int_{\R^d} e^{-\frac{|x|}{4}}\,\d x  \;;
\end{equation}
\begin{equation} \label{eq:logint_appo3}
\int_{\{e^{-|x|}<\,\rho_t<1\}} |\log\rhoT(x)|^2\,\rhoT(x)\,\d x \;\leq\, \int_{\R^d} |x|^2 \rhoT(x)\,\d x  \;.
\end{equation}
The r.h.s. of \eqref{eq:logint_appo1} is finite by \eqref{eq:rho_bounded}. 
The r.h.s. of \eqref{eq:logint_appo3} has finite supremum over $t\in(0,T)$ by Corollaries \ref{cor:expectation_y}, \ref{cor:expectation_xt}. This concludes the proof.
\end{proof}

\begin{proof}[Proof of Theorem \ref{th:integrability_gradlog}]
We refer to Theorem 7.4.1 in \cite{BKRS}. It ensures that \eqref{eq:integrability_gradlog} is a consequence of two integrability conditions:
\begin{align}
    \int_0^T \!\!\int_{\R^d} |\nabla V(x)|^2\,\rhoT(x)\,\d x\,\d t \,&<\infty\;,\\[4pt]
    \int_0^T \!\!\int_{\R^d} \log^2\max(1,|x|)\,\rhoT(x)\,\d x\,\d t \,&<\infty
\end{align}
which in turn follow from Corollaries \ref{cor:expectation_y}, \ref{cor:expectation_xt}.
\end{proof}

\begin{proof}[Proof of Proposition \ref{prop:integrability_log2}]
Since $\rhob$ is solution of a suitable Fokker-Planck equation with drift $\frac{1}{\lambda}\,\langle\nablay V\rangle_t$ (Theorem \ref{th:FP_marg}), Corollary 7.3.8 in \cite{BKRS} ensures that 
\begin{equation} \label{eq:rho2_bounded}
    \|\rhob\|_{L^\infty(\Rb\!\times(0,T))} \,<\infty  \;,
\end{equation}
provided $\rhoib$ is bounded on $\Rb$ (assumption \textup{(B2)}) and
\begin{equation}
    \int_0^T \!\!\int_{\Rb} \big|\langle\nablay V\rangle_t(x_2)\big|^{d+3}\rhoTb(x_2)\;\d x_2\,\d t \,<\infty\:,
\end{equation}
which is true by Corollaries \ref{cor:expectation_y}, \ref{cor:expectation_xt} and Jensen inequality.
The proof is then concluded by miming the Proof of Proposition \ref{prop:integrability_log} above.
%
\end{proof}

\begin{proof}[Proof of Theorem
\ref{th:integrability_gradlog2}]
Being $\rhob$ solution of a suitable Fokker-Planck equation, we may refer to Theorem 7.4.1 in \cite{BKRS}. The latter ensures that \eqref{eq:integraibility_gradlog2} follows from two integrability conditions:
\begin{align}
    \int_0^T \!\!\int_{\Rb}\! \big|\langle\nablay V\rangle_t(x_2)\big|^2\,\rhoTb(x_2)\,\d x_2\,\d t\,&<\infty \;,\\ 
    \int_0^T \!\!\int_{\Rb}\! \log^2\max(1,|x_2|)\,\rhoTb(x_2)\,\d x_2\,\d t \,&<\infty 
\end{align}
which in turn follow from Corollaries \ref{cor:expectation_y}, \ref{cor:expectation_xt}.
\end{proof}

\subsection{Expectations of polynomials with respect to $\rhoKa\rhotb\,$: proof of Theorem \ref{th:condexpectation_norm}} \label{subsec:proof_condexpectations}
Assumption \textup{(A2)} ensures polynomial bounds for the derivatives of $V$.
We show that the conditional expectation of a polynomial in $|x_1|$ with respect to the measure $\rhoKa(x_1|x_2)$ is bounded by a polynomial in $|x_2|$. Then from Subsection \ref{subsec:proof_expectations} we already know that this quantity has uniformly bounded expectations with respect to the measure $\rhotb$ for all $t>0\,$, $\lambda\geq1\,$, hence we can prove Theorem \ref{th:condexpectation_norm}. 
In particular the assertion about derivatives of the effective potential $\Veff$ follows as they can be expressed in terms of conditional expectation of products of derivatives of $V$.

\begin{proposition} \label{prop:condexpectation_poly}
Suppose that $V$ verifies assumptions \textup{(A2)} for $\nu=0$ and \textup{(A3)}.
Then for every $r\in[0,\infty)$ there exist $s_r=s_r(V)\in [0,\infty)$ and two finite non-negative constants $C_{0,r}=C_{0,r}(\beta_1,V)\,$, $C_{1,r}=C_{1,r}(V)$ such that for every $x_2\in\Rb$
\begin{equation} \label{eq:condexpectation_poly}
    \int_{\Ra} |x_1|^r\,\rhoKa(x_1|x_2)\;\d x_1 \;\leq\; C_{0,r} + C_{1,r} \,|x_2|^{s_r}\;.
\end{equation}
In particular by assumptions \textup{(A2)} for $\nu=0$ and \textup{(A3)} there exist $a_1,a_2\in(0,\infty)$, $a_0\in[0,\infty)$, $m_1,m_2\in[2,\infty)$, $\gamma_0,\gamma_1,\gamma_2\in[0,\infty)$ such that for every $(x_1,x_2)\in\Rd$
\begin{equation} \label{eq:V_ub}
    a_1\,|x_1|^2 + a_2\,|x_2|^2 - a_0 \,\leq\, V(x_1,x_2) \,\leq\, \gamma_1\,|x_1|^{m_1} + \gamma_2\,|x_2|^{m_2} + \gamma_0\;,
\end{equation}
then a suitable choice for $s_r,\,C_{0,r},\,C_{1,r}$ is given by:
\begin{equation}
    s_r \,\equiv\, r\,\frac{m_2}{2} \;,\quad
    C_{1,r} \,\equiv\, \bigg(\frac{2\gamma_2}{a_1}\bigg)^{\!\frac{r}{2}} \,,\quad
    C_{0,r} \,\equiv\, \frac{(\beta_1\gamma_1)^{\frac{1}{m_1}}\,\Gamma(\frac{r+1}{2})}{2\,\big(\frac{\beta_1a_1}{2}\big)^{\!\frac{r+1}{2}}\,\Gamma(1+\frac{1}{m_1})}\,e^{\beta_1(a_0+\gamma_0)}\;.
\end{equation}
\end{proposition}

\begin{proof}
Let $\Delta>0$ and  consider the set
\begin{equation} \label{eq:Ay}
    A_{x_2} \,\equiv\, \left\{ x_1\in\Ra \Big|\; |x_1|^2 \geq \Delta\,|x_2|^{m_2} \right\} \;.
\end{equation}
for every $x_2\in\Rb\,$.
Integrating over its complementary set we have
\begin{equation}
\label{eq:int_Ac} 
    \int_{\Ra\setminus A_{x_2}}\!\! |x_1|^r\,\rhoKa(x_1|x_2)\;\d x_1 \;\leq\; \Delta^\frac{r}{2}\, |x_2|^{\frac{m_2}{2}\,r} \;.
\end{equation}
Now we evaluate the contribution of the integral over $A_{x_2}\,$. By assumption \textup{(A3)}
\begin{equation} \label{eq:int_Anum}
\begin{split}
    \int_{A_{x_2}} \!|x_1|^r\,e^{\,-\beta_1\,V(x_1,x_2)}\,\d x_1 \,&\leq\, \int_{A_{x_2}}\! |x_1|^r\,e^{\,-\beta_1\,(a_1\,|x_1|^2 \,+\, a_2\,|x_2|^2  \,-\, a_0)} \,\d x_1 \\[4pt]
    &\leq\,
    I_r\  e^{\,-\beta_1\,\left(\frac{a_1}{2}\,\Delta\,|x_2|^{m_2}\,+\,a_2\,|x_2|^2\right)} 
\end{split} \end{equation}
where we set
\begin{equation} \begin{split}
I_r \,\equiv\, \int_{\Rb} |x_1|^r\,e^{\,-\beta_1\,\left(\frac{a_1}{2}\,|x_1|^2 \,-\, a_0\right)}\,\d x_1 \,=\, \frac{|S_{d_1-1}|}{2}\;\frac{\Gamma\big(\frac{r+1}{2}\big)}{\big(\frac{\beta_1a_1}{2}\big)^{\frac{r+1}{2}}}\;e^{\beta_1a_0} \;.
\end{split} \end{equation}
On the other hand, by assumption \textup{(A2)} with $\nu=0$,
\begin{equation} \label{eq:int_Aden}
\begin{split}
\int_{\Ra} e^{\,-\beta_1\,V(x_1,x_2)}\,\d x_1 \,&\geq\, 
\int_{\Ra} e^{\,-\beta_1\,\left(\gamma_1\,|x_1|^{m_1}\,+\,\gamma_2\,|x_2|^{m_2}\,+\,\gamma_0\right)}\,\d x_1 \\[4pt]
&=\, J_r\ e^{\,-\beta_1\gamma_2\,|x_2|^{m_2}} 
\end{split} \end{equation}
where we set
\begin{equation}
    J_r \,\equiv\, \int_{\Ra} e^{\,-\beta_1\,\left(\gamma_1\,|x_1|^{m_1}\,+\,\gamma_0\right)}\,\d x_1 \,=\, |S_{d_1-1}|\; \frac{\Gamma\big(1+\frac{1}{m_1}\big)}{(\beta_1\gamma_1)^{-\frac{1}{m_1}}}\, e^{-\beta_1\gamma_0}\;.
\end{equation}
By inequalities \eqref{eq:int_Anum}, \eqref{eq:int_Aden} it follows that:
\begin{equation} \label{eq:int_A}
\begin{split}
\int_{A_{x_2}}|x_1|^r\,\rhoKa(x_1|x_2)\,\d x_1 \,&=\, \frac{\int_{A_{x_2}}|x_1|^r\,e^{\,-\beta_1\,V(x_1,x_2)}\,\d x_1}{\int_{\Ra}e^{\,-\beta_1\,V(x_1,x_2)}\,\d x_1} \\[4pt]
&\leq\, \frac{I_r}{J_r}\;e^{\,-\beta_1\,\left(\frac{a_1}{2}\,\Delta\,|x_2|^{m_2}\,+\,a_2\,|x_2|^2 \,-\,\gamma_2\,|x_2|^{m_2}\right)} \\[4pt]
&\leq\, \frac{I_r}{J_r}
\end{split}\end{equation}
where the last inequality holds true for any $\Delta>\frac{2\gamma_2}{a_1}\,$.
Finally, summing inequalities \eqref{eq:int_Ac}, \eqref{eq:int_A} concludes the proof.
\end{proof}

\begin{remark}
If the potential $V(x_1,x_2)$ grows faster than quadratically in $x_1$, the exponent $s_r$ in Proposition \ref{prop:condexpectation_poly} can be improved. 
Precisely if there are $n_1\in[2,\infty)\,$, $b_2,\,b_1,\,b_0\in[0,\infty)$ such that
\begin{equation}
    V(x_1,x_2) \,\geq\, b_1\,|x_1|^{n_1} - b_2\,|x_2|^{m_2} - b_0
\end{equation}
for all $(x_1,x_2)\in\R^d\,$, then inequality \eqref{eq:condexpectation_poly} holds true with
\begin{equation}
    s_r \,\equiv\, \frac{m_2}{n_1}\,r 
\end{equation}
changing also the constants $C_{1,r}$, $C_{0,r}$ accordingly. 
Indeed the proof of Proposition \ref{prop:condexpectation_poly} can be suitably modified, starting by considering the set
\begin{equation}
    \widetilde A_{x_2} \,\equiv\, \left\{ x_1\in\Ra \Big|\; |x_1|^{n_1} \geq \Delta\,|x_2|^{m_2} \right\} \;.
\end{equation}
\end{remark}

\begin{proof}[Proof of Theorem \ref{th:condexpectation_norm}]
Equation \eqref{eq:expectation_DV_rhoKrhot} follows combining assumption \textup{(A2)} with Proposition \ref{prop:condexpectation_poly} and Corollary \ref{cor:expectation_y}.\\
Equation \eqref{eq:expectation_DF} for $|\nu|=0$ follows from assumptions \textup{(A2), (A3)} which guarantee:
\begin{equation}
    a_1\,|x_1|^2 + a_2\,|x_2|^2 - a_0 \,\leq\, V(x_1,x_2) \,\leq\, \gamma_1 |x_1|^{r_1} + \gamma_2 |x_2|^{r_2} + \gamma_0 \;.
\end{equation}
Therefore:
\begin{equation} \label{eq:Veff_bounds}
    \Veff(x_2) \,=\, -\frac{1}{\beta_1}\log \int_{\Ra}\!e^{-\beta_1 V(x_1,x_2)}\,\d x_1\, \begin{cases}
    \,\leq\, C_0 + \gamma_2\,|x_2|^{r_2} \\[4pt]
    \,\geq\, C_1 + a_2\,|x_2|^2 
    \end{cases}
\end{equation}
where $C_0\equiv\gamma_0-\frac{1}{\beta_1}\int_{\Ra}e^{-\beta_1\gamma_1|x_1|^{r_1}}\d x_1\,$, $C_1\equiv-a_0-\frac{1}{\beta_1}\int_{\Ra}e^{-\beta_1a_1|x_1|^2}\d x_1$ are finite constants. Inequalities \eqref{eq:Veff_bounds} combined with Corollary \ref{cor:expectation_y} ensure that
\begin{equation}
    \sup_{t>0,\lambda\geq1}\int_{\Rb}|F(x_2)|^s\,\rhotb(x_2)\,\d x_2 \,<\infty \;.
\end{equation}
Finally, equation \eqref{eq:expectation_DF} for $|\nu|\geq1$ is an application 
of the multivariate Fa\`a Di Bruno formula (see \cite{Faadibruno} and references therein):
\begin{equation} \label{eq:DF} \begin{split}
    D_{x_2}^\nu \Veff(x_2) \,=& \sum_{\pi\in\mathcal P([\nu])}\! \frac{(-1)^{|\pi|}(|\pi|-1)!}{\beta_1}\,\prod_{A\in\pi}\,\sum_{\sigma\in\mathcal P(A)} (-\beta_1)^{|\sigma|}\;\cdot\\[4pt]
    &\cdot\, \int_{\Ra}\prod_{B\in\sigma}D_{x_2}^{[B]}V(x_1,x_2)\,\rhoKa(x_1|x_2)\;\d x_1
\end{split} \end{equation}
where: for a multi-index $\nu=(\nu_1,\dots,\nu_{d_2})\in\N^{d_2}\setminus\{0\}\,$, $[\nu]$ denotes the set where each number from $1$ to $d_2$ is repeated in a distinguishable way according to its multiplicity encoded in $\nu$, i.e, $\big\{1',\dots,1^{(\nu_1)},\,\dots,\,d_2',\dots,d_2^{(\nu_{d_2})}\big\}\,$; conversely for a set $B\subseteq[\nu]\,$, $[B]$ denotes the multi-index $\big(|i:1^
{(i)}\in B|,\,\dots,\,|i:d_2^
{(i)}\in B|\big)\in\N^{d_2}\,$; $\mathcal P(A)$ denotes the set of partitions of a set $A$.
Assuming without loss of generality $s\geq1$, equation \eqref{eq:expectation_DF} for $|\nu|\geq1$ then follows from expressions \eqref{eq:DF} using Jensen inequality and equation \eqref{eq:expectation_DV_rhoKrhot}.
\end{proof}

\subsection{Regularity of $\rho\,$: proof of the regularity part of Theorem \ref{th:existence_uniq_reg}} \label{subsec:proof_regularity}

In this and all the following Subsections we will denote
\begin{equation}
    A \,\equiv\, (\Lambda\beta)^{-1} \,=\, \begin{pmatrix} \frac{1}{\beta_1}\,I_{d_1} & 0 \\ 0 & \frac{1}{\lambda\,\beta_2}\,I_{d_2} \end{pmatrix} \,,\quad 
    b(x)\,\equiv\, \Lambda^{-1}\,\nabla V(x) \,=\, \begin{pmatrix}\nablax V(x) \\ \frac{1}{\lambda}\,\nablay V(x) \end{pmatrix} \,.
\end{equation}
The Fokker-Planck operator \eqref{eq:Ldual_matrix} rewrites as
\begin{equation} \label{eq:Ldual_short}
    \Ldual \varphi \,=\, \nabla\cdot(A\,\nabla\varphi) \,-\, b\cdot\nabla\varphi \;.
\end{equation}
This compact notation is convenient due to the generality of results.

In the present Subsection we prove the regularity of $\rho$ following the argument sketched in \cite{JKO}. Instead of the standard heat kernel used there, we introduce
\begin{equation} \label{eq:G_def}
    G_t(x) \,\equiv\, \frac{1}{(4\pi\, t)^{\frac{d}{2}}\,(\det A)^\frac{1}{2}}\, \exp\bigg(\!\!-\frac{x^TA^{-1}x}{4t}\bigg)
\end{equation}
for $x\in\R^d\,$, $t>0\,$.
\begin{remark}
$G_t(x)$ is the fundamental solution of the diffusion equation
\begin{equation} \label{eq:G_diffusioneq}
    \partial_t G_t(x) \,=\, \nabla\cdot(A\,\nabla G_t(x))\;.
\end{equation}
Precisely we have:
\begin{align} \label{eq:Gnabla}
    &\nabla G_t(x) \,=\, -\frac{A^{-1}x}{2t}\,G_t(x)\;, \\[4pt]
    &\nabla\cdot(A\,\nabla G_t(x)) \,=\, \bigg(\!\!-\frac{d}{2t}+\frac{A^{-1}x}{4t^2}\bigg)\,G_t(x) \,=\, \partial_t G_t(x) \;.
\end{align}
Moreover, 
for all $f\in L^1(\mathbb R^d)$ we have:
\begin{equation}
\|f*G_\delta-f\|_{L^1(\R^d)}\,\xrightarrow[t\to0]\,0 \;,
\end{equation}
where $*$ denotes the convolution with respect to space variables only:
\begin{equation}
(f*G_t)(x) \,\equiv\, \int_{\R^d} G_t(x-y)\,f(y)\,\d y \;.
\end{equation}
\end{remark}

\begin{lemma} \label{lem:G_homo}
Let $p\geq1$. For all $t>0$ we have
\begin{align} \label{eq:G_homo}
    \|G_t \|_{L^p(\R^d)} \,&=\, t^{-\frac{d}{2}\,(p-1)}\,\|G_1\|_{L^p(\R^d)} \;, \\[4pt]
    \label{eq:Gnabla_homo}
    \|\nabla G_t \|_{L^p(\R^d)} \,&=\, t^{-\frac{d}{2}\,(1-\frac{1}{p})-\frac{1}{2}}\,\|\nabla G_1\|_{L^p(\R^d)} \;.
\end{align}
\end{lemma}

\begin{proof}
Direct computation using \eqref{eq:G_def}, \eqref{eq:Gnabla} and performing the change of variable $x'=x/\sqrt{t}$.
\end{proof}

\begin{lemma} \label{lem:LSU}
Let us denote
\begin{equation}
    (f \oast G)_{t\,}(x) \,\equiv\, \int_{-\infty}^{\,t} (f_s*G_{t-s})(x)\, \d s \;.
\end{equation}
Let $p>1$. There exist constants $c_1=c_1(p,d),\,c_2=c_2(p,d)\in(0,\infty)$ such that:
\begin{equation} \label{eq:LSU1}
    \|\partial_t(f\oast G)\|_{L^p(\R^d\times\R)} \,+\, 
    \sum_{i,j=1}^d \|\partial_{x_i}\partial_{x_j}(f\oast G)\|_{L^p(\R^d\times\R)} \,\leq\, c_1\, \|f\|_{L^p(\R^d\times\R)}
\end{equation}
for every $f\in L^p(\R^d\!\times\R)$, and:
\begin{equation} \label{eq:LSU2} 
    \|\partial_t(\phi* G)\|_{L^p(\R^d\times(0,\infty))} \,+ \sum_{i,j=1}^d \|\partial_{x_i}\partial_{x_j}(\phi* G)\|_{L^p(\R^d\times(0,\infty))} \,\leq\, c_2\, \sum_{i=1}^d \|\partial_{x_i}\phi\|_{L^p(\R^d)}
\end{equation}
for every $\phi\in W^{1,p}(\R^d)\,$.
\end{lemma}

\begin{proof}
We refer to \cite{LSU} Chapter IV Section 3, where the estimates (3.1), (3.2) coincide\footnote{Actually (3.2) in \cite{LSU} is expressed in terms of the Slobodeckij seminorm and is stronger than \eqref{eq:LSU2} here.}  respectively with \eqref{eq:LSU1}, \eqref{eq:LSU2} provided the matrix $A$ is replaced by the identity in the definition of $G$. 
The general case of a symmetric positive definite matrix $A$ can be deduced by performing the change of variable $y'=A^{-\frac{1}{2}}y$ in the convolution.
Notice that in the case of interest for the present paper $A$ is diagonal and the constants $c_1,c_2$ do not depend on its entries.
\end{proof}

\begin{remark} \label{rem:KLdiss}
The weak formulation \eqref{eq:FPdef} of FP equation extends to the following:
\begin{equation} \label{eq:FPdef_time}
\int_{\R^d}\varphi_t(y)\,\rhoT(y)\,\d y \,- \int_{\R^d}\varphi_{t_0}(y)\,\rho_{t_0}(y)\,\d y
\,= \int_{t_0}^t \int_{\R^d} \!\big(\partial_s\varphi_s \,+\, \Ldual\varphi_s\big)(y)\, \rhos(y)\,\d y\,\d s \;,
\end{equation}
for every $t>t_0>0\,$, for every $\varphi\in C^{2,1}(\Rd\times(0,\infty))\,$ such that $\supp \varphi_s \subset K$ compact subset of $\R^d$ for all $s>0$.
See, e.g., Proposition 6.1.2 and following remarks in \cite{BKRS}.
\end{remark}

\begin{proof}[Proof of Theorem \ref{th:existence_uniq_reg} part i]
Let $\eta\in C^\infty_c(\R^d)\,$, $\delta\in(0,1)\,$. Let $x\in\R^d$, $0<t_{-1}<t_0<t_1<t<T'<T$ and set
\begin{equation}
    \varphi_s(y) \,\equiv\,  G_{\delta+t-s}(x-y)\;\eta(y)
\end{equation}
for all $y\in\R^d$, $s\in(0,t]\,$. Since $\varphi\in C^\infty(\R^d\times(0,t])$, $\supp\varphi_s\subseteq\supp\eta\,$, identity \eqref{eq:FPdef_time} holds true.
Computing $\Ldual\varphi$ and using \eqref{eq:G_diffusioneq}, one obtains:
\begin{equation} \label{eq:FP_Otto} \begin{split}
\big((\eta\rhoT)*G_\delta\big)(x) \,=\; &\big((\eta\,\rho_{t_0})*G_{\delta+t-t_0}\big)(x) \;+ \\[4pt]
&+ \int_{t_0}^t \big((\Ldual\eta\,\rhos)*G_{\delta+t-s}\big)(x)\;\d s \;+\\[4pt]
&+ \int_{t_0}^t \big((\M\eta\,\rhos)*\nabla
G_{\delta+t-s}\big)(x)\;\d s \;,
\end{split} \end{equation}
setting
\begin{equation}
    \M\eta(y) \,\equiv\, -2A\,\nabla\eta(y) \,+\, b(y)\,\eta(y) \;.
\end{equation}
As $\delta\to0$ we have $(\eta\rhoT)*G_\delta\to\eta\rhoT$ in $L^1(\R^d)$ and also in $L^1(\R^d\times(t_0,T))$ by dominate convergence (this will be clear adapting the following argument with $p=1$). Thus:
\begin{equation} \label{eq:FP_Otto2} \begin{split}
(\eta\rhoT)(x) \,=\;\, &\big((\eta\,\rho_{t_0})*G_{t-t_0}\big)(x) \;\,+\\[4pt]
&+ \int_{t_0}^t \big((\Ldual\eta\,\rhos)*G_{t-s}\big)(x)\,\d s \;\,+\\[4pt]
&+ \int_{t_0}^t \big((\M\eta\,\rhos)*\nabla
G_{t-s}\big)(x)\,\d s
\end{split} \end{equation}
for almost every $(x,t)\in\R^d\times(t_0,T)\,$.\\
Now let $p\in(1,\frac{d}{d-1})\,$. Using Young's convolution inequality in \eqref{eq:FP_Otto2} we find:
\begin{equation} \begin{split}
\|\eta\rhoT\|_{L^p(\R^d)} \,\leq\;\, &\|\eta\;\rho_{t_0}\|_{L^1(\R^d)}\,\|G_{t-t_0}\|_{L^p(\R^d)} \;+ \\[4pt]
&+ \int_{t_0}^t \|\Ldual\eta\,\rhos\|_{L^1(\R^d)}\;\|G_{t-s}\|_{L^p(\R^d)}\;\d s \;+\\[4pt]
&+ \int_{t_0}^t \|\M\eta\,\rhos\|_{L^1(\R^d)}\;\|\nabla
G_{t-s}\|_{L^p(\R^d)}\;\d s \;.
\end{split} \end{equation}
Then, using Lemma \ref{lem:G_homo} 
we obtain:
\begin{equation} \label{eq:Otto00_appo1} \begin{split}
\|\eta\rhoT\|_{L^p(\R^d)} \,\leq\;\, &\|\eta\|_{\infty}\,\|G_{1}\|_{L^p(\R^d)}\, \big(t-t_0\big)^{-\frac{d}{2}\,(p-1)} \;+ \\[4pt]
&+\, \|\Ldual\eta\|_{\infty}\, \|G_{1}\|_{L^p(\R^d)}\; \frac{\big(t-t_0\big)^{1-\frac{d}{2}\,(p-1)}}{1-\frac{d}{2}\,(p-1)} \;+\\[4pt]
&+\, \|\M\eta\|_{\infty}\,\|\nabla
G_{1}\|_{L^p(\R^d)}\;\frac{\big(t-t_0\big)^{\frac{1}{2}-\frac{d}{2}\,(1-\frac{1}{p})}}{\frac{1}{2}-\frac{d}{2}\,(1-\frac{1}{p})} \;,
\end{split} \end{equation}
since our choice of $p$ guarantees $1-\frac{d}{2}\,(p-1)>0$ and $\frac{1}{2}-\frac{d}{2}\,(1-\frac{1}{p})>0\,$.
Observing that $\|f\|_{L^p(\R^d\times(t_1,T))}^p = \int_{t_1}^T\|f_t\|_{L^p(\R^d)}^p\d t\,$, inequality \eqref{eq:Otto00_appo1} shows that 
$\eta\,\rho\in L^p(\R^d\times(t_1,T))\,$. By arbitrarity of $\eta,\,t_1,\,T$, one concludes
\begin{equation} \label{eq:Otto00_appo4}
    \rho\in L^p_\textit{loc}(\R^d\times(0,\infty))
\end{equation}
(without knowing a priori the continuity of $\rho$). 
In \cite{JKO} a \textit{bootstrap argument} that uses equation \eqref{eq:FP_Otto2} iteratively together with Lemmas \ref{lem:G_homo}, \ref{lem:LSU} is invoked. Let us detail the first steps.
With the notation introduced in Lemma \ref{lem:LSU}, equation \eqref{eq:FP_Otto2} rewrites as
\begin{equation} \label{eq:FP_Otto3} \begin{split}
   (\eta\rhoT)(x) \,=\;\, & \big((\eta\,\rho_{t_0})*G_{t-t_0}\big)(x) \,+\, \big((\Ldual\eta\;\rho\,\chi)\oast G\big)_{t\,}(x) \;+\\[4pt]
   &+\,\sum_{k=0}^d\big((\M_k\!\eta\;\rho\,\chi)\oast \partial_{x_k} G\big)_{t\,}(x) \,+\, R_t(x)
\end{split} \end{equation}
where: $\chi\in C^\infty(\R)$ is such that $\chi_s=1$ for $t_0\leq s\leq T'\,$, $\chi_s=0$ for $s\leq t_{-1}$ or $s\geq T$, and we set
\begin{equation}
    R_t(x) \,\equiv\, \int_{t_{-\!1}}^{t_0} \big((\Ldual\eta\,\rhos\,\chi_s)*G_{t-s} \,+\, (\M\eta\,\rhos\,\chi_s)*\nabla G_{t-s} \big)(x)\,\d s \;,
\end{equation}
$\M\eta=(\M_k\!\eta)_{k=1}^d\,$, $x=(x_k)_{k=1}^d\in\R^d\,$.
We claim that $\partial_{x_i}(\eta\,\rho)$ for $i=1,\dots,d$ exist in $L^p_\textit{loc}(\R^d\times(t_1,T'))$.
The first and the last term on the r.h.s. of \eqref{eq:FP_Otto3}
cannot cause any problem because the kernel $G$ remains far from its singularity. Let us focus on the other terms. 
By Young's convolution inequality and Lemma \ref{lem:G_homo}, we have:
\begin{equation} \label{eq:Otto10_appo1} \begin{split}
    \big\|\big((\Ldual\eta\;\rho\,\chi)\oast \partial_{x_i}G\big)_t\big\|_{L^p(\R^d)} \,&\leq\,
    \int_{t_{-\!1}}^t \|(\Ldual\eta\,\rhos \chi_s)*\partial_{x_i}G_{t-s}\|_{L^p(\R^d)}\,\d s \\[4pt]
    &\leq\, 
    \|\Ldual\eta\|_{\infty}\, \|\partial_{x_i}G_1\|_{L^p(\R^d)}\, \frac{(t-t_{-1})^{\frac{1}{2}-\frac{d}{2}\,(p-1)}}{\frac{1}{2}-\frac{d}{2}\,(p-1)}\;,
\end{split} \end{equation}
hence $(\Ldual\eta\,\rho\,\chi)\oast \partial_{x_i}G\in L^p(\R^d\times(t_1,T'))$.
By Lemma \ref{lem:LSU}, we have:
\begin{equation} \label{eq:Otto10_appo2} \begin{split}
    \|(\M_k\!\eta\;\rho\,\chi)\oast \partial_{x_i}\partial_{x_k}G\|_{L^p(\R^d\times(t_1,T'))} \,&\leq\,
    c_1\, \| \M_k\!\eta\;\rho\,\chi \|_{L^p(\R^d\times\R)} \\[4pt]
    &\leq\, c_1\, \|\chi\|_\infty\,\| \M_k\!\eta\;\rho \|_{L^p(\R^d\times(t_{-1},T))} \;,
\end{split} \end{equation}
which is finite since we know \eqref{eq:Otto00_appo4}. 
Therefore \eqref{eq:Otto10_appo1}, \eqref{eq:Otto10_appo2} used in \eqref{eq:FP_Otto3} entail that it exists $\partial_{x_i}(\eta\,\rho) \in L^p(\R^d\times(t_1,T'))\,$. By arbitarity of $\eta,\,t_1,\,T'$ it follows
\begin{equation}
    \partial_{x_i}\rho\in L^p_\textit{loc}(\R^d\times(0,\infty))\;.
\end{equation}
Now we claim that $\partial_t(\eta\,\rho)$ exists in $L^p_\textit{loc}(\R^d\times(t_1,T'))$. 
By Lemma \ref{lem:LSU} we have:
\begin{equation} \label{eq:Otto20_appo1} \begin{split}
    \big\|\partial_t\big((\Ldual\eta\;\rho\,\chi)\oast G\big)\big\|_{L^p(\R^d\times(t_1,T'))} \,&\leq\, c_1\,\|\chi\|_\infty\,\|\Ldual\eta\,\rho\|_{L^p(\R^d\times(t_1,T'))} \;,
\end{split} \end{equation}
\begin{equation} \label{eq:Otto20_appo2} \begin{split}
    \big\|\partial_t\big((\M_k\!\eta\;\rho\,\chi)\oast \partial_{x_k} G\big)\big\|_{L^p(\R^d\times(t_1,T'))} \,&\leq\, c_1\,\|\chi\|_\infty\,\|\partial_{x_k}(\M_k\!\eta\,\rho)\|_{L^p(\R^d\times(t_1,T'))}\;,
\end{split} \end{equation}
which are finite by \eqref{eq:Otto00_appo4}, \eqref{eq:Otto10_appo2}. Using the previous inequalities in \eqref{eq:FP_Otto3} proves the claim. By arbitrarity of $\eta,\,t_1,\,T'$ it follows
\begin{equation}
    \partial_t\rho\in L^p_\textit{loc}(\R^d\times(0,\infty)) \;.
\end{equation}
In the same way one proves that $\partial_{x_i}\partial_{x_j}\rho\in L^p_\textit{loc}(\R^d\times(0,\infty))$. Iterating this argument up to space-time derivatives of any order (since the coefficients of $\Ldual$, $\M$ are smooth) one obtains
\begin{equation}
    \rho \in W^{r,p}_\textit{loc}(\R^d\times(0,\infty))
\end{equation}
for all $r\in\N\,$. 
Then general Sobolev inequalities entail $\rho\in C^{\infty}(\R^d\times(0,\infty))\,$, concluding the proof.
%
\end{proof}

\begin{proof}[Proof of Theorem \ref{th:existence_uniq_reg} part ii]
Let $t_0\to0$ in equation \eqref{eq:FP_Otto2}. By property \eqref{eq:FPdef_continuity} and uniform continuity of $G$ on compact subsets of $\R^d\times(0,\infty)$ we have:
\begin{equation} \label{eq:FP_Otto20} \begin{split}
(\eta\rhoT)(x) \,=\; &\big((\eta\rhoi)*G_{t}\big)(x) \,+\, \int_{0}^t \big((\Ldual\eta\,\rhos)*G_{t-s}\big)(x)\,\d s \;+\\[2pt]
&+ \int_{0}^t \big((\M\eta\,\rhos)*\nabla
G_{t-s}\big)(x)\,\d s\;.
\end{split} \end{equation}
Then Young's convolution inequality and Lemma \ref{lem:G_homo} entail
\begin{equation} \begin{split}
\|\eta\rhoT \,-\, (\eta\rhoi)*G_{t}\|_{L^1(\R^d)}  \,&\leq\, \|\Ldual\eta\|_\infty\, \|G_1\|_{L^1(\R^d)}\; t \,+\,
\|\M\eta\|_\infty\,\|\nabla
G_1\|_{L^1(\R^d)}\; 2\sqrt{t} \\[4pt]
&\xrightarrow[t\to0]{}\,0\;.
\end{split} \end{equation}
On the other hand $(\eta\rhoi)*G_t\to\eta\rhoi$ in $L^1(\R^d)$ as $t\to0\,$, hence
\begin{equation}
    \|\eta\rhoT \,-\, \eta\rhoi\|_{L^1(\R^d)} \,\xrightarrow[t\to0]{}\, 0\;.
\end{equation}
Now, consider a sequence $\eta_N\in C^\infty_c(\R^d)$ such that $\eta_N(x)=1$ for $|x|\leq N$, $0\leq\eta_N\leq1\,$. We have:
\begin{equation} \label{eq:Ottoconv_appo}
    \|\rhoT-\rhoi\|_{L^1(\R^d)} \,\leq\, \frac{S}{N} + \|\eta_N\rhoT\,-\,\eta_N\rhoi\|_{L^1(\R^d)} + \frac{S_0}{N} \;\xrightarrow[t\to0]{}\; \frac{S+S_0}{N}
\end{equation}
where:
\begin{equation}
    S \,\equiv\, \sup_{t\in(0,T)} \int_{\R^d}|x|\,\rhoT(x)\,\d x \;,\quad
    S_0 \,\equiv\, \int_{\R^d}|x|\,\rhoi(x)\,\d x
\end{equation}
are finite by Proposition \ref{prop:expectation_vlambda} and assumption \textup{(B3)} respectively.
Inequality \eqref{eq:Ottoconv_appo} concludes the proof by arbitrarity of $N\in\N\,$.
\end{proof}

\subsection{Regularity of $\rhob\,$: proof of Theorem \ref{th:regularity2}} \label{subsec:proof_regularity2}
In this Subsection the marginal $\rhoTb(x_2)=\int_{\Ra}\rhoT(x_1,x_2)\,\d x_1$ is shown to be in $C^\infty(\Rb\times(0,\infty))$. By Theorem \ref{th:existence_uniq_reg} we already know that $\rho$ is smooth, hence suitable pointwise estimates for its derivatives would be sufficient to conclude. Another way would be to appeal to the fact that $\rhob$ is itself weak solution of a Fokker-Planck equation (Theorem \ref{th:FP_marg}), but the drift coefficient of this equation involves $\langle\nablay V\rangle_t(x_2)=\int_{\Ra}\nablay V(x_1,x_2)\,\rhoT(x_1,x_2)\,\d x_1$ which a priori is also not known to be smooth.
We will modify the \textit{bootstrap argument} of the previous Subsection (sketched in \cite{JKO}), obtaining that both $\rhob$, $\langle\nablay V\rangle_t$ are smooth. Precisely we consider the class of functions $\eta\in\mathcal C$ defined by the following two conditions:
\begin{itemize}
    \item[i.] $\eta\in C^\infty(\R^d)\,$;
    \item[ii.] for every multi-index $\nu\in\N^d$ there exist $r_1,\,r_2,\,\gamma_0,\, \gamma_1\,,\gamma_2\in[0,\infty)$ such that
    \begin{equation}
        |D^\nu\eta(x_1,x_2)| \,\leq\, \gamma_0 + \gamma_1\,|x_1|^{r_1} + \gamma_2\,|x_2|^{r_2}
    \end{equation}
    for all $(x_1,x_2)\in\R^d\,$.
\end{itemize}
Setting
\begin{equation}
   (\I_t\eta)(x_2) \,\equiv\, \int_{\Ra}\eta(x_1,x_2)\,\rhoT(x_1,x_2)\,\d x_1\;,
\end{equation}
we will prove that
\begin{equation}
 \I\eta\in C^\infty(\Rb\!\times(0,\infty))   \quad\forall\,\eta\in \mathcal C \;.
\end{equation}
In particular it will follow that $\rhoTb(x_2)$ and $\langle\nablay V\rangle_t(x_2)$ are in $C^\infty(\Rb\!\times(0,\infty))$ (by taking respectively  $\eta\equiv1$, $\eta\equiv\nablay V$).

\begin{remark} \label{rem:bootstrap_Cclass}
In order to build a \textit{bootstrap argument}, it is essential to notice that the class of functions $\mathcal C$ is closed under sum, product and differentiation.
Moreover by assumptions \textup{(A1), (A2)} the coefficient $b=\Lambda^{-1}\nabla V$ - which defines the differential operators $\Ldual,\,\M$ (cf. \eqref{eq:Ldual_short}, \eqref{eq:M}) - belongs to $\mathcal C$.
We conclude that if $\eta\in \mathcal C$, then also $\Ldual\eta,\,\M_k\!\eta\in\mathcal C$ for all $k=1,\dots,d\,$.
\end{remark}

\begin{remark} \label{rem:bootstrap_Cintegr}
If $\eta\in\mathcal C$, by Proposition \ref{prop:expectation_vlambda} we know that
\begin{equation}
    \sup_{t\in(0,T)}\int_{\R^d}|\eta(x)|\,\rhoT(x)\,\d x\,<\infty \;.
\end{equation}
In particular $\I_t\eta$ is well defined and belongs to $L^1(\Rb)\,$.
\end{remark}
 
\begin{remark} \label{rem:bootstrap_Gprod}
Since the diffusion matrix $A=(\Lambda\beta)^{-1}$ is diagonal, the kernel $G$ introduced in \eqref{eq:G_def} writes as the product
\begin{equation}
    G_t(x_1,x_2) \,=\, G_t^\textup{\tiny(1)}(x_1)\;G_t^\textup{\tiny(2)}(x_2)\;,
\end{equation}
where
\begin{align}
    G_t^{\textup{\tiny(1)}}(x_1) \,&\equiv\, \bigg(\frac{\beta_1}{4\pi\,t}\bigg)^{\!\!\frac{d_1}{2}}\,\exp\bigg(\!\!-\frac{\beta_1}{4\,t}\;|x_1|^2\bigg) \;, \\[4pt]
    G_t^{\textup{\tiny(2)}}(x_1) \,&\equiv\, \bigg(\frac{\beta_2\,\lambda}{4\pi\,t}\bigg)^{\!\!\frac{d_2}{2}}\,\exp\bigg(\!\!-\frac{\beta_2\,\lambda}{4\,t}\;|x_2|^2\bigg) \;.
\end{align}
Clearly $\int_{\Ra}G_t^\textup{\tiny(1)}(x_1)\,\d x_1 = \int_{\Rb}G_t^\textup{\tiny(2)}(x_2)\,\d x_2 = 1$ for all $t>0\,$.
For $f,g\in L^1(\Rb)$ let us denote
\begin{equation}
    (f *_2 g)(x_2) \,\equiv\, \int_{\Rb} g(x_2-y_2)\,f(y_2)\,\d y_2 \;.
\end{equation}
If $\eta\in\mathcal C$, by Fubini theorem we get:
\begin{align}
    &\int_{\Ra} ((\eta\rhoT)*G_\tau)(x_1,x_2)\;\d x_1 \,=\, \big((\I_t\eta)\,*_2\, G_\tau^\textup{\tiny(2)}\big)(x_2) \;,\\[4pt]
    &\int_{\Ra} ((\eta\rhoT)*\partial_1 G_\tau)(x_1,x_2)\;\d x_1 \,=\, 0 \;,\\[4pt]
    &\int_{\Ra} ((\eta\rhoT)*\partial_2 G_\tau)(x_1,x_2)\;\d x_1 \,=\, \big((\I_t\eta)\,*_2\, \partial_2 G_\tau^\textup{\tiny(2)}\big)(x_2) \;,
\end{align}
where $\partial_1,\,\partial_2$ denote any partial derivative w.r.t. one variable in $x_1,\,x_2$ respectively.
\end{remark}

\begin{proof}[Proof of Theorem \ref{th:regularity2} part i.]
Let $\eta\in\mathcal C$. We claim that $\eta$ can be plugged into equation \eqref{eq:FP_Otto2}.
To prove this, consider a sequence $\psi_N\in C^\infty_c(\R^d)$ such that $\psi_N(x)=1$ for $|x|\leq N$ and $\psi_N,\,\nabla\psi_N,\,\Hess\psi_N$ are uniformly bounded for all $N\in\N$. Since $\eta\,\psi_N\in C^\infty_c(\R^d)$, it can be plugged into equation \eqref{eq:FP_Otto2} in place of $\eta$. Letting $N\to\infty$ our claim can be proved by dominate convergence using Young's convolution inequality, Remark \ref{rem:bootstrap_Cintegr}, and Lemma \ref{lem:G_homo}. 
Now, integrating equation \eqref{eq:FP_Otto2} with respect to $x_1\in\Ra$ and using Remark \ref{rem:bootstrap_Gprod} we obtain:
\begin{equation} \label{eq:FP_Ottomarg} \begin{split}
\I_t\eta (x_2) \,=\;\, &\big((\I_{t_0}\eta) \,*_2\, G_{t-t_0}^\textup{\tiny(2)}\big)(x_2) \;+ \\[4pt]
&+ \int_{t_0}^t \big((\I_s\Ldual\eta) \,*_2\, G_{t-s}^\textup{\tiny(2)}\big)(x_2)\;\d s \;\,+\\[4pt]
&+ \int_{t_0}^t \big((\I_s\M_2\!\eta) \,*_2\, \nablay\,
G_{t-s}^\textup{\tiny(2)}\big)(x_2)\;\d s
\end{split} \end{equation}
for almost every $(x_2,t)\in\Rb\!\times(t_0,T)$,
setting:
\begin{equation}
    \M_2\!\eta \,\equiv\, -\frac{2}{\beta_2\lambda}\,\nablay\eta \,+\, \frac{1}{\lambda}\,\eta\,\nablay V\;.
\end{equation}
Let $p\in(1,\frac{d_2}{d_2-1})$. Using Young's convolution inequality and Lemma \ref{lem:G_homo} (with $d_2,\,G^\textup{\tiny(2)}$ in place of $d,\,G$) into equation \eqref{eq:FP_Ottomarg} we find:
\begin{equation} \begin{split}
    \| \I_t\eta \,\|_{L^p(\Rb)} \,\leq\;\,& \|\I_{t_0}\eta\,\|_{L^1(\Rb)}\; \|G_1^\textup{\tiny(2)}\|_{L^p(\Rb)}\,(t-t_0)^{-\frac{d_2}{2}(p-1)} \,\;+\\[4pt]
    &+ \int_{t_0}^t \|\I_s\Ldual\eta\,\|_{L^1(\Rb)}\;\|G_1^\textup{\tiny(2)}\|_{L^p(\Rb)}\,(t-s)^{-\frac{d_2}{2}(p-1)}\,\d s \,\;+\\[4pt]
    &+ \int_{t_0}^t \|\I_s\M_2\!\eta\,\|_{L^1(\Rb)}\;\|\nablay G_1^\textup{\tiny(2)}\|_{L^p(\Rb)}\,(t-s)^{-\frac{d_2}{2}(1-\frac{1}{p})-\frac{1}{2}}\,\d s\;,
\end{split} \end{equation}
hence, by Remark \ref{rem:bootstrap_Cintegr} and our choice of $p$, we have
\begin{equation} \begin{split}
    \| \I_t\eta \,\|_{L^p(\Rb)} \,\leq\;\,& \|\I_{t_0}\eta\,\|_{L^1(\Rb)}\; \|G_1^\textup{\tiny(2)}\|_{L^p(\Rb)}\,(t-t_0)^{-\frac{d_2}{2}(p-1)} \,\;+\\[4pt]
    &+ \sup_{s\in(0,T)}\|\Ldual\eta\,\rhos\|_{L^1(\R^d)}\;\|G_1^\textup{\tiny(2)}\|_{L^p(\Rb)}\,\frac{(t-t_0)^{1-\frac{d_2}{2}(p-1)}}{1-\frac{d_2}{2}(p-1)} \,\;+\\[4pt]
    &+ \sup_{s\in(0,T)}\|\M_2\!\eta\,\rhos\|_{L^1(\Rb)}\;\|\nablay G_1^\textup{\tiny(2)}\|_{L^p(\Rb)}\,\frac{(t-t_0)^{\frac{1}{2}-\frac{d_2}{2}(1-\frac{1}{p})}}{\frac{1}{2}-\frac{d_2}{2}(1-\frac{1}{p})} 
\end{split} \end{equation}
and the r.h.s. is finite.  Since $\|\I\eta\|_{L^p(\R^d\times(t_1,T))}^p = \int_{t_1}^T\|\I_t\eta\|_{L^p(\R^d)}^p\d t\,$, it follows that
\begin{equation}
    \I\eta \in L^p(\Rb\!\times(t_1,T)) \;.
\end{equation}
We now invoke a \textit{bootstrap argument} that iteratively uses equation \eqref{eq:FP_Ottomarg} together with Lemmas \ref{lem:G_homo}, \ref{lem:LSU} (with $d_2,\,*_2,\,\oastb ,\,G^\textup{\tiny(2)}$ in place of $d,\,*,\,\oast,\,G$) to prove that
\begin{equation}
    \I\eta \in W^{r,p}(\Rb\!\times(t_1,T))
\end{equation}
for all $r\in\N\,$. We skip the details, which are similar to the previous Subsection: let us only notice that the here argument can be iterated up to space-time derivatives of any order thanks to Remarks \ref{rem:bootstrap_Cclass}, \ref{rem:bootstrap_Cintegr}.
Finally general Sobolev inequalities entail that $\I\eta\in C^\infty(\Rb\!\times(0,\infty))\,$.
\end{proof}

\begin{proof}[Proof of Theorem \ref{th:regularity2} part ii.]
It follows immediately by Theorem \ref{th:existence_uniq_reg}, since
$\|\rhoTb-\rhoib\|_{L^1(\Rb)} \leq \|\rhoT-\rhoi\|_{L^1(\R^d)}\,$.
\end{proof}

\subsection{Auxiliary results} \label{subsec:proof_KLdiss}

This last subsection of the Appendix is devoted to the proof of Propositions \ref{prop:KLdissipation}, \ref{prop:KL2dissipation} and the Lemmas used in the proof of Theorem \ref{th:conditional}.\\
The expression of entropy dissipation given by Proposition \ref{prop:KLdissipation} is essentially an application of the weak FP equation \eqref{eq:FPdef} for $\rho$ to the \textit{test function} $\varphi\equiv\log(\rho/\rhoK)\,$. This is possible thanks to the regularity and integrability results given by Theorems \ref{th:existence_uniq_reg}, \ref{th:expectation_norm_lapla}, \ref{th:integrability_gradlog}, and Proposition \ref{prop:integrability_log}.\\
Similarly, Proposition \ref{prop:KL2dissipation} gives an expression for the marginal entropy dissipation, relaying on the fact that the marginal $\rhob$ satisfies FP equation \eqref{eq:FP_marg}. This time one would like to choose $\varphi\equiv\log(\rhob/\rhoKb)$ as a test function. The proof is based on regularity and integrability results given by Theorems \ref{th:regularity2}, \ref{th:expectation_norm_lapla}, \ref{th:integrability_gradlog2}, and Proposition 
\ref{prop:integrability_log2}.

\begin{remark}
We will rewrite the weak formulation \eqref{eq:FPdef_time} of FP equation.
Let $\varphi\in C^{2,1}(\Rd\times(0,\infty))\,$ such that $\supp \varphi_t \subset K$ compact subset of $\R^d$ for all $t>0$.
The contribution in \eqref{eq:FPdef_time} coming from the diffusion term of $\Ldual$ can be integrated by parts:
\begin{equation} \label{eq:diffusion_byparts}
    \int_{\R^d} \nabla\cdot(A\,\nabla\varphi_s(x))\, \rhos(x)\,\d x \,=\, - \int_{\R^d} A\,\nabla\varphi_s(x)\cdot \nabla\!\rhos(x)\,\d x
\end{equation}
for almost every $s\in(t_0,t)\,$, since $\rho_s\in W^{1,1}(\Rd)$ by Theorem \ref{th:integrability_gradlog} and $\varphi_s\in C^2_c(\R^d)\,$.
Using \eqref{eq:Ldual_short}, \eqref{eq:diffusion_byparts} into \eqref{eq:FPdef_time} we obtain:
\begin{equation} \label{eq:FPdef_bi} \begin{split}
    &\int_{\R^d}\varphi_t(x)\,\rho_t(x)\,\d x \,- \int_{\R^d}\varphi_{t_0}(x)\,\rho_{t_0}(x)\,\d x \,=\\[4pt]
    &= \int_{t_0}^t \int_{\R^d} \big(\partial_s\varphi_s - \left(A\,\nabla\log\rhos+\,b\right)\cdot\nabla\varphi_s\big)(x)\; \rho_s(x)\;\d x\,\d s \,.
\end{split} \end{equation}
\end{remark}

\begin{proof}[Proof of Proposition \ref{prop:KLdissipation}] $\rho\in C^\infty(\R^d\!\times\!(0,\infty))$ by Theorem \ref{th:existence_uniq_reg}. Consider a sequence $\psi_N\in C^\infty_c(\R^d)$ such that $\psi_N(x)=1$ for $|x|\leq N$ and $\psi_N,\,\nabla\psi_N$ are uniformly bounded for all $N\in\N\,$. Set
\begin{equation}
    \varphi_{N,t}(x)\,\equiv\, \psi_N(x)\,\log\frac{\rhoT(x)}{\rhoK(x)} \;.
\end{equation}
By Theorem \ref{th:existence_uniq_reg} there exists a vanishing sequence $(t_n)_{n\in\N}$ such that $\rho_{t_n}(x)\to\rhoi(x)$ for a.e. $x\in\R^d$ as $n\to\infty\,$.
$\varphi_N\in C^\infty(\R^d\!\times\!(0,\infty))\,$, $\supp\varphi_{N,t}\subseteq\supp\psi_N$ compact subset of $\R^N$ for all $t>0\,$, therefore identity \eqref{eq:FPdef_bi} applies:
\begin{equation} \label{eq:KLdiss_appo1} \begin{split}
    &\int_{\R^d}\varphi_{N,t}(x)\,\rhoT(x)\,\d x \,- \int_{\R^d}\varphi_{N,t_n}(x)\,\rho_{t_n}(x)\,\d x \,=\\[4pt]
    &= \int_{t_n}^t \int_{\R^d} \big(\partial_s\varphi_{N,s} - \left(A\,\nabla\log\rhos+\,b\right)\cdot\nabla\varphi_{N,s}\big)(x)\; \rhos(x)\;\d x\,\d s
\end{split} \end{equation}
for every $n,\,N\in\N$.
We will let $n\to\infty$ first.
By Proposition \ref{prop:integrability_log} the sequences $|\log\rho_{t_n}|\,\rho_{t_n}\,$, $|\log\rhoK|\,\rho_{t_n}$ are uniformly bounded, hence by dominate convergence (as $\psi_N\in L^1(\R^d)$) we have:
\begin{equation}
    \int_{\R^d}\varphi_{N,t_n}(x)\,\rho_{t_n}(x)\,\d x \,\xrightarrow[n\to\infty]{}\, \int_{\R^d}\psi_N(x)\,\log\frac{\rhoi(x)}{\rhoK(x)}\,\rhoi(x)\,\d x \;.
\end{equation}
On the r.h.s. of \eqref{eq:KLdiss_appo1}, by Fubini theorem and fundamental theorem of Calculus we have:
\begin{equation} \begin{split}
    &\int_{t_n}^t\!\int_{\R^d}\partial_s\varphi_{N,s}(x)\;\rho_{s}(x)\,\d x\,\d s \,=\, \int_{t_n}^t\!\int_{\R^d}\psi_N(x)\,\partial_s\rho_{s}(x)\,\d x\,\d s \,=\\[4pt] &=\,\int_{\R^d}\psi_N(x)\,(\rhoT(x)-\rho_{t_n}(x))\,\d x \,\xrightarrow[n\to\infty]{}\, \int_{\R^d}\psi_N(x)\,(\rhoT(x)-\rhoi(x))\,\d x \;.
\end{split} \end{equation}
Now let $N\to\infty$. By Proposition \ref{prop:integrability_log} and Remark \ref{rem:integrability_logI} we have dominate convergence in the following:
\begin{equation}
\int_{\R^d}\varphi_{N,t}(x)\,\rhoT(x)\,\d x \,\xrightarrow[N\to\infty]{}\, \int_{\R^d}\log\frac{\rhoT(x)}{\rhoK(x)}\,\rhoT(x)\,\d x\;;
\end{equation}
\begin{equation}
\int_{\R^d}\psi_N(x)\,\log\frac{\rhoi(x)}{\rhoK(x)}\,\rhoi(x)\,\d x \,\xrightarrow[N\to\infty]{}\, \int_{\R^d}\log\frac{\rhoi(x)}{\rhoK(x)}\,\rhoi(x)\,\d x\;.
\end{equation}
Also:
\begin{equation}
    \int_{\R^d}\psi_N(x)\,(\rhoT(x)-\rhoi(x))\,\d x \,\xrightarrow[N\to\infty]{}\, \int_{\R^d}(\rhoT(x)-\rhoi(x))\,\d x \,=\, 0 \;.
\end{equation}
Finally:
\begin{equation} \begin{split}
    &\int_{0}^t\! \int_{\R^d} (A\,\nabla\log\rhos+\,b)(x)\cdot\nabla\varphi_{N,s}(x)\, \rhos(x)\,\d x\,\d s \;=\\[4pt]
    &=\,\int_{0}^t\! \int_{\R^d} (A\,\nabla\log\rhos+\,b)(x)\cdot\nabla\psi_N(x)\,\log\frac{\rhos(x)}{\rhoK(x)}\, \rhos(x)\,\d x\,\d s \,\ +\\[4pt]
    &\quad +\int_{0}^t\! \int_{\R^d} (A\,\nabla\log\rhos+\,b)(x)\cdot\nabla\log\frac{\rhos(x)}{\rhoK(x)}\;\psi_N(x)\, \rhos(x)\,\d x\,\d s \;\xrightarrow[N\to\infty]{}\\[4pt]
    &\xrightarrow[N\to\infty]{}\, 0\,+ \int_{0}^t\! \int_{\R^d} (A\,\nabla\log\rhos+\,b)(x)\cdot\nabla\log\frac{\rhos(x)}{\rhoK(x)}\, \rhos(x)\,\d x\,\d s \;,
\end{split}
\end{equation}
where dominate convergence is ensured by Theorems \ref{th:expectation_norm_lapla}, \ref{th:integrability_gradlog}, and Proposition \ref{prop:integrability_log}.
The previous computations show that letting $n\to\infty$ first and then $N\to\infty$ in \eqref{eq:KLdiss_appo1} we obtain:
\begin{equation} \begin{split}
    &\int_{\R^d}\log\frac{\rhoT(x)}{\rhoK(x)}\,\rhoT(x)\,\d x \,- \int_{\R^d}\log\frac{\rhoi(x)}{\rhoK(x)}\,\rhoi(x)\,\d x \,=\\[4pt]
    &= -\int_{0}^t\! \int_{\R^d} (A\,\nabla\log\rhos+\,b)(x)\cdot\nabla\log\frac{\rhos(x)}{\rhoK(x)}\; \rhos(x)\;\d x\,\d s
\end{split} \end{equation}
which concludes the proof.
\end{proof}

\begin{proof}[Proof of Proposition \ref{prop:KL2dissipation}]
By Theorem \ref{th:FP_marg}, $\rhob$ satisfies a Fokker-Planck equation.
In particular, for $\varphi\in C^{2,1}(\Rb\times(0,\infty))\,$ such that $\supp \varphi_t \subset K$ compact subset of $\Rb$, and for $t>t_0>0\,$,
identity \eqref{eq:FP_marg} extends to the following one:
\begin{equation} \begin{split}
&\int_{\Rb}\varphi_t(x_2)\,\rhoTb(x_2)\,\d x_2 \,- \int_{\Rb}\varphi_{t_0}(x)\,\rho_{t_0}^\textit{\tiny(2)}(x_2)\,\d x_2
\,=\\[4pt]
&=\, \int_{t_0}^t\int_{\Rb} \!\bigg(\partial_s\varphi_s\,+\, \frac{1}{\lambda}\,\langle\Ldualy\rangle_{s}\,\varphi_s\bigg)(x_2)\, \rhosb(x_2)\,\d x_2\,\d s 
\end{split} \end{equation}
(see Proposition 6.1.2 and following remarks in \cite{BKRS}).
Integrating by parts the diffusion term of $\langle\Ldualy\rangle_s$ as in Remark \ref{rem:KLdiss}, we obtain:
\begin{equation} \label{eq:FPmarg_bi} \begin{split}
    &\int_{\Rb}\varphi_t(x_2)\,\rhoTb(x_2)\,\d x_2 \,- \int_{\Rb}\varphi_{t_0}(x_2)\,\rho_{t_0}^\textit{\tiny(2)}(x_2)\,\d x_2 \,=\\[4pt]
    &= \int_{t_0}^t \int_{\Rb} \bigg(\partial_s\varphi_s - \frac{1}{\lambda}\,\bigg(\frac{1}{\beta_2}\,\nablay\log\rhosb\,+\,\langle\nablay V\rangle_s\bigg)\cdot\nablay\varphi_s\bigg)(x_2)\, \rhosb(x_2)\;\d x_2\,\d s \,.
\end{split} \end{equation}
Now let us consider a sequence $\psi_N\in C^\infty_c(\Rb)$ such that $\psi_N(x_2)=1$ for $|x_2|\leq N$ and $\psi_N,\,\nabla\psi_N$ are uniformly bounded for all $N\in\N\,$. Set
\begin{equation}
    \varphi_{N,t}(x_2)\,\equiv\, \psi_N(x_2)\,\log\frac{\rhoTb(x_2)}{\rhoKb(x_2)} \;.
\end{equation}
By Theorem \ref{th:regularity2} $\rhob\in C^\infty(\Rb\!\times\!(0,\infty))\,$, hence 
identity \eqref{eq:FPmarg_bi} applies to $\varphi\equiv\varphi_N\,$.
We let $t_0\to0$ along a suitable sequence such that $\rho_{t_0}\to\rhoi$ a.e. (Theorem \ref{th:regularity2}), then we let $N\to\infty$. Proceeding as in the previous Proof of Proposition \ref{prop:KLdissipation}, the integrability results of Theorems \ref{th:expectation_norm_lapla}, \ref{th:integrability_gradlog2}, and Proposition \ref{prop:integrability_log2} guarantee that:
\begin{equation} \begin{split}
    &\int_{\Rb}\log\frac{\rhoTb(x_2)}{\rhoKb(x_2)}\,\rhoTb(x_2)\,\d x_2 \,- \int_{\Rb}\log\frac{\rhoib(x_2)}{\rhoKb(x_2)}\,\rhoib(x_2)\,\d x_2 \,=\\[4pt]
    &= - \frac{1}{\lambda}\,\int_{0}^t\! \int_{\Rb}  \!\bigg(\frac{1}{\beta_2}\,\nablay\log\rhosb\,+\,\langle\nablay V\rangle_s\bigg)(x_2)\cdot\nablay\log\frac{\rhosb(x_2)}{\rhoKb(x_2)}\, \rhosb(x_2)\;\d x_2\,\d s 
\end{split} \end{equation}
concluding the proof.
\end{proof}

The following lemmas formalise the claims used in the proof of Theorem \ref{th:conditional}.

\begin{lemma} \label{lem:int_der}
Under the hypothesis of Theorem \ref{th:conditional}, for a.e. $x_2\in\R^2\,$, $t>0$
\begin{equation}
    \int_{\Ra} \nablay\rhoTa(x_1|x_2)\,\d x_1 \,=\, 0 \;.
\end{equation}
\end{lemma}

\begin{proof}
This is equivalent to prove that 
\begin{equation} \label{eq:appo1_intder}
    \int_{\Rb} \psi(x_2)\, \bigg[\int_{\Ra}\! \nablay\rhoTa(x_1|x_2)\,\d x_1\bigg] \,\d x_2 \,=\, 0 
\end{equation}
for all $\psi\in C^\infty_c(\Rb)$, a.e. $t>0\,$.
By Corollary \ref{cor:integrability_gradlog1}, $\rhoTa\in W^{1,1}\big(\R^d,\rhoTb\d x_1\d x_2\big)$ which is included in $ W^{1,1}\big(\Rb,\psi\,\d x_1\d x_2)\,$. Therefore by Fubini theorem and integration by parts we have
\begin{equation} \label{eq:appo2_intder} \begin{split}
    &\int_{\Rb} \!\psi(x_2)\, \int_{\Ra}\! \nablay\rhoTa(x_1|x_2)\,\d x_1 \,\d x_2 \,=\,
    \int_{\R^d} \psi(x_2)\, \nablay\rhoTa(x_1|x_2)\,\d x \,=\\[4pt]
    &=\,- \int_{\R^d} \nablay\psi(x_2)\, \rhoTa(x_1|x_2)\,\d x \,=\,
    -\int_{\Rb} \!\nablay\psi(x_2) \,\d x_2 \,.
\end{split} \end{equation}
since $\int_{\Ra}\! \rhoTa(x_1|x_2)\,\d x_1=1$ for every $x_2\in\Rb\,$. As $\psi$ has compact support, the r.h.s. of \eqref{eq:appo2_intder} vanishes, concluding the proof.
\end{proof}

\begin{lemma} \label{lem:div_int}
Under the hypothesis of Theorem \ref{th:conditional}, let $\Phi\!:\R^d\to\Rb$ such that $\Phi\in L^2(\rhoT)\,$, $\nablay\cdot\Phi\in L^1(\rhoT)\,$ for all $t>0\,$. Then for a.e. $t>0$
\begin{equation} \label{eq:div_int} \begin{split}
    &\int_{\Rb}\!\nablay\rhoTb(x_2)\,\cdot\,\int_{\Ra}\! \Phi(x)\,\rhoTa(x_1|x_2)\,\d x_1\;\d x_2 \,=\\[4pt]
    & =- \int_{\Rb}\!\rhoTb(x_2)\int_{\Ra}\!\!\big(\nablay\cdot\Phi(x)\,\rhoTa(x_1|x_2)
    \,+\, \Phi(x)\cdot\nablay\rhoTa(x_1|x_2)\big)\,\d x_1\,\d x_2 \;.
\end{split} \end{equation}
\end{lemma}

\begin{proof}
To shorten the notation set $f(x)\equiv\Phi(x)\rhoTa(x_1|x_2)\,$, $g(x)\equiv\rhoTb(x_2)\,$. Observe that $f\,g\,$, $f\cdot\nablay g\,$, $g\,\nablay\cdot f$ belong to $L^1(\R^d)\,$ for a.e. $t>0$, indeed:
\begin{equation} \begin{split}
    &\int_{\R^d}\!|f\cdot\nablay g|(x)\,\d x \,=\, \int_{\R^d}\!\big|\Phi(x)\cdot\nablay\log\rhoTb(x_2)\big|\,\rhoT(x)\,\d x \,\leq \\[4pt]
    &\leq \bigg(\int_{\R^d}\!|\Phi(x)|^2\rhoT(x)\,\d x\bigg)^{\!\!\frac{1}{2}} \bigg( \int_{\R^d}\!\big|\nablay\log\rhoTb(x_2)\big|^2\rhoT(x)\,\d x\bigg)^{\!\!\frac{1}{2}}\,<\infty \;;\\[4pt]
\end{split} \end{equation}
\begin{equation} \begin{split}
    &\int_{\R^d}\!|g\,\nablay\cdot f|(x)\,\d x \,=\, \int_{\R^d}\!\big|\nablay\cdot\Phi(x)\,\rhoTa(x_1|x_2) \,+\, \Phi(x)\cdot\nablay\rhoTa(x_1|x_2)\big|\,\rhoTb(x_2)\,\d x \,\leq\\[4pt]
    &\leq \int_{\R^d}\!|\nablay\cdot\Phi(x)|\,\rhoT(x)\,\d x \,+ \bigg(\int_{\R^d}\!|\Phi(x)|^2\rhoT(x)\,\d x\bigg)^{\!\!\frac{1}{2}} \bigg(\int_{\R^d}\big|\nablay\log\rhoTa(x_1|x_2)\big|^2\rhoT(x)\,\d x\bigg)^{\!\!\frac{1}{2}}\\[4pt]
    &<\infty \;.
\end{split} \end{equation}
Now consider a sequence $\varphi_N\in C^\infty_c(\R^d)$ such that $\varphi_N(x)=1$ for $|x|\leq N$ and $\varphi_N$, $\nablay\varphi_N$ are uniformly bounded for $N\in\N\,$. Integration by parts gives
\begin{equation}
    \int_{\R^d}\!\nablay\cdot(f\,g)(x)\;\varphi_N(x)\,\d x \,=\, - \int_{\R^d}\!(f\,g)(x)\cdot\nablay\varphi_N(x)\,\d x \;.
\end{equation}
Letting $N\to\infty$ by dominate convergence we obtain
\begin{equation}
    \int_{\R^d}\!\nablay\cdot(f\,g)(x)\,\d x \,=\, 0 \;,
\end{equation}
which rewrites as
\begin{equation}
    \int_{\R^d}\!(f\cdot \nablay g)(x)\,\d x \,=\,
    - \int_{\R^d}\!(g\,\nablay\cdot f)(x)\,\d x \;.
\end{equation}
Identity \eqref{eq:div_int} finally follows by Fubini theorem.
\end{proof}

\textbf{Acknowledgments.} It is a pleasure to thank Giulio Tralli, Antoine Bodin, Pierluigi Contucci, Jorge Kurchan, Alberto Parmeggiani, and Angelo Vulpiani for discussions on various aspects of this work. The work of D.A. was supported by Swiss National Science Foundation grant no 200020 182517. D.A. and E.M. are members of GNFM INdAM.

\bibliographystyle{unsrt}
\bibliography{refs}


\end{document}